\documentclass{article}

\usepackage{amsmath}
\usepackage{amsthm}
\usepackage{amssymb}
\usepackage{mathrsfs}
\usepackage{mathtools}
\numberwithin{equation}{section} 
\usepackage{slashed}
\usepackage{braket}
\usepackage{physics}
\usepackage{mdframed}
\usepackage{dsfont} % \mathds{C,1,R,.....}

\usepackage{graphicx,float}
\usepackage{pgf,tikz,pgfplots,pdfpages}
\pgfplotsset{compat=1.15}
\usepackage[compat=1.0.0]{tikz-feynman}
\usetikzlibrary{decorations.markings}
\usetikzlibrary{decorations.pathreplacing,calligraphy}
\usetikzlibrary{decorations.pathmorphing}
\usetikzlibrary{arrows}
\tikzset{snake it/.style={decorate, decoration=snake}}
\usepackage{circuitikz}
\usetikzlibrary{snakes}

\usepackage{xcolor,tcolorbox}

\usepackage[a4paper]{geometry}

\usepackage{fancybox} %for shadowing texts via\shadowbox{}
\usepackage{fancyhdr}
\usepackage{afterpage}
\usepackage{multicol}
\usepackage{tabularx}
\usepackage{cellspace}
\usepackage[toc,page]{appendix}
\usepackage{array}

\usepackage{hyperref}
\hypersetup{
    colorlinks,
    linkcolor={blue!50!black},
    citecolor={blue!50!black},
    urlcolor={blue!80!black}
}
\usepackage{url}

\usepackage{lmodern}
\usepackage{pifont}  % for check marks
\usepackage{relsize}
\usepackage{enumitem}
\usepackage{imakeidx}

\usepackage{comment}
\usepackage{MnSymbol}

\newcommand{\bvec}[1]{\mathbf{#1}} % define bold vectors
\newcommand{\tz}{\text{\large Tr}}
\newcommand{\pd}{\mathcal{D}}
\newcommand{\gz}{\mathcal{Z}}
\newcommand{\lag}{\mathcal{L}}
\renewcommand{\exp}{\text{exp}}
\newcommand{\eff}{\text{\tiny eff}}
\newcommand{\tl}{\text{\tiny TL}}
\newcommand{\cw}{\text{\tiny CW}}
\newcommand{\rings}{\text{\tiny rings}}
\newcommand{\aop}{\hat{\mathcal{A}}}
\newcommand{\zsym}{$\mathbb{Z}_2$}
\newcommand\numberthis{\addtocounter{equation}{1}\tag{\theequation}}

\title{On Finite Temperature Quantum Field Theory \\
  \vspace{2mm} \large From Theoretical Foundations To Electroweak Phase Transition}

\author{
Mohamed Aboudonia and Csaba Balazs \thanks{School of Physics and Astronomy, Monash University, Melbourne 3800 Victoria, Australia. \\
Emails: \texttt {mohamed.aboudonia@monash.edu},\qquad \texttt{csaba.balazs@monash.edu}}
}

\date{}

\begin{document}

\maketitle
\tableofcontents

\pagebreak

\begin{abstract} 
In the immediate aftermath of the Big Bang, the universe existed in an extremely hot, dense state in which particle interactions occurred not in vacuum but within a thermal medium. Under such conditions, the standard framework of quantum field theory (QFT) requires a finite-temperature extension, wherein propagators—and hence the fundamental structure of the theory—are modified to reflect thermal background effects. These thermal modifications are central to understanding the nature of electroweak symmetry breaking (EWSB) as a high-temperature phase transition, potentially leading to qualitatively different vacuum structures for the Higgs field as the universe cooled.
Finite-temperature corrections naturally regulate ultraviolet divergences in propagators, hinting at a possible route toward ultraviolet completion. However, these same thermal effects exacerbate infrared pathologies and can lead to imaginary contributions to the effective potential, particularly when analyzing metastable or multi-vacuum configurations. Additional theoretical challenges, such as gauge dependence and renormalization scale ambiguity, further obscure the precise characterization of the electroweak phase transition—even in minimal extensions of the Standard Model (SM).
This review presents the theoretical foundations of finite-temperature QFT with an emphasis on how different field species respond to thermal effects, identifying the bosonic sector as the primary source of key theoretical subtleties. We focus particularly on the scalar extension of the SM, which offers a compelling framework for realizing first-order electroweak phase transitions, electroweak baryogenesis, and accommodating dark matter candidates depending on the underlying $\mathbb{Z}_2$ symmetry structure.
\end{abstract} 

\section{Introduction}

The baryon asymmetry of the universe (BAU), the nature of dark matter (DM), and the origin of electroweak symmetry breaking (EWSB) are central open questions at the interface of particle physics and cosmology. Although each problem can be formulated independently, a suggestive commonality is that all are tied to the physics of the early universe. In particular, the electroweak epoch at $T\!\sim\!100~\mathrm{GeV}$ ($t\!\sim\!10^{-11}\,\mathrm{s}$, order of magnitude) provides a natural stage on which new dynamics could have operated, potentially linking the generation of the BAU to the dynamics of the electroweak phase transition (EWPT) and to the properties of additional states that might also address DM.
Electroweak baryogenesis (EWB) realizes Sakharov’s conditions within the electroweak plasma: baryon number violation through sphaleron transitions, C and CP violation, and a departure from thermal equilibrium provided by a first-order EWPT \cite{morrissey2012electroweak,cohen1993progress,garbrecht2020there,rubakov1996electroweak,trodden1999electroweak}. In a first-order transition, bubbles of the broken phase nucleate and expand in the surrounding symmetric plasma. Charge-transport dynamics across the bubble walls can source CP-asymmetric densities that, in the presence of weak sphalerons in the symmetric phase, are converted into net baryon number; inside the bubbles, sphaleron processes are exponentially suppressed, preserving the asymmetry \cite{morrissey2012electroweak}. The qualitative viability of EWB therefore hinges on the thermodynamic nature of the EWPT and on the associated real-time transport and sphaleron rates. 
Within the minimal Standard Model (SM), non-perturbative calculations showed that for the observed Higgs mass $m_h = 125~\mathrm{GeV}$ the EWPT is a crossover rather than first order, precluding the necessary out-of-equilibrium dynamics \cite{kajantie1996there}. In particular, the usual proxy criterion $v_c/T_c \gtrsim 1$ (with $v_c$ the Higgs expectation value at the critical temperature $T_c$) cannot be satisfied. This result motivates extensions of the Higgs sector and/or the particle content of the SM. A robust intuition is that additional bosonic degrees of freedom coupled to the Higgs can induce sizable thermal cubic terms in the finite-temperature effective potential, strengthening the transition and rendering it first order over portions of parameter space \cite{espinosa2012strong,choi1993real,o2007minimal,ham2005electroweak,ramsey2020electroweak,carena2020electroweak,papaefstathiou2022electro,profumo2007singlet,profumo2015singlet}. Other extensions can simultaneously accommodate DM phenomenology by providing a stable, weakly coupled relic (e.g.\ a gauge-singlet scalar) with the correct relic abundance \cite{chiang2018standard,alanne2014strong,chiang2021electroweak,xiao2023dilution,barger2009complex,branco1998electroweak}.\\

\noindent
Predictive control over such scenarios requires a framework that consistently treats quantum and thermal effects in the early universe. At high temperatures, the dynamics of interest are not those of the zero-temperature vacuum, but of a thermal medium near (or out of) equilibrium. Accordingly, one replaces vacuum expectation values by ensemble averages and employs finite-temperature quantum field theory (FTQFT). Two complementary formulations are standard: the imaginary-time (Matsubara) formalism, in which Euclidean time is compactified with (anti)periodic boundary conditions for (fermions) bosons and energies are discretized into Matsubara frequencies; and the real-time (Schwinger--Keldysh) formalism, which organizes contour-ordered correlators and is suited to dynamical, non-equilibrium processes \cite{quiros1998,landsman1987real}. At a structural level, the temperature-dependent parts of loop amplitudes are ultraviolet (UV) finite, however the temperature-independent parts are still UV-divergent, which are renormalized by the same counterterms as at $T=0$. On the other hand, the infrared (IR) physics becomes more delicate: thermal populations enhance long-wavelength bosonic modes, electric fields are Debye screened, and the magnetostatic sector of non-Abelian gauge theories remains non-perturbative \cite{laine2017basics}. These features underpin the use of resummation schemes (e.g.\ ring/daisy resummations for scalar sectors and hard-thermal-loop (HTL) resummations for gauge sectors) and motivate effective-theory approaches that systematically separate thermal scales \cite{laine2017basics,arnold1993effective,espinosa1993nature,carrington1992effective}.\\

\noindent
Even with these tools, several theoretical systematics must be confronted. First, the finite-order perturbative effective potential is gauge dependent; physical observables are gauge independent, but intermediate quantities used as proxies (e.g.\ $v_c/T_c$ derived from $V_{\rm eff}$) can inherit gauge artifacts unless care is taken \cite{patel2011baryon,andreassen2015,garny2012gauge,balui2025gauge}. Second, predictions retain a residual renormalization-scale dependence when computed at finite order; renormalization-group improvement and judicious choices of the scale (often $\mu \sim T$) mitigate, but do not eliminate, this uncertainty \cite{Papaefstathiou_2021,lewicki2024impacttheoreticaluncertaintiesmodel}. Third, the high-temperature expansion and truncations implicit in resummation can introduce spurious imaginary parts (when field-dependent masses turn tachyonic) and miss subleading, potentially relevant contributions; improved schemes based on gap equations or variational resummations aim to reduce these effects. Finally, IR sensitivity associated with bosonic zero Matsubara modes is an inherent challenge; while daisy and HTL resummations capture the dominant contributions, the magnetostatic sector in non-Abelian theories remains genuinely non-perturbative, necessitating effective descriptions or lattice input for complete control \cite{espinosa1993nature,carrington1992effective,arnold1992phase,arnold1993effective}.\\

\noindent
The phenomenology of a first-order EWPT spans several frontiers. At colliders, modifications of the Higgs potential can appear through altered Higgs self-interactions, exotic decays, and deviations in couplings; new bosonic states invoked to strengthen the transition can be searched for directly or indirectly \cite{ramsey2020electroweak,carena2020electroweak,papaefstathiou2022electro,profumo2007singlet,forslund2022high,liu2021probing,no2014probing}. Cosmologically, a strong first-order transition can source a stochastic background of gravitational waves through bubble collisions, sound waves, and magnetohydrodynamic turbulence; the predicted spectra are model dependent but potentially testable in the next generation of interferometers \cite{huber2008gravitational,caprini2018cosmological}.\\

\noindent
This review has three objectives. First, we summarize the foundations of FTQFT with an emphasis on the structural modifications of two-point functions in a thermal medium. We present both imaginary-time and real-time formalisms, clarify their domains of applicability, and highlight how UV renormalization and IR physics reorganize at finite temperature. Second, we assemble the principal sources of theoretical uncertainty---gauge and scale dependence, IR sensitivity, and spurious imaginary parts---survey standard remedies (ring/daisy resummations,  and renormalization-group improvement), and quantify the residual systematics where possible. Third, we apply these tools to a minimal and widely studied extension of the SM: the real singlet scalar extension (RxSM). This framework illustrates how additional bosonic degrees of freedom can induce a strong first-order EWPT, potentially realize electroweak baryogenesis in concert with CP-violating sources, and -in some incarnations- supply a dark-matter candidate \cite{chiang2018standard,alanne2014strong,chiang2021electroweak,xiao2023dilution,barger2009complex}. 
Our aim throughout is to provide a transparent audit of theoretical uncertainties behind the common approximations. We conclude by mapping the theory space onto current and near-future experimental programs: precision measurements of Higgs properties and multi-boson processes at colliders, direct searches for additional scalars, and gravitational-wave probes of a first-order EWPT. The combined scrutiny across these arenas offers a realistic path to confirm or exclude broad classes of electroweak-scale explanations for BAU and DM.

%########################################################%
%########################################################%

\section{Quantum Field Theory in a Thermal Bath}\label{sec2}

In both classical and quantum descriptions, interactions are sensitive to the state of the system, either vacuum or medium. In the classical (tree-level) limit of QFT, fields propagate without loop corrections, yielding UV-finite amplitudes. In the full quantum theory, fluctuations around the vacuum modify correlation functions: 
n-point functions receive loop corrections, and ultraviolet (UV) divergences appear because local interactions in a continuum theory integrate over arbitrarily short distances (large loop momenta). In some sense, this is related to 
the uncertainty principle, where the fluctuations of high-momentum modes are allowed at short distances
$\delta x\to 0$, while the energy–momentum remains conserved at interaction vertices (ends). Renormalization absorbs the UV divergences into counterterms, and the renormalization group tracks the residual scale dependence, keeping the physical observables finite. By contrast, infrared (IR) divergences come from the continuum of long-wavelength (soft/collinear) modes; they cancel in inclusive (IR-safe) quantities once one sums over all the degenerate real and virtual processes as shown by Bloch–Nordsieck and  Kinoshita–Lee–Nauenberg in the vacuum QFT. 
\\[5pt]
In a thermal bath, correlation functions are no longer vacuum expectation values but ensemble averages, so fields are weighted by the statistical distribution of the medium. This immediately changes the structure of propagators: the medium picks out a preferred rest frame (reducing Lorentz symmetry), and time acquires thermal boundary conditions that discretize energies in equilibrium. We can then expect two broad consequences. First, the short-distance (ultraviolet) behavior is governed by locality and therefore should mirror the zero-temperature theory, suggesting that temperature does not introduce new UV counterterms, while ordinary renormalization continues to control high-momentum sensitivity. Second, thermal populations enhance long-wavelength modes, so the small-momentum behavior is more delicate: inclusive, physically defined observables will remain finite, but perturbative predictions can become sensitive to how one organizes the soft sector. In practice, this often means choosing scales adapted to the temperature and, where needed, resuming the most important medium effects. The next sections develop these points carefully, showing how the imaginary-time and real-time formalisms implement thermal averaging and make these expectations precise.
\\[5pt]

\noindent
Before diving into details, we highlight how thermal averaging modifies vacuum QFT at the level of the propagator.
In a thermal bath, on–shell excitations in the medium exchange energy–momentum with probes, so correlation functions
are ensemble averages. In a relativistic QFT, the particle number is not generally conserved, which suggests the appropriate ensemble to be the grand canonical ensemble, 
\begin{align}
\expval{\hat{\mathcal{O}} } = \frac{1}{\mathcal{Z}} \tz \left[ \hat{\mathcal{O}} e^{-\beta (\hat{H} - \mu \hat{Q})}\right] \label{grand}
\end{align}
Where, $\beta, \hat{H}, \mu, \hat{Q}$ are the temperature, Hamiltonian, chemical potential and the charge. $\mathcal{Z} = \tz \left[e^{-\beta (\hat{H} - \mu \hat{Q})}\right]$ is the partition function. For a generic field $\chi(\vec{x},t)$, which could be a scalar, vector or fermionic field, the two-point correlator is given by,
\begin{align}
G(x,y) &= \expval{T\{ \chi(t,\vec{x}) \chi(t',\vec{y})\}},\\
&= \Theta(t-t') G^>(x,y) + \Theta(t'-t) G^<(x,y).
\end{align}
Here, $G^>(x,y) = \langle\chi(t,\vec{x}) \chi(t',\vec{y})\rangle$ and $G^<(x,y) = \pm \langle\chi(t',\vec{y}) \chi(t,\vec{x})\rangle$ are the forward and backward propagators, and the plus minus sign is for bosonic and fermionic fields, respectively.  In vacuum QFT, Heisenberg's time evolution of the operator is given from
\begin{align}
\chi(t,\vec{x}) = e^{it H} \chi (0,\vec{x}) e^{-itH} \label{qfttimeevolution}
\end{align}
Neglecting the chemical potential in Eq.\eqref{grand} and comparing it to Eq.\eqref{qfttimeevolution}, it seems that in a thermal bath, we can get an equivalent of Heisenberg's operator evolution of the operator
\begin{align}
\chi(\tau,\vec{x}) = e^{\tau H} \chi (0,\vec{x}) e^{-\tau H}. \label{thermalevolution}
\end{align}
where we made a Wick rotation of $t\to -i\tau$ with $\tau\in[0,\beta]$, and $\beta\equiv 1/T$, which is the imaginary-time (Matsubara) formalism that we will discuss shortly. In that formalism, the density matrix is conserved due to the thermal equilibrium, so $\mu=0$ is justified. Using Eq.\eqref{thermalevolution} together with the trace property $\tz[AB] = \tz[BA]$, the backward propagator becomes
\begin{align}
G^<(\vec{y},\vec{x},t',t) &= \frac{1}{\mathcal{Z}} \tz \left[e^{-\beta H} \chi(t',\vec{y}) \chi(t,\vec{x}) \right]\\
&= \frac{1}{\mathcal{Z}} \tz \left[\chi (t',\vec{y}) e^{-\beta H} e^{\beta H} \chi(t,\vec{x}) e^{-\beta H} \right]\\
&= \frac{1}{\mathcal{Z}} \tz \left[ \chi(t',\vec{y}) e^{-\beta H}\chi(t+\beta,\vec{x}) \right] \\
&= \frac{1}{\mathcal{Z}} \tz \left[e^{-\beta H}\chi(t+\beta,\vec{x})  \chi(t',\vec{y}) \right]\\
&=\pm G^>(\vec{x},\vec{y};t+\beta,t'). \label{periodicity_cond}
\end{align}

\noindent
The plus sign characterizes the bosonic fields, while the minus sign is for the fermionic fields as they obey Grassmann algebra where $\psi_\alpha(\vec{x};t) \psi_\beta(\vec{y};t') = - \psi_\alpha(\vec{y};t') \psi_\beta(\vec{x};t)$. The (anti) periodicity condition in Eq.\eqref{periodicity_cond}
allows us to write the correlator as a Fourier series in the time component
\begin{align}
G(\tau,\vec{x}) &= \frac{1}{\beta}\sum_{n=-\infty}^\infty G(i\omega_n,\vec{x}) e^{-i\omega_n \tau}. \label{fourier_repres}
\end{align}
Using Eq.\eqref{fourier_repres} in Eq.\eqref{periodicity_cond} then yields
\begin{align}
\frac{1}{\beta}\sum_{n=-\infty}^\infty G(i\omega_n,\vec{x}) e^{-i\omega_n \tau} = \pm \frac{1}{\beta}\sum_{n=-\infty}^\infty G(i\omega_n,\vec{x})e^{-i\omega_n (\tau+\beta)},
\end{align}
which leads to quantization of the Matsubara modes as,
\begin{align}
\omega_n = \begin{cases}2n\pi \beta^{-1},\hfill &\text{For Bosons.}\\[5pt] (2n+1)\pi \beta^{-1},\hfill &\text{For Fermions.}\end{cases} \label{matsubaraas}
\end{align}

\noindent
Clearly, quantization of the Matsubara modes is rooted in tracing over the canonical ensemble, which, in turn, will convert the time component integral in the loop integral into a discrete sum over the Matsubara modes.
\begin{align}
  \int \frac{d^4 p}{(2\pi)^4}\ \longrightarrow\
  \sumint_{p}\ \equiv\ T\sum_{n=-\infty}^{\infty} \int \frac{d^{3}p}{(2\pi)^3},
  \qquad p=(i\omega_n,\vec{p}),
\end{align}
where $\omega_n$ is field sensitive as in Eq.\eqref{matsubaraas}.
 In addition, the propagator in Eq.\eqref{fourier_repres} gains temperature dependence in the thermal bath. Consequently, the self-energy loop contribution is expected to add a temperature-dependent correction to the tree-level such that $m_{\text{\tiny eff}}^2(T) = m_0^2 + f(T)$. This thermal screening is central to symmetry restoration at high temperature, as will be detailed in
the following section.

\subsection{Imaginary-time Formalism}

The path–integral formulation provides a natural framework for treating quantum field theory as a continuum of fields. 
Concepts that are introduced axiomatically in the canonical approach—such as time ordering and the operator formalism—arise here in a unified way from the integral epresentation and source differentiation. At the same time, it offers a transparent bridge between classical and quantum descriptions: quantum fluctuations are organized as expansions around classical field configurations (saddle points). The central object is the generating functional $\gz$ [Check chapter 14 in \cite{schwartz2014quantum} for a simple review], 
\begin{align}
\gz = \int \pd \chi_1 \cdots \pd \chi_n \,\, e^{iS[\chi_1 \cdots \chi_n]} \label{generatingf}
\end{align}
which encodes all information about the theory and its fields ($\phi,\psi,A$). Where all physical observables and n-point correlation functions are extracted from it,
\begin{align}
\expval{0|T\{\chi_1 \cdots \chi_2\}|0} &= \int \pd \chi\,\cdots \pd \chi_n\,\, e^{iS[\chi_1 \cdots \chi_n]} \,\, \chi_1 \cdots \chi_n,\label{npointfunc}\\
&= (-i)^n \frac{1}{\gz[0]} \frac{\partial^n \gz}{\partial J_1 \cdots \partial J_n}.\label{propagatorZ}
\end{align}
The practical difficulty resides in evaluating the path integral in the generating functional, which
is generally intractable once interactions are present and, for many theories, impossible to carry out in closed form. Wick's theorem in QFT allows for evaluating that path integral by splitting the action into free and interaction parts whenever the interaction is small compared to the free case, and consequently generating all the higher-order quantum corrections in a self-consistent way. is equivalent, order by order, to the Feynman diagrammatic approach, provided the couplings are small enough for perturbation theory to converge. In thermal equilibrium the appropriate object is the density operator, $\rho = e^{-\beta H}$, which can be split into free and interaction parts that obey the Bloch equation. 
\begin{align}
\rho &= e^{-\beta H} = e^{-\beta(H_0 + H_I)},\\
&= \rho_0 \rho_I.
\intertext{The time evolution of the interaction density is then given by,}
\frac{\partial \rho_I}{\partial \tau} &= \rho_0^{-1} H_0 \rho - \rho_0^{-1}(H_0 +H_I)\rho,\\
&=- \rho_0^{-1} H_I \rho_0 \, \rho_I,\\
&= -H_I(\tau) \rho_I
\end{align}
Hence, the field's evolution in a thermal bath obeys,
\begin{align}
\chi(\tau) = e^{\tau H_0} \chi(0) e^{\tau H_0},
\end{align}
which is equivalent to Heisenberg's operator evolution in Eq.\eqref{qfttimeevolution} if we evaluated the time-integral in the action along the negative imaginary axis $t\to -i\tau$, where $0<\tau<\beta$. Hence the name imaginary-time approach.
\begin{figure}[htb!]
\centering
\begin{tikzpicture}
\draw[->] (-4,0) -- (4,0);
\draw[->] (0,-3) -- (0,3);
\draw[color=blue, line width = 2pt] (0,0) -- (0,-2);
\begin{scriptsize}
\draw[] (3.8,-.3) node {Re[$t$]};
\draw[] (-.5,2.9) node {Im[$t$]};
\draw[] (0.4,-2) node {$-i\beta$};
\end{scriptsize}
\end{tikzpicture}
\caption{Time contour in the imaginary time formalism for the thermal equilibrium case.}
\label{imag-time-contour}
\end{figure}
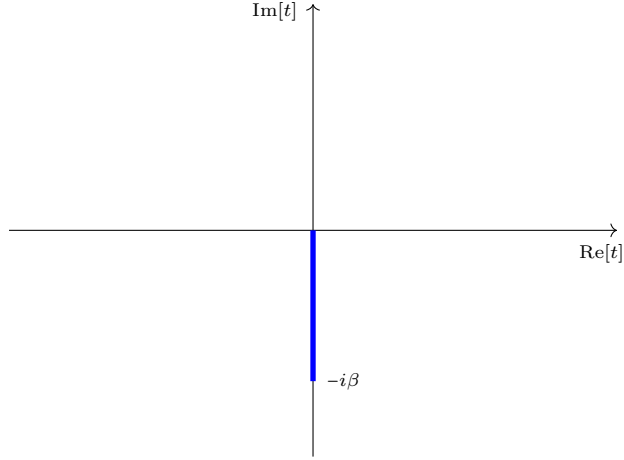

\noindent
The boundedness of the time component in the action integral in the generating functional Eq.\eqref{generatingf},
resulting from the trace over the thermal bath, 
implies the (anti) periodicity of the participating fields in the theory, and the quantization of their time component. Both together shift the Lagrangian from Minkowski space ($t,\vec{x}$) to the Euclidean space ($-i\tau, \vec{x}$) due to the Wick rotation of the time component to the imaginary time $\tau$ as depicted in Fig.\ref{imag-time-contour},
\begin{align}
\lag_M &= \left(\partial_t \chi \right)^2 - \left(\partial_i \chi \right)^2  - V(\chi),\\
\intertext{Wick-rotating $t\to -i\tau$,}
\lag_M \to - \lag_E &= \left (\partial_\tau \chi\right)^2 + \left(\partial_i  \chi\right)^2 + V(\chi),
\intertext{and the generating functional in Eq.\eqref{generatingf} becomes,}
\gz &= \int_{\chi_i(\beta) = \pm \chi_i(0)} \pd \chi_1 \cdots \pd \chi_n \,\, \exp\left\{-S_E[\chi_1 \cdots \chi_n]\right\},\\
&= \int_{\chi_i(\beta) = \pm \chi_i(0)} \pd \chi_1\cdots \chi_n \,\, \exp\left\{-\int_0^\beta d\tau \int d^3\vec{x} \,\,\lag_E\right\}.\label{thermalnfunc}
\end{align}
An interesting phenomenon in this approach is that the higher quantum corrections are exponentially suppressed in the thermal background, so that for high temperature, $\beta\ll1$, the generating function will converge, leading to a natural regularization of the UV-divergences. Yet, a remnant UV-divergence will survive as the temperature goes to zero, $\beta \to \infty$, where the Minkowski representation is restored.
According to Eq.\eqref{npointfunc} and Eq.\eqref{thermalnfunc}, a generic propagator in a thermal background using the imaginary-time approach is given by
\begin{align}
\expval{0|T\{\chi_1\chi_2\}|0} = \int \pd\chi\,\, e^{-\int_0^\beta d\tau \int d^3x\,\, \lag_E} \chi_1 \chi_2.\label{timeordered}
\end{align}
For a free real scalar, the Lagrangian in Euclidean space is
\begin{align}
\lag_E = \frac{1}{2} (\partial_\tau \phi)^2 + \frac{1}{2}(\nabla \phi)^2 + \frac{1}{2}m^2 \phi^2.
\end{align}
Then, the free generating functional becomes
\begin{align}
\gz_0[J] &= \int \pd \phi\,\, \exp\left\{-\frac{1}{2}\int_0^\beta d\tau \int  d^3x \,\phi (\partial_\tau^2 + \nabla^2 + m^2)\phi - 2J(\vec{x})\phi\right\},\\
&= \gz_0[0] \,\, \exp\left\{\frac{1}{2}\int_0^\beta d\tau \int d^3x d^3 y J(\vec{x}) G_S(x-y) J(\vec{y})\right\},\label{scalarimag}
\end{align}
where $G_S(x-y)$ is the inverse of the operator ($\partial_\tau^2+\nabla^2+m^2$)\footnote{Gaussian integral implies: $\int d\vec{q} e^{-\frac{1}{2} \vec{p} A\vec{p} + \vec{B} \vec{P}} = \sqrt{\frac{(2\pi)^n}{det\, A}} e^{\frac{1}{2} \vec{J}^T A^{-1} \vec{J}}$},
\begin{align}
(\partial_\tau^2+\nabla^2+m^2)G_S(x-y) = \delta(\tau_x - \tau_y) \delta^{(3)}(\vec{x} -\vec{y}).
\end{align}
Integrating both sides over$\frac{1}{(2\pi)^4}\int d\tau \,\,e^{-i\omega_n(\tau_x -\tau_y)} \int d^3p \,\,e^{-i\vec{p}\cdot(\vec{x} -\vec{y} )}$, returns the scalar propagator in momentum space,
\begin{align}
\widetilde{G}_S(i\omega_n,\vec{p}) = \frac{1}{E_p^2 + \omega_n^2} = \frac{1}{\abs{\vec{p}}^2 + m^2 + \omega_n^2}.\label{scalarprop}
\end{align}
where, $\omega_n = 2n\pi T$ for bosons. Similarly, a fermion field in Euclidean space is characterized by
\begin{align}
\lag_E &= \overline{\psi} \big(\gamma^0_E\,\partial_\tau+\vec{\gamma}_E \cdot\vec{\nabla} + m\big)\psi,
\end{align}
and the corresponding generating functional is,
\begin{align}
\gz_0[J] &= \int \pd \phi\,\, \exp\left\{-\int_0^\beta d\tau \int  d^3x\,\overline{\psi}\big(\gamma^0_E\,\partial_\tau+\vec{\gamma}_E \cdot\vec{\nabla} + m\big)\psi - \overline{\psi}J(\vec{x})\psi\right\},\label{imag_ferm_action}\\
&= \gz_0[0]\,\, \exp\left\{-\int_0^\beta d\tau \int d^3x d^3y J^T(\vec{x}) G_F(x-y) J(\vec{y})\right\},
\end{align}
such that
\begin{align}
\big(\gamma^0_E\,\partial_\tau+\vec{\gamma}_E \cdot\vec{\nabla} + m\big)G_F(x-y) = \delta(\tau_x - \tau_y) \delta^{(3)}(\vec{x} -\vec{y}),\label{fermioneom}
\end{align}
which in momentum space produces the fermion propagator,
\begin{align}
\widetilde{G}_F(i\omega_n,\vec{p}) &= \frac{1 }{i\omega_n \gamma^0 + \vec{p} \cdot \vec{\gamma} + m},\label{fermionprop}
\end{align}
with $\omega_n = (2n+1)\pi T$. From Eq.\eqref{scalarprop} and Eq.\eqref{fermionprop}, we can already see two main differences from the corresponding vacuum QFT case: firstly, the propagator in the Euclidean space is no longer singular at the $p = \pm m$ region that shows up in vacuum QFT. Secondly, it becomes temperature dependent, which is a vital property to understand symmetry restoration at high temperature where the mass term in the Lagrangian becomes positive ($T > \abs{m}$), forcing the global minimum to be the symmetric one, $\expval{\chi} = 0$. Yet, the imaginary-time approach is physically confusing due to the integration over the imaginary time contour. In addition, it is only valid to study dynamics taking place at equilibrium, which makes it not suitable to study phenomena that take place out of equilibrium, such as gravitational waves propagating during the electroweak phase transition (EWPT).  For such scenarios, we have to rely on the real-time approach. Despite that, the imaginary-time approach is widely adopted, whenever adequate, for its simplicity, as we just discussed.

\subsection{Real-time formalism}\label{realformalism}

A more directly physical route is to retain real time and evolve along a contour in the complex
time plane, which naturally accommodates departures from thermal equilibrium. In the real–time
(or Schwinger–Keldysh/closed–time–path) formalism, the statistical state need not be an
equilibrium ensemble; instead, its evolution is governed by the quantum Liouville (von Neumann)
equation
\begin{align}
i\frac{\partial \rho(t)}{\partial t} = [H,\rho(t)].\label{realtimeform}
\end{align}
When the density operator commutes with the Hamiltonian (or equivalently adopts a Boltzmann
form, the real–time formalism reproduces thermal equilibrium. Crucially, it also encompasses non-equilibrium situations when this condition is not met. The solution of Eq.\eqref{realtimeform} shows that the thermal density evolves similarly to the vacuum QFT operators,
\begin{align}
\rho(t) &= e^{-iHt}\rho(0) e^{iHt},\\
&= U(t,0)\,\rho(0)\,U(0,t).\label{unitrity}
\end{align}
Where, $U(t,t')$ is the unitary time evolution,
\begin{align}
i\frac{\partial U(t,t')}{\partial t} = H(t)U(t,t'),
\end{align}
and it satisfies the conditions:
\begin{align}
U(t,t) &= 1,\label{prop1}\\
U(t_1,t_2)U(t_2,t_1) &= 1,\\
U(t_1,t_2)U(t_2,t_3) &= U(t_1,t_3),\qquad t_1>t_2>t_3.\label{prop3}
\end{align}
These identities are self-explanatory: the first states that the system is not changing as time is not passing, the second indicates that the inverse of a process shall take us back to the initial state, and the third implies that successive transitions from $t_3 \to t_2$ followed by a transition from $t_2 \to t_1$ is equivalent to a single large transition from $t_3\to t_1$. Taking the density matrix to be,
\begin{align}
\rho(0) = \frac{e^{-\beta H}}{\tz\left[e^{-\beta H}\right]}
\end{align}
since it has a unit trace. Then, using the property in Eq.\eqref{unitrity}, we can rewrite the density matrix as
\begin{align}
\rho(0) = \frac{U(T-i\beta,T)}{\tz U(T-i\beta,T)},
\end{align}
and consequently, the time evolution of the density matrix becomes,
\begin{align}
    \rho(t) = \frac{U(t,0)\,U(T-i\beta,T)\,U(0,t)}{\tz U(T-i\beta,T)}
\end{align}

\begin{figure}[htb!]
\centering
\begin{tikzpicture}
\draw[->] (-4,0) -- (4,0);
\draw[->] (0,-3) -- (0,3);
\draw[color=blue, line width = 2pt, ->] (-3,0) -- (1,0);
\draw[color=blue, line width = 2pt] (1,0) -- (3,0);
\draw[color=blue, line width = 2pt] (3,0) -- (3,-1);
\draw[color=blue, line width = 2pt, ->] (3,0) -- (3,-.5);
\draw[color=blue, line width = 2pt] (3,-.5) -- (3,-1);
\draw[color=blue, line width = 2pt, ->] (3,-1) -- (-1,-1);
\draw[color=blue, line width = 2pt] (-1,-1) -- (-3,-1);
\draw[color=blue, line width = 2pt, ->] (-3,-1) -- (-3,-1.5);
\draw[color=blue, line width = 2pt] (-3,-1.5) -- (-3,-2);
\begin{scriptsize}
\draw[] (-0.5,3) node {Im[$T$]};
\draw[] (4.5, 0) node {Re[$t$]};
\draw[] (-3,0.3) node {$T$};
\draw[] (3, 0.3) node {$T'$};
\draw[] (3,-1.3) node {$t$};
\draw[] (-3.3,-1) node {$t'$};
\draw[] (-3.5,-2) node {$T-i\beta$};
\end{scriptsize}
\end{tikzpicture}
\caption{The real-time contour: A forward real--time branch $\mathcal C_1\!: T\to T'$, a forward
segment past the insertion to $t$ ($\mathcal C_2$), a backward real--time branch $\mathcal C_3\!: t\to t'$,
and a short imaginary--time leg $\mathcal C_4\!: t'\to T-i\beta$. This closed-time contour encodes
both the causal evolution and, when the imaginary leg is included, the choice of an initial thermal
state.}
\label{real-time-contours}
\end{figure}
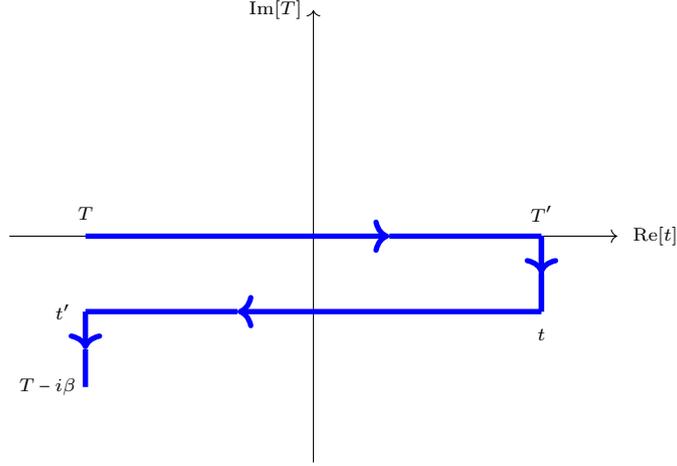

\noindent
Correspondingly, the thermal expectation of any field operator becomes,
\begin{align}
    \expval{\chi}_\beta &= \tz[\rho(t) \chi],\\
    &= \frac{\tz U(t,0)\,U(T-i\beta,T)\,U(0,t) \, \chi}{\tz U(T-i\beta,T)},
    \intertext{using properties in Eq.\eqref{prop1} : Eq.\eqref{prop3}, we get}
    \expval{\chi}_\beta &= \frac{\tz U(T-i\beta,t')\, U(t',t)\, U(t,T')\, \chi\, U(T',T)}{\tz U(T-i\beta,T)\, U(T,T')\, U(T',T)}.
\end{align}
This construction describes evolution along a real-time contour that begins at an initial time $T$ (taken large and negative), evolves forward to a later time $T'$, includes the insertion of an operator $\chi$, and then continues to a turning point at time $t$. The contour is subsequently
reflected and traced backward from $t$ to an earlier time $t'$, after which it is extended along a vertical Euclidean segment to $T-i\beta$. Thus, the path consists of four distinct branches in the complex time plane (see Fig.\ref{real-time-contours}).
The price of the real–time (closed–time–path) formalism is a doubling of degrees of freedom: instead of a single propagator one obtains a $2\times 2$ matrix of contour–ordered two–point functions, each associated with a branch of the time contour. This is evident from Eq.\eqref{timeordered}, where the propagator in real-time formalism will be given from,
\begin{align}
    \braket{0|T_c{\chi_1\chi_2}|0}= \int \pd \chi e^{-\int_c dt \int d^3x\,\, \lag_M} \chi_1\chi_2.
\end{align}
Here, we are using the traditional Minkowski Lagrangian, but the time integral is now taken across the complex contours extending from $-\infty:\infty$ such that
\begin{align}
    \int_c dt = \int_{-\infty}^\infty dt_+ - \int_{-\infty}^\infty dt_-.
\end{align}
For a scalar field, the generating functional is evaluated as in Eq.\eqref{scalarimag},
\begin{align}
\gz_0[J] &= \gz_0[0]\,\, \exp\left\{-i\int dt\,dt'\, \int d^3x d^3 y J^T(\vec{x},t) G_S(x-y) J(\vec{y},t')\right\},
\end{align}
where $G_S(x-y)$ is the inverse of the operator,
\begin{align}
(\Box + m^2)G_S(x-y) = - \delta_c(t-t')\delta^{(3)}(\vec{x} -\vec{y}).
\end{align}
In momentum space, this differential equation returns four propagators depending on the two branches $C_+,\,C_-$
\begin{align}
    G^S_{++}(p) &= \frac{1}{p^2-m^2+i\varepsilon}- 2in\pi n_B(p^0)\delta(p^2-m^2),\qquad ~~~t,t' \in C_+ \label{gpp}\\
    G^S_{+-}(p) &= -2i\pi\Big(\Theta(-p^0) + n_B(p^0) \Big)\delta(p^2-m^2),\qquad \qquad t\in C_+,\, t'\in C_-\\
    G^S_{-+}(p) &= -2i\pi\Big(\Theta(p^0) + n_B(p^0) \Big)\delta(p^2-m^2),\qquad \qquad ~~t\in C_-,\, t'\in C_+\\
    G^S_{--}(p) &= -\frac{1}{p^2-m^2-i\varepsilon}- 2in\pi n_B(p^0)\delta(p^2-m^2),\qquad ~t,t' \in C_-\label{gmm}
\end{align}
In this framework, the thermal contributions are cleanly disentangled from the vacuum part of the
theory. In particular, all ultraviolet (UV) divergences of finite–temperature QFT are precisely the
remnants of the zero–temperature sector; they are removed by the same counterterms fixed in
vacuum QFT, with the explicitly temperature–dependent pieces remaining UV–finite. At the same
time, the formalism makes transparent the origin of the severe infrared (IR) sensitivity: in the
bosonic sector, the zero Matsubara mode and Bose enhancement, $n_B(E)\sim T/E$  for $E\to 0$ amplifies soft contributions over an extended low–momentum region, necessitating appropriate resummations (e.g. ring/daisy).
The fermionic sector is handled analogously, but with antiperiodic boundary conditions in
imaginary time. Its two–point function is therefore obtained from the corresponding equation of motion with the Dirac operator and
antiperiodic temporal boundary conditions, leading to a thermal propagator that is IR–safe at
leading order. Its corresponding equation of motion will be given, similar to Eq.\eqref{fermioneom}, by
\begin{align}
(i\slashed{\partial} -m )G_F(x-y) = -\delta_c(t-t')\delta^{(3)}(\vec{x} -\vec{y}).
\end{align}
which, in momentum space, produces four similar propagators depending on the two branches $C_+,\,C_-$
\begin{align}
    G^F_{++}(p) &= \frac{1}{\slashed{p} -m +i\varepsilon}+ 2in\pi n_F(p^0)\delta(p^2-m^2),\qquad ~~~~~t,t' \in C_+\\
    G^F_{+-}(p) &= 2i\pi\Big(-\Theta(-p^0) + n_F(p^0) \Big)\delta(p^2-m^2),\qquad \qquad t\in C_+,\, t'\in C_-\\
    G^F_{-+}(p) &= 2i\pi\Big(-\Theta(p^0) + n_F(p^0) \Big)\delta(p^2-m^2),\qquad \qquad ~~ t\in C_-,\, t'\in C_+\\
    G^F_{--}(p) &= -\frac{1}{\slashed{p} - m -i\varepsilon} + 2in\pi n_F(p^0)\delta(p^2-m^2),\qquad  ~~~~t,t' \in C_-
\end{align}
Consequently, there is no fermionic zero Matsubara mode and the Fermi–Dirac distribution,
$n_F(E)\sim  1/2$ as $E\to 0$, does not generate the same IR enhancement as for bosons.

%%%%%%%%%%%%%%%%%%%%%%%%%%%%%%%%%%%%%%%%%%%%%%%%%%%%%%%%%%
%%%%%%%%%%%%%%%%%%%%%%%%%%%%%%%%%%%%%%%%%%%%%%%%%%%%%%%%%%

\section{Fields responses to Thermal Bath }

Building on the discussion of two–point correlation functions in real and imaginary time, this
section develops the finite–temperature effective potential for the field content relevant to our electroweak phase–transition (EWPT) analysis in section \ref{EWPTsec}. 
Our focus there will be on the
thermodynamics and phase structure of the EWPT itself. Since we are
interested in the critical regime where thermal equilibrium is a good approximation, we adopt the
imaginary–time (Matsubara) formalism.

A key advantage of the imaginary–time framework is that it admits a diagrammatic treatment
closely parallel to vacuum QFT. We begin from the Euclidean generating functional for a generic
field $\chi(x)$ and show how the usual Feynman rules are recovered, with the sole modification that
propagators reflect the (anti)periodic boundary conditions in Euclidean time. We then apply this
construction to each sector of the theory that enters our EWPT study. The central object is the
finite–temperature effective potential, $V_\eff(\chi,T)$, which encodes the free energy density as a
function of background fields, determines the thermal ground state, and governs the evolution of
order parameters. In what follows we make explicit the relation between $V_\eff$ and the generating
functional, and we provide its diagrammatic representation to the loop orders used later.\\[5pt]
For a generic field $\chi(x)$, the generating functional in the imaginary-time formalism is given by
\begin{align}
\gz[J] &= \int \pd \chi \,\, \exp\left\{-S_E[\chi]+ \int_0^\beta d^4x \, \, \chi \,J\right\}. \label{scalargeneratingz}
\end{align}
Here, we use $\int_0^\beta d^4 x = \int_0^\beta d\tau \int_\Omega d \vec{x}$. The generating functional $\gz[J]$ can be evaluated through expanding around the source $J$, similar to the vacuum QFT
\begin{align}
\gz[J]&= \gz[0]\left(1 - \int_0^\beta d^4 x  J(x) \chi(x)+\frac{1}{2} \int_0^\beta d^4x\int_0^\beta d^4 y J(x) J(y) \chi(x) \chi(y) +\cdots \right),\\
&= \gz[0] \sum_{n=0}^n \frac{(-1)^n}{n!} \int_0^\beta d^4x_1 \cdots d^4x_n \,\,J(x_1) \cdots J(x_n) \,\,G^n(x_1,\cdots, x_n).
\end{align}
Where, $\gz[0] = \int \pd \chi \,\, \exp\left\{ -S_E[\chi]\right\}$. On the one hand, this structure indicates that $\gz[J]$ encompasses all the one-particle irreducible (1PI) n-point correlation functions, $G^n(x_i)$. On the other hand, $G^n(x_i)$ contains all possible permutations of field pairs according to Wick's theorem, from which some of them are causally disconnected bubbles. These unphysical diagrams were shown in vacuum QFT to be removed term by term via dividing by the free generating functional $\gz[0]$. So, the generating functional $W[J]$ only contains the physically connected contributions by construction,
\begin{align}
W[J] &= \ln \gz[J],
\intertext{such that,}
\frac{\partial W[J]}{\partial J} &= \frac{1}{\gz[J]} \frac{\partial \gz[J]}{\partial J} = \expval{\chi}_J.
\end{align}
 Using this connected generating functional, we can now obtain the quantity of our interest, which is the effective potential, $V_\eff$, that basically represents the Gibbs free energy (ground state of the system) in the imaginary-time formalism. One further step is required before that, which is to define the effective action from the connected functional $W[J]$ through Legendre's transformation,
\begin{align}
\Gamma^E_\eff[\chi] =\int_0^\beta d^4x \,\, \chi(x) J(x) - W[J] \label{effectiveaction}
\end{align}
By definition, $\Gamma_\eff^E[\chi]$ is the effective action in the Euclidean space obtained from  $S_E[\phi]$ via integrating out the heavy modes. So, it is also given by
\begin{align}
\Gamma^E_\eff[\chi] = \int_0^\beta d^4 x \,\, \lag^E_\eff
\end{align}
Under translation invariance, $\chi(x) = \chi_c$, where the field is constant everywhere, the kinetic term in $\lag^E_\eff$ would vanish, and the effective action becomes the spacetime integral of the effective potential,
\begin{align}
\Gamma^E_\eff[\chi_c] = \int_0^\beta d^4x \,\, V^E_\eff(\chi_c).\label{effectiveactionapprox}
\end{align}
From Eq.\eqref{effectiveaction}, the effective potential in the absence of external source current is
\begin{align}
V^E_\eff(\chi_c) = -\frac{T}{\Omega} \ln \gz[0].\label{freeenergy}
\end{align}
This is the central object for the EWPT analysis in the remaining parts.

\subsection{Scalar Fields}

\begin{figure}[htb!]
\centering
\begin{tikzpicture}
\draw[dashed,line width = 1.5pt] (-1,1) -- (0,0);
\draw[dashed,line width = 1.5pt]  (-1,-1) -- (0,0);
\draw[dashed,line width = 1.5pt]  (1,0) circle (1);
\draw[dashed,line width = 1.5pt] (3,2) -- (4,1);
\draw[dashed,line width = 1.5pt]  (5,2) -- (4,1);
\draw[dashed,line width = 1.5pt] (3,-2) -- (4,-1);
\draw[dashed,line width = 1.5pt]  (5,-2) -- (4,-1);
\draw[dashed,line width = 1.5pt]  (4,0) circle (1);
\draw[dashed,line width = 1.5pt] (6,2) -- (7,1);
\draw[dashed,line width = 1.5pt]  (8,2) -- (7,1);
\draw[dashed,line width = 1.5pt] (6,-2) -- (7,-1);
\draw[dashed,line width = 1.5pt]  (8,-2) -- (7,-1);
\draw[dashed,line width = 1.5pt] (9,1) -- (8,0);
\draw[dashed,line width = 1.5pt]  (9,-1) -- (8,0);
\draw[dashed,line width = 1.5pt]  (7,0) circle (1);
\begin{scriptsize}
\draw[] (2.5,0) node {$+$};
\draw[] (5.5,0) node {$+$};
\draw[] (9.5,0) node {$+ \quad \cdots\cdots $};
\end{scriptsize}
\end{tikzpicture}
\caption{The sum of the 1-loop scalar contribution to the effective potential representing the diagrammatic equivalent of equation (\ref{eq22}).}
\label{ScalarLoopsSum}
\end{figure}
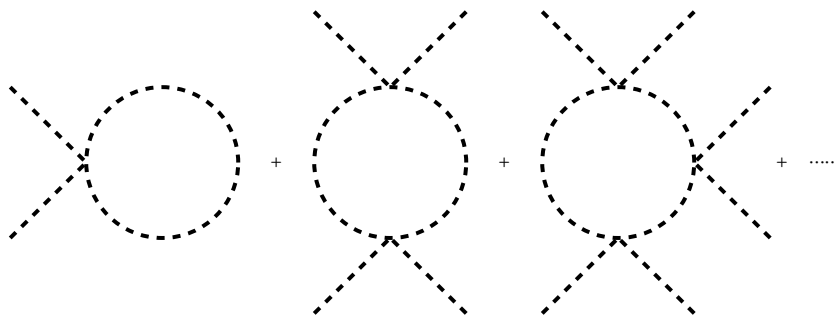

A real scalar field in Euclidean space is given by the action
\begin{align}
S_E[\phi] &=  \int_0^\beta d^4 x \,\, \frac{1}{2}(\partial_\tau \phi)^2 + \frac{1}{2} (\vec{\nabla} \phi)^2 + V_{\text{\tiny TL}}(\phi),\\
&=  \int_0^\beta d^4 x \,\, \frac{1}{2} \phi \left(- \partial_\tau^2 - \nabla^2 \right) \phi + V_{\text{\tiny TL}}(\phi).
\end{align}
 In the second line, we integrated by parts to bring the action into the Gaussian integral form needed to evaluate the path integral in the generating functional\footnote{The boundary term resulting from integration by parts is vanishing over time using the fields periodicity in thermal bath: $\phi(\beta) = \phi(0)$, and the boundary conditions of the field and its derivative at infinity.}. 
Using the background field method ($\phi \rightarrow \phi_c + \eta$), the scalar-filed generating functional in Eq.\eqref{scalargeneratingz} becomes
\begin{align}
\mathcal{Z}[0] &= \int \pd \phi \,\,\exp\left\{-\int d_0^\beta \eta \frac{1}{2} [\partial_\tau^2 -\nabla^2  + m^2(\phi_c)]\eta + V_\tl(\phi_c) \right\},
\end{align}
where we expanded $V_\tl(\phi_c+\eta)$ and used $V'(\phi_c) = 0$, and $V''(\phi_c) = m^2(\phi_c).$ Using the Gaussian integral property
\begin{align}
\mathcal{Z}[0] &=\exp\{-\beta \Omega V_\tl(\phi_c)\} \,\, \text{det}[-\partial_\tau^2- \nabla^2 + m^2(\phi_c)]^{-1/2} + \mathcal{Z}^{(n\geq 2)}[0]. \label{scalarz0}
\end{align}
Where, $\mathcal{Z}^{(n\geq 2)}[0]$ corresponds to interaction terms $\eta^3,\,\eta^4$ that contribute to $n\geq 2$ loop order. Here we focus on the one-loop correction to the tree-level potential, so the $n\geq 2 $ corrections are not included. Consequently, the effective potential at one-loop becomes
\begin{align}
V_\eff(\phi_c) &= V_\tl (\phi_c) +\frac{T}{2\Omega} \tz \Big[ \ln(-\partial_\tau^2 - \nabla^2 + m^2(\phi_c) )\Big],\label{freeenergytrace}\\
&= V_\tl(\phi_c)  + \frac{T}{2}\sum_{n=-\infty}^\infty\int\frac{d^3 p}{(2\pi)^3} \ln[\omega_n^2 + \abs{\vec{p}}^2 + m^2(\phi_c)],\label{summedscalarint}\\
&= V_\tl(\phi_c) + V_B^{(1)}(\phi_c,T) 
\end{align}
In the second line, we traded the trace for an integral over space-states and Fourier transformed the integrand to momentum space, taking into account the field's periodicity over the time component. $V^{(1)}(\phi_c,T)$ describes one scalar loop with n-external legs, as the integrand is just the inverse of the scalar propagator in the imaginary-time approach (Check Eq.\eqref{scalarprop}). So, by construction, 
the effective potential contains all the 1PI-loop corrections as it is directly related to the effective action (connected generating functional, $W[J]$). This could be seen already from the $V^{(1)}(\phi_c,T)$ using the identity,
\begin{align}
\ln[\omega_n^2 + \abs{\vec{p}}^2 &+ m^2(\phi_c)] = \ln\left[ \omega_n^2 + \abs{\vec{p}}^2 + m_\circ^2\right] + \sum_{l=1}^\infty \frac{(-1)^{l+1}}{l} \left( \frac{\frac{1}{2} \lambda \phi_c^2}{\omega_n^2 + \abs{\vec{p}}^2 + m^2_\circ}\right)^l,\label{eq11}
\intertext{then,}
V_B^{(1)}(\phi_c,T) &= \frac{T}{2} \sum_{n=-\infty}^\infty  \int \frac{d^3 p}{(2\pi)^3}\sum_{l=1}^\infty \frac{(-1)^{l+1}}{l} \left( \frac{\frac{1}{2} \lambda \phi_c^2}{\omega_n^2 + \abs{\vec{p}}^2 + m^2_\circ}\right)^l.\label{eq22}
\end{align}
Where the field-independent term has been dropped. This describes a loop with $n$ propagators and vertices and $2n$ legs. Diagrammatically, it is equivalent to Figure \ref{ScalarLoopsSum}, which was derived in \cite{quiros1998} but for the propagator of the vacuum QFT case.
To evaluate the summed integral in Eq.\eqref{summedscalarint}, we first evaluate the infinite sum using the property,
\begin{align}
\sum_{n=-\infty}^\infty \frac{1}{y^2 + n^2} = \frac{\pi}{y} \coth(\pi y).\label{bosonicsum}
\end{align}
Then,
\begin{align}
\frac{d V_B^{(1)}(\phi_c, T)}{dm^2} &= \frac{T}{2} \int \frac{d^3p}{(2\pi)^3} \,\, \sum_{n=-\infty}^\infty  \frac{1}{E^2_p + \omega_n^2},\\
&=\int \frac{d^3p}{(2\pi)^3} \,\, \left( \frac{ 1 + 2 n_B(E_p)}{2\,E_p}  \right).
\intertext{Integrating over $dm^2$ using $dm^2 = 2E_pdE_p$ from the relativistic energy relation,}
V_B^{(1)}(\phi_c, T) &=  \int \frac{d^3p}{(2\pi)^3} \left[ \frac{E_p}{2} + T \ln\left( 1 - e^{-\frac{E_p}{T}}\right)\right]. \label{distengles}
\end{align}

\noindent
So, the temperature-dependent part has been disentangled from the temperature-independent part in $V^{(1)}(\phi_c,T)$, where the latter is nothing but the famous 1PI-Coleman-Weinberg (CW) loop correction. This can be made clear using the identity,
\begin{align}
\frac{E_p}{2} = -\frac{i}{2}\int_{-\infty}^\infty \frac{dz}{2\pi} \ln(-z^2 + E_p^2 - i\varepsilon),
\end{align}
where the integration on the RHS can be evaluated using the residue theorem. Then, the temperature-independent term becomes
\begin{align}
\int \frac{d^3 p}{(2\pi)^3} \frac{E_p}{2} &= -\frac{i}{2}\int \frac{d^4p}{(2\pi)^4} \ln(-(p^0)^2 + E_p^2 - i\varepsilon),\\
&= \frac{1}{2}\int \frac{d^4p_{\text{\tiny E}} }{(2\pi)^4} \ln(p_{\text{\tiny E}}^2 + m^2(\phi_c) ).\label{cwpotential}
\end{align}
Where in the second line, we make a Wick rotation to Euclidean space ($p^0 \to ip^0_{\text{\tiny E}}$). The result in Eq.\eqref{cwpotential} is the 1-loop CW quantum correction, which is temperature independent. It is straightforward to note that the temperature-dependent part of $V^{(1)}(\phi_c,T)$ is not UV-divergent, since the UV-modes in the integral will be naturally regulated by the logarithmic function, $\lim\limits_{p\to \infty} \ln(1-e^{-E_p/T}) = 0$. On the other hand, the temperature-independent part is clearly UV-divergent. Yet, it is just the usual CW-contribution which would be regulated identically to the vacuum QFT where the UV-divergence will be regulated by some scale, $\bar\mu$, which could be eliminated later via RGE. So, the complete 1PI-thermal correction, $V^{(1)}(\phi_c,T)$ can be written as
\begin{align}
    V_B^{(1)}(\phi_c,T) = V^{(1)}_{B,\cw}(\phi_c,\bar\mu) + V^{(1)}_{B,\beta}(\phi_c,T)
\end{align}
The effective potential finally becomes
\begin{align}
    V_\eff(\phi_c,T) = V_\tl + V^{(1)}_{B,\cw}(\phi_c,\bar\mu) + V^{(1)}_{B,\beta}(\phi_c,T)\label{bosoniccontribu}
\end{align}
\noindent
 Obviously, in the $T\longrightarrow 0$ limit, we restore the usual QFT structure, which is UV-divergent. Yet, at high temperatures, the thermal contribution $V^{(1)}_{B,\beta}(\phi_c,T)$  dominates numerically while the UV renormalization is controlled by the $T=0$ counterterms, so the overall effective potential can be guaranteed to be UV-finite $V_\eff(\phi_c,T) \approx V_\tl + V^{(1)}_{B,\beta}(\phi_c,T)$. Unfortunately, another divergence badly emerges at high temperatures in the IR-range ($\ln(1- e^{-\omega/T}) = \ln (0))$. Securing this severe IR-divergence has to be done through careful treatment of the bosonic thermal mass, as it clearly becomes worse at higher temperatures. We will return to this challenge later in section \ref{IRproblems}.

\subsection{Fermion Fields}
Similarly to the bosonic fields, the n-point correlation functions of a queue of n-fermionic operators, $\aop = \psi\overline{\psi} \cdots,$ have to be weighted by the thermal corrections in a thermal bath,
\begin{align}
\expval{\aop} = \tz\left[e^{-\beta \hat{H}}\aop\right].
\end{align}
Fermions follow Fermi–Dirac statistics, which is encoded in the operator (canonical) formalism by the anticommutation relations, and in the path-integral formalism, it’s implemented via Grassmann fields. Either way, fermions are found to obey anticommutation relations [Check Appendix \ref{grassmanalgebra}], 
\begin{align}
\psi(x,\tau) &= -\psi(x,\tau+\beta).\label{fou112}
\end{align}

\noindent
Where, at finite temperature, the Kubo–Martin–Schwinger (KMS) condition then enforces antiperiodic imaginary-time boundary conditions for fermions, yielding odd Matsubara frequencies and the Fermi–Dirac distribution.
Following the rules of the Grassmann algebra, we can proceed to evaluate the fermions' contributions to the effective potential in a fashion similar to the scalars in the previous section. The fermion's action in Euclidean space is (check Eq.\eqref{imag_ferm_action})
\begin{align}
S_{\text{\tiny E}}[\psi,\overline{\psi}] = \int_0^\beta d^4x\,\, \overline{\psi}(\gamma^0_E \partial_\tau + \gamma^k_E\partial_k + m) \psi.  
\end{align}
The free generating functional will then be
\begin{align}
\mathcal{Z}[0] &= \int_{\psi(\beta) = -\psi(0)}\pd \psi \pd \overline{\psi} \,\, \exp{\left\{- \int_0^\beta d^4x \,\,\overline{\psi}(\gamma^0_E \partial_\tau + \gamma^k_E\partial_k + m) \psi\right\}}\label{fermion_generating_func}
\end{align}

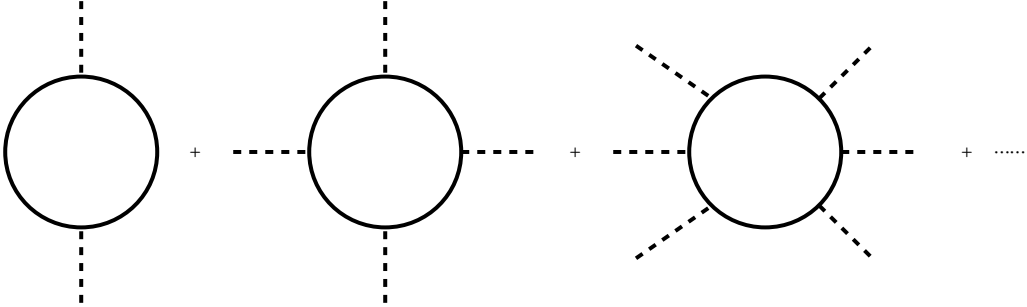
\begin{figure}[htb!]
\centering
\begin{tikzpicture}
\draw[dashed,line width = 1.5pt] (0,2) -- (0,1);
\draw[dashed,line width = 1.5pt]  (0,-2) -- (0,-1);
\draw[line width = 1.5pt]  (0,0) circle (1);
\draw[dashed,line width = 1.5pt] (2,0) -- (3,0);
\draw[dashed,line width = 1.5pt]  (5,0) -- (6,0);
\draw[dashed,line width = 1.5pt] (4,2) -- (4,1);
\draw[dashed,line width = 1.5pt]  (4,-2) -- (4,-1);
\draw[line width = 1.5pt]  (4,0) circle (1);
\draw[dashed,line width = 1.5pt] (7,0) -- (8,0);
\draw[dashed,line width = 1.5pt]  (10,0) -- (11,0);
\draw[dashed,line width = 1.5pt] (7.3,1.41) -- (8.3,0.71);
\draw[dashed,line width = 1.5pt]  (9.71,0.71) -- (10.41,1.41);
\draw[dashed,line width = 1.5pt] (7.3,-1.41) -- (8.3,-0.71);
\draw[dashed,line width = 1.5pt]  (9.71,-0.71) -- (10.41,-1.41);
\draw[line width = 1.5pt]  (9,0) circle (1);
\begin{scriptsize}
\draw[] (1.5,0) node {$+$};
\draw[] (6.5,0) node {$+$};
\draw[] (12,0) node {$+ \quad \cdots\cdots $};
\end{scriptsize}
\end{tikzpicture}
\caption{The diagrammatic representation of the 1-loop fermion contributions to the effective potential obtained from the imaginary time approach at thermal equilibrium.}
\label{fermion-img-time}
\end{figure}

\noindent
Similarly to Eq.\eqref{scalarz0}, keeping in mind the Grassmann properties of the fermionic field, the fermions generating functional becomes
\begin{align}
\mathcal{Z}[0] &= \sqrt{\text{det}[(\gamma^0_E \partial_\tau + \gamma^k_E\partial_k + m)(-\gamma^0_E \partial_\tau +\gamma^k_E\partial_k + m)]}. \label{fermionsfreeZ}
\end{align}
Where the integral in Eq.\eqref{fermion_generating_func} is computed through Eq.\eqref{gaussian_fermion}\footnote{The  $\text{det} \,[\gamma^0_E \partial_\tau + \gamma^k_E\partial_k + m]$ is written as $\sqrt{\text{det}\, [\gamma^0_E \partial_\tau + \gamma^k_E\partial_k + m)(-\gamma^0_E \partial_\tau + \gamma^k_E\partial_k + m]}$ to reduce the spinor-valued operator into a scalar one.}. Consequently, the fermionic effective potential becomes
\begin{align}
V_\eff^F(\phi_c) &= \frac{T}{2\Omega} \tz \Big[ \ln(\gamma^0_E \partial_\tau + \gamma^k_E\partial_k + m)(-\gamma^0_E \partial_\tau +\gamma^k_E\partial_k + m)\Big],\\
&= (-4N_c) \frac{T}{2}\sum_{n=-\infty}^\infty\int\frac{d\vec{p}}{(2\pi)^3} \ln[\omega_n^2 + \abs{\vec{p}}^2 + m^2(\phi_c)],\label{summedfermionint}
\end{align}
This is very similar to the bosonic result in Eq.\eqref{summedscalarint} with a few changes. The minus sign is an artifact of the Grassmann properties of the fermion fields. And an extra $4N_c$ factor reflecting the fact that fermions come in doublets (fermion and antifermion) with two possible spin states for each one of them and with color number $N_c$. In addition, the sum now runs only over the odd values of $n$.
Similar to the argument leading to Eq.\eqref{eq11} and Eq.\eqref{eq22} in the scalar sector, the second term in Eq.\eqref{summedfermionint}, is now referring to a complete 1PI-loop correction to the TL-fermion potential that is diagrammatically equivalent to the ones in Fig.\ref{fermion-img-time}.

\noindent
Following the steps developed in the scalar field part, we can separate the loop correction to the TL-fermion potential into temperature-dependent and independent contributions through evaluating the infinite sum in Eq \eqref{summedfermionint}. In \cite{laine2017basics}, there is a useful relation connecting the bosonic sum and the fermionic sum  given by,
\begin{align}
\sum_F(T) = 2\sum_B \left(T/2\right) - \sum_B(T).\label{sumrel}
\end{align}
Which directly leads to
\begin{align}
  \sum_F \ln[\omega_n^2 + \abs{\vec{p}}^2 + m^2(\phi_c)]  = \frac{1-n_F(E_p)}{2TE_p}.
\end{align}
Where, the fermion distribution function, according to Eq.\eqref{sumrel} is related to the boson distribution function through,
\begin{align}
n_F(E_p) = - [2n_B(2E_p) - n_B(E_p)].
\end{align}
As a result, the fermions' contribution to the effective potential becomes\footnote{Here we omitted the integration over $dm^2$ performed in Eq.\eqref{distengles} and only focused on the fermionic loop-thermal contribution.},
\begin{align}
V_F^{(1)}(\phi_c, T) &=  (-4N_c) \int \frac{d^3p}{(2\pi)^3} \left[ \frac{E_p}{2} + \frac{T}{2}\ln\left( 1 + e^{-\frac{E_p}{T}}\right)\right],\\
&=n_fV^{(1)}_{F,\cw}(\phi_c,\Lambda) + n_fV^{(1)}_{F,\beta}(\phi_c,T)\label{fermionveff}
\end{align}
Where $n_f=-4N_c$ refers to the fermion degrees of freedom in this case, as we only consider a particle and antiparticle case ($\psi,\overline{\psi}$). this result is quite promising as it shows that the fermionic effective potential will not suffer from IR-divergence as $T\rightarrow \infty$, at least at LO, unlike bosonic contributions. This is because fermions do not have vanishing Matsubara modes, unlike bosons, as we will see later.

\subsection{Vector Gauge Fields}\label{vectorfields}

The gauge–field Lagrangian can be written as
\begin{align}
\lag_{\mathrm{gauge}}
= -\,\frac{1}{4}\,W^{a}_{\mu\nu} W^{a\,\mu\nu}
   - \frac{1}{4}\,B_{\mu\nu} B^{\mu\nu},\label{weaklag}
\end{align}
where $W^{a}_{\mu\nu}$ and $B_{\mu\nu}$ are the field–strength tensors of $\mathrm{SU}(2)_L$ and $\mathrm{U}(1)_Y$,
\begin{align}
W^a_{\mu\nu} &= \partial_\mu W^a_\nu - \partial_\nu W^a_\mu + g f^{abc} W^b_\mu W^c_\nu,\\
B_{\mu\nu} &=\partial_\mu B_\nu - \partial_\nu B_\mu.
\end{align}
where $f^{abc}$ is the structure constant characterizing the non-commutativity of the non-abelian gauge fields. A salient difference between gauge fields and matter fields (scalars or fermions) is the presence of
redundant components in their field representation. By Wigner’s classification of unitary
irreducible representations of the Poincaré group, a spin–1 particle carries three physical
polarizations if massive and two if massless. The vector field $W_\mu\,(B_\mu)$ therefore introduces
components that are not independent physical degrees of freedom: for a massless gauge boson,
the time–like and longitudinal components are gauge artifacts, while for a massive vector the
longitudinal mode becomes physical but the time–like component remains constrained.
This redundancy manifests itself in quantization. In the canonical approach, one finds a primary
constraint on the time component of the canonical momentum, $\pi_\mu$,
\begin{align}
\pi_\mu &= \frac{\partial \lag}{\partial (\partial_0 W_\mu)} = - W^a_{0\mu}.\label{canonicalrelat}
\end{align}
A similar result is also obtained for $B_\mu$ field. Consequently, we cannot construct the corresponding commutation relation between the field and its canonical momentum,
\begin{align}
 [W^a_\mu(x), \pi^b_\nu(y)] = i\delta^a_b\,\, g^{\mu\nu}\, \delta(x-y)   
\end{align}
as it is clearly violated by $\pi_0$ that has a null result due to the antisymmetry of $W^a_{\mu\nu}$. 
In the path–integral formulation, the same issue appears as a non–non-invertible quadratic operator: the inverse propagator has zero modes and the functional integral is ill–defined. In vacuum QFT, The free generating functional is given by,
\begin{align*}
\gz[0]
= \int \mathcal{D}W^a\,\mathcal{D}B_\mu\;
\exp\!\Bigg\{\, i\!\int d^4x\,
\Big[
&\frac{1}{2}\,W^{a}_{\mu}\big(-\Box\, g^{\mu\nu}+\partial^\mu\partial^\nu\big)W^{a}_{\nu}\\
&+\frac{1}{2}\,B_{\mu}\big(-\Box\, g^{\mu\nu}+\partial^\mu\partial^\nu\big)B_{\nu}
\Big]\Bigg\}.
\numberthis\label{weakkernel}
\end{align*}
Which is obtained starting from the weak gauge–field Lagrangian in Eq.~\eqref{weaklag}, then integrating the kinetic terms by parts to obtain the quadratic operator appearing in Eq.~\eqref{weakkernel}. The cubic
and quartic self–self-interactions are omitted here since we require only the free generating functional.
As written, however, the path integral cannot be evaluated: the quadratic kernel $ -\Box\,  _{\mu\nu}+\partial_\mu\partial_\nu$  is non-invertible, so its functional determinant is singular. This reflects the redundant (gauge) degrees of freedom present in vector fields: infinitely many field configurations related by gauge transformations yield the same physics, making the Gaussian kernel degenerate. In canonical language, the same issue appears as a primary constraint for the time component
($W^a_0,\, B_0$), which prevents the naive imposition of equal–time commutators. The standard cure —both canonically and in the path integral— is to fix the gauge so that the kinetic operator becomes invertible. In a covariant $R_\xi$ gauge one adds the gauge–fixing functionals
\begin{align}
\lag_{\text{GF}}
= -\,\frac{1}{2\xi_W}\,(\partial_\mu W^{a\,\mu})^2
  -\,\frac{1}{2\xi_B}\,(\partial_\mu B^\mu)^2.\label{gaugefixinglag}
\end{align}
The total gauge–fixed Lagrangian for the free theory is therefore
\begin{align}
\mathcal L_{\text{gauge+GF}}
= -\frac{1}{4}\,W^a_{\mu\nu}W^{a\,\mu\nu}
  -\frac{1}{4}\,B_{\mu\nu}B^{\mu\nu}
  -\frac{1}{2\xi_W}\,(\partial_\mu W^{a\,\mu})^2
  -\frac{1}{2\xi_B}\,(\partial_\mu B^\mu)^2.
\label{weakgaugefix}
\end{align}
In particular, in Feynman gauge (\(\xi_W=\xi_B=1\)) the canonical momenta for the weak fields
become
\begin{align}
\pi^{a}_\mu
\;\equiv\;
\frac{\partial\lag_{\text{gauge+GF}}}{\partial(\partial_0 W^{a\,\mu})}
= -\,W^{a}_{0\mu}\;-\;\frac{1}{\xi_W}\,g_{\mu 0}\,\partial_\lambda W^{a\,\lambda},
\label{newcanonicalrelat}
\end{align}
and analogously for $B_\mu$ with $\xi_B$. Thus, the problematic time component is no longer constrained to vanish:
$\pi^{a}_{ 0}=-(1/\xi_W)\,\partial_\lambda W^{a\,\lambda}$,
while the spatial momenta $\pi^{a}_{ i}=-W^{a}_{0i}$ are unchanged (since $g_{i0}=0$). This gauge fixing eliminates the redundancy, renders the quadratic operator invertible, and provides well–defined propagators for use in the generating functional. Where the canonical quantization can proceed: with the gauge–fixing terms in place the
equal–time commutators for the weak gauge fields take the standard canonical form
\begin{align}
[W^a_\mu (t,\vec{x}), \pi^b_\nu(t,\vec{y})] &= i \delta^a_b\, g_{\mu \nu} \, \delta^{(3)}(\vec{x} - \vec{y})\label{commutation1}\\
[W^a_\mu (t,\vec{x}), W^b_\nu(t,\vec{y})] &= [\pi^a_\mu (t,\vec{x}), \pi^b_\nu(t,\vec{y})] = 0.\label{commutaion2}
\end{align}
And on the path integral side, the GF term modifies the fields' quadratic operators to be
\begin{align}
    K_W^{\mu\nu} &=-\Box\,g^{\mu\nu}+\Big(1-\frac{1}{\xi_W}\Big)\partial^\mu\partial^\nu,\\
    K_B^{\mu\nu} &=-\Box\,g^{\mu\nu}+\Big(1-\frac{1}{\xi_B}\Big)\partial^\mu\partial^\nu,
\end{align}
which are safely invertible for generic $\xi_{W,B}$.\\

Obviously, the gauge–fixing terms restore a well–defined canonical and path–integral quantization: the
Euler–Lagrange equation for the auxiliary field enforces the gauge condition (Lorenz gauge in the simplest case) $(\partial_\mu W^{a\,\mu}=0$ and $\partial_\mu B^\mu=0$.
While these terms preserve Lorentz invariance, they explicitly break gauge invariance, so the
Covariant quantization is carried out in a Hilbert space with an indefinite metric, which necessarily contains unphysical (longitudinal and time–like) polarizations. In other words, the gauge fixing terms change the theory in such a way that allows for non-physical states to occupy Hilbert space. This could be seen from the commutation relations in Eq.\eqref{commutation1} and Eq.\eqref{commutaion2}, where the spatial derivative of Eq.\eqref{commutaion2} gives a vanishing result,
\begin{align}
\frac{\partial}{\partial y^i}[W^a_\mu (t,\vec{x}), W^b_\nu(t,\vec{y})] = [W^a_\mu (t,\vec{x}), \frac{\partial}{\partial y^i}W^b_\nu(t,\vec{y})] = 0.\label{modifiedcomm}
\end{align}
Setting $\pi^a_\nu = \partial_0W^a_\nu$ in Eq.\eqref{commutation1} we find that the spatial-derivative parts will vanish according to Eq.\eqref{modifiedcomm} leading to
\begin{align}
[W^a_\mu (t,\vec{x}),\partial_0 W^b_\nu(t,\vec{y})] = -i \delta^a_b\, g_{\mu \nu} \, \delta^{(3)}(\vec{x} - \vec{y}).
\end{align}
The $-\,g_{\mu\nu}$ implies that the time–like component carries the opposite sign relative to spatial components, signalling an indefinite norm. Which means that the corresponding Hilbert space encompasses unphysical states. In the Abelian gauge theories, this was fixed by restricting the physical states to satisfy the Gupta-Bleuler condition to project out the unphysical sector,
\begin{align}
\expval{\text{phys}'|\partial_\mu (A_+^\mu + A_-^\mu)|\text{phys}}=0,\label{guputa}
\end{align}
where the positive–frequency part of the gauge–fixing operator annihilates physical states,
which removes the time–like/longitudinal photon polarizations from the spectrum and restores a
positive–definite inner product on the physical subspace.  However, this is only valid for abelian fields as $\partial_\mu A^\mu$ satisfies the free Klein-Gordon equation, unlike the non-abelian gauge field $W^a_\mu$.  This indicates that there is still some missing contribution/symmetry ( BRST symmetry) necessary to restore the global symmetry of the theory. So the condition in Eq.\eqref{guputa} has to be replaced by another one of the form (Check chapter 4 in \cite{das2023finite} for the detailed proof),
\begin{align}
\expval{\text{phys}'|\hat{\mathcal{O}}_{\text{\tiny BRST}}|\text{phys}}=0.\label{brst}
\end{align}
Here, $\hat{\mathcal{O}}_{\text{\tiny BRST}}$ is the sum of the gauge fixing term and the missing contribution that cancels the unphysical states created by breaking the gauge invariance. This is best illustrated in the path integral formulation, where we can use the identity that the path integral is unaffected by field redefinition.
The generating functional of the $W^a_\mu$ field is
\begin{align}
    \gz[0] = \int \pd W^a_\mu \exp\left\{i\int d^4 x \,\, \lag(W^a_\mu)\right\}.\label{gaugefixingZ}
\end{align}
Here, $W^{a\,\mu}$ is gauge transformed as,
\begin{align}
    W^{a\,\mu} \rightarrow W^{a\,\mu} + \partial^\mu \eta^a + f^{abc} W^{b~\mu} \eta^c = W^{a\,\mu} + D^\mu \eta^a,\label{gaugetransformation}
\end{align}
such that $\eta$ is an infinitesimal transformation and $D_\mu \eta^a = \partial_\mu \eta^a + f^{abc} W^b_\mu \eta^c$ is the covariant derivative. Now we can make use of the path integral identity
\begin{align}
    f[W_\mu] = \int \pd \eta^a \exp\left\{-i \int d^4 x \, \frac{1}{2\xi} (\partial_\mu W^{a\,\mu} - \partial^\mu D_\mu \eta^a)^2\right\},
\end{align}
to write Eq.\eqref{gaugefixingZ} as
\begin{align}
     \gz[0]  &= \int \pd \eta^a \pd W_\mu \,\,\frac{1}{f[W_\mu]} \,\, \exp\left\{i\int d^4 x \,\, \lag(W^a_\mu)  - \frac{1}{2\xi} (\partial_\mu W^{a\,\mu} - \partial^\mu D_\mu \eta^a)^2\right\},\label{Wredifining}\\
    &= \int \pd \eta^a \pd W^a_\mu \,\,\frac{1}{f[W_\mu]} \,\, \exp\left\{i\int d^4 x \,\, \lag(W^a_\mu)  - \frac{1}{2\xi} (\partial_\mu W^{a\,\mu})^2\right\},
\end{align}
where in the second step we redefined the gauge field $W^a_\mu$ according to Eq.\eqref{gaugetransformation}. This way, introducing $f[W_\mu]$ to the generating functional is absorbed in the numerator of Eq.\eqref{Wredifining} via field redefinition. This, in a way, is similar to dividing by $\gz[0]$ to get rid of the disconnected bubbles we mentioned earlier. We can make this clearer by integrating $f[W_\mu]$ using the Gaussian trick, since it is quadratic in $\eta^a$,
\begin{align}
    f[W] &= \frac{\text{const.}}{\sqrt{\text{det} (\partial^\mu D_\mu)^2}}\\
    &= \text{const.}\,\, \text{det} (\partial^\mu D_\mu)^{-1}.\label{ghosts}
\end{align}
Hence, the path integral over the gauge orbits $\eta$ returns an output similar to the path integral of the fermionic fields that obey the Grassmann algebra despite their bosonic nature. So, we can rewrite Eq.\eqref{ghosts} as a path integral over Grassmann-like variables,
\begin{align}
    \text{det} (\partial^\mu D^{ab}_\mu) = \int \pd \overline{c}\pd c\,\, \exp\left\{i\int d^4x\,\, \overline{c}^a (-\partial^\mu D_\mu^{ab}) c^b \right\}.\label{gohstsinsertion} 
\end{align}
Where $c^a$ is the Faddeev-Popov ghosts. They are named ghosts due to their mixed properties illustrated above: obeying the Grassmann algebra despite being of a bosonic nature. Then the overall contribution to the generating functional of the non-abelian gauge fields becomes
\begin{align}
    \gz[0] &= \int \pd \overline{c} \pd c \pd W^a_\mu \,\,\exp\left\{i\int d^4 x \,\, \lag_\mathrm{EW}(W^a_\mu,\xi,c) \right\}.
\end{align}
Where,
\begin{align}
    \lag_\mathrm{EW}(W^a_\mu,\xi,c) &= \lag_\mathrm{gauge} + \lag_\mathrm{GF} + \lag_\mathrm{ghost},
    \intertext{where $\lag_\mathrm{ghost}$ can be read of Eq.\eqref{gohstsinsertion}}
    \lag_{\text{\tiny ghost}} &= - \overline{c}^a \partial^\mu D_\mu^{ab} c^a. \label{nonabelianghosts}
\end{align}
And $\lag_\mathrm{gauge},\,\lag_\mathrm{GF}$ are given in Eq.\eqref{weaklag} and Eq.\eqref{gaugefixinglag}. Now the corresponding operator of the Gupta-Behealar condition in the abelian gauge fields would be
\begin{align}
  \hat{\mathcal{O}}_{\text{\tiny BRST}} =  \lag_\mathrm{GF} + \lag_\mathrm{ghost}\label{brst2}
\end{align}
for the non-abelian gauge theories. Eq.\eqref{nonabelianghosts} explains why we do not get similar ghost contributions in abelian theories. As for abelian theories $D_\mu \to \partial_\mu$, the ghost fields satisfy the free Klein-Gordon equation, $\Box c = 0$, leading to a vanishing $\lag_\text{\tiny ghost}$. \\

\noindent
We now extend the construction to finite temperature following the same bosonic scheme used for
scalars, but including the full electroweak gauge sector and ghosts. In the imaginary time
(Matsubara) Formalism, the fields are periodic in Euclidean time $\tau\in[0,\beta]$:
\begin{align}
W_\mu^a(\tau+\beta,\vec{x})=W_\mu^a(\tau,\vec{x}),\qquad
B_\mu(\tau+\beta,\vec{x})=B_\mu(\tau,\vec{x}),
\end{align}
and, although Grassmann, the Faddeev–Popov ghosts are also \emph{periodic} so as to cancel the
unphysical gauge modes:
\begin{align}
c^a(\tau+\beta,\vec{x})=c^a(\tau,\vec{x}),\qquad
\bar c^{\,a}(\tau+\beta,\vec{x})=\bar c^{\,a}(\tau,\vec{x}).
\end{align}
The Euclidean generating functional for the $\mathrm{SU}(2)_L\times \mathrm{U}(1)_Y$ free (quadratic) theory in
covariant $R_\xi$ gauges is
\begin{align*}
\gz[0]
= \int_{\substack{W(\beta)=W(0)\\ B(\beta)=B(0)\\ c(\beta)=c(0)}}
\pd \overline{c}\,\pd c\,\pd W\,&\pd B\,\,
\exp\Bigg\{-\!\int_{0}^{\beta}\! d\tau \int d^3x\;
\Big[
\frac{1}{4}\,W^a_{\mu\nu}W^{a}_{\mu\nu}
+\frac{1}{2\xi_W}(\partial_\mu W^{a}_{\mu})^2
+\frac{1}{4}\,B_{\mu\nu}B_{\mu\nu}\\
&+\frac{1}{2\xi_B}(\partial_\mu B_{\mu})^2
+\bar c^{\,a}(-\partial_\mu D^{ab}_{\mu})c^{\,b}
\Big]\Bigg\},
\numberthis\label{gaugeZfinal}
\end{align*}
where all indices are Euclidean and $(D^{ab}_\mu=\delta^{ab}\partial_\mu+f^{acb}W^c_\mu$). Keeping only the quadratic (free) part and integrating by parts, the action can be written as
\begin{align}
S_E^{(2)}
= \int_{0}^{\beta}\! d\tau \int d^3x\;
\Big[
\frac{1}{2}\,W^{a}_{\mu}\,K_W^{\mu\nu}\,W^{a}_{\nu}
+\frac{1}{2}\,B_{\mu}\,K_B^{\mu\nu}\,B_{\nu}
+\bar c^{\,a}(-\partial^2_E)\,c^{\,a}
\Big],
\end{align}
with the Euclidean kernels
\begin{align}
K_V^{\mu\nu} \;=\; -\,\partial_E^2\,\delta^{\mu\nu}
+\Big(1-\frac{1}{\xi_V}\Big)\partial^\mu\partial^\nu,
\qquad
\partial_E^2 \equiv \partial_\tau^2+\nabla^2,
\qquad V\in\{W,B\}.
\end{align}
Fourier–expanding the periodic fields, one obtains the diagonal quadratic form in momentum space,
\begin{align*}
S_E^{(2)}
= \sum_{n}\!\int\!\frac{d^3p}{(2\pi)^3}\,
\Bigg\{
&\frac{1}{2}\,W^{a}_{\mu}(-p)\,\Big[p_E^2\,\delta^{\mu\nu}
-\Big(1-\frac{1}{\xi_W}\Big)p^\mu p^\nu\Big]\,W^{a}_{\nu}(p)\\
&+\frac{1}{2}\,B_{\mu}(-p)\,\Big[p_E^2\,\delta^{\mu\nu}
-\Big(1-\frac{1}{\xi_B}\Big)p^\mu p^\nu\Big]
 B_{\nu}(p) +\bar c^{\,a}(-p)\,p_E^2\,c^{\,a}(p)
\Bigg\}, \numberthis\label{momentumZ}
\end{align*}
where $p_E^2=\omega_n^2+\abs{\vec{p}}^2$.
This structure makes explicit the role of the Faddeev–Popov ghosts. In a covariant gauge, the
vector fields carry, besides the two physical transverse polarizations, a longitudinal and a time–like component. A naive evaluation of the free energy would therefore overcount degrees of freedom (for massless vectors, yielding twice the blackbody value) \cite{laine2017basics, das2023finite}. The ghost determinant
precisely cancels the unphysical longitudinal/time–like contribution of the gauge fields, leaving only the physical transverse modes. In the Abelian sector, the ghost is free and decouples (no interactions), but its contribution at the quadratic level still enforces the same cancellation
pattern. Evaluating the Gaussian integrals in Eq.\eqref{momentumZ}, following Eq.\eqref{freeenergytrace} and Eq.\eqref{fermionsfreeZ},  and applying the standard trace–log relations for periodic bosonic Matsubara sums, one recovers the expected Stefan–Boltzmann free energy for $2$ polarizations per massless gauge boson (times the group multiplicity), in agreement with BRST condition in Eq.\eqref{brst} and Eq.\eqref{brst2}.\\

\noindent
A final remark is that, in a thermal medium, the Lorentz symmetry $O(1,3)$ is reduced to spatial
rotations $O(3)$ by the presence of the plasma rest frame with four–velocity $u^\mu$ (in the plasma
frame $u^\mu=(1,\vec{0})$. As a result, gauge–field correlators naturally decompose into
transverse, longitudinal, and pure–gauge components that behave differently at finite temperature. This is analogous to the Zeeman splitting of spin multiplets in an external
magnetic field: here, the thermal bath selects the temporal direction. In particular, at leading
order in perturbation theory, the longitudinal (electric) sector acquires a thermal Debye mass, whereas the transverse (magnetic) sector does not. To make this explicit, work in Euclidean momentum space with $p^\mu=(\omega_n,\vec{p})$, and define
$\tilde p^\mu \equiv p^\mu-(p\!\cdot\!u)\,u^\mu$ so that $\tilde p^\mu u_\mu=0$.
A convenient set of projectors is
\begin{align}
P_T^{\mu\nu}(p) &= \delta^{\mu i}\delta^{\nu j}\!\left(\delta^{ij}-\hat p^{\,i}\hat p^{\,j}\right),
\quad
P_G^{\mu\nu}(p) = \frac{p^\mu p^\nu}{p_E^2},\quad
P_L^{\mu\nu}(p) = \delta^{\mu\nu}- P_T^{\mu\nu}(p) - P_G^{\mu\nu}(p),
\label{eq:projectors}
\end{align}
where $i,j$ are spatial indices and $\hat p^{\,i}\equiv p^i/|\vec{p}|$. These obey
$ P_T+P_L+P_G=\delta^{\mu\nu},\,\, P_X P_Y=\delta_{XY}P_X$ for $X,Y\in\{T,L,G\}$, and consequently, they diagonalize the quadratic form in covariant gauges. We therefore change variables linearly,
\begin{align}
W_\mu^{a} \;\to\; W_{\mu}^{a\,T}+W_{\mu}^{a\,L}+W_{\mu}^{a\,G},
\qquad
W_{\mu}^{a\,X} \equiv P^{X}_{\mu\nu} W^{a\,\nu},\quad X\in\{T,L,G\},
\end{align}
and analogously for the Abelian field \(B_\mu\). The Jacobian is constant and can be absorbed into
the overall normalization. Starting from the quadratic Euclidean action in momentum space in  Eq.~\eqref{momentumZ}, the decomposition with \eqref{eq:projectors} gives (no mixing among \(T,L,G\))
\begin{align*}
S_E^{(2)} = \sum_{n}\,\int\,&\frac{d^3p}{(2\pi)^3}\Bigg\{
\frac{1}{2}\,W^{a\,T}_{\mu}(-p)\,p_E^2\,W^{a\,T}_{\mu}(p)
+\frac{1}{2}\,W^{a\,L}_{\mu}(-p)\,p_E^2\,W^{a\,L}_{\mu}(p)
+\frac{1}{2}\,W^{a\,G}_{\mu}(-p)\,\frac{p_E^2}{\xi_W}\,W^{a\,G}_{\mu}(p)\\
&+\bar c^{\,a}(-p)\,k_E^2\,c^{\,a}(p) +\frac{1}{2}\,B^{T}_{\mu}(-p)\,p_E^2\,B^{T}_{\mu}(p)
+\frac{1}{2}\,B^{L}_{\mu}(-p)\,p_E^2\,B^{L}_{\mu}(p)
+\frac{1}{2}\,B^{G}_{\mu}(-p)\,\frac{p_E^2}{\xi_B}\,B^{G}_{\mu}(p)
\Bigg\}.\numberthis \label{modifiedZ0}
\end{align*}
Adding sources in each sector and inverting the quadratic forms, the free Euclidean propagators are obtained to be
\begin{align}
\langle W^{a\,T}_{\mu} W^{b\,T}_{\nu}\rangle(p) &= \delta^{ab}\,\frac{P^{T}_{\mu\nu}(p)}{p_E^2},
&
\langle W^{a\,L}_{\mu} W^{b\,L}_{\nu}\rangle(p) &= \delta^{ab}\,\frac{P^{L}_{\mu\nu}(p)}{p_E^2},
&
\langle W^{a\,G}_{\mu} W^{b\,G}_{\nu}\rangle(p) &= \delta^{ab}\,\xi_W\,\frac{p_\mu p_\nu}{p_E^4},\\
\langle B^{T}_{\mu} B^{ }_{\nu}{}^{T}\rangle(p) &= \frac{P^{T}_{\mu\nu}(p)}{p_E^2},
&
\langle B^{L}_{\mu} B^{ }_{\nu}{}^{L}\rangle(p) &= \frac{P^{L}_{\mu\nu}(p)}{p_E^2},
&
\langle B^{G}_{\mu} B^{ }_{\nu}{}^{G}\rangle(p) &= \xi_B\,\frac{p_\mu p_\nu}{p_E^4}.
\end{align}
Equivalently, recombining the sectors,
\begin{align}
\langle W^{a}_{\mu} W^{b}_{\nu}\rangle(p)
&= \delta^{ab}\left[\frac{P^{T}_{\mu\nu}(p)}{p_E^2}+\frac{P^{L}_{\mu\nu}(p)}{p_E^2}
+\xi_W\,\frac{p_\mu p_\nu}{p_E^4}\right],\label{completepropag}\\
\langle B_{\mu} B_{\nu}\rangle(p) &= \frac{P^{T}_{\mu\nu}(p)}{p_E^2}+\frac{P^{L}_{\mu\nu}(p)}{p_E^2}
+\xi_B\,\frac{p_\mu p_\nu}{p_E^4}
\end{align}
and the (periodic) ghost propagator is
\begin{align}
\langle \bar c^{\,a} c^{\,b}\rangle(p) = \frac{\delta^{ab}}{p_E^2},
\end{align}
with bosonic Matsubara frequencies despite its Grassmann nature, which ensures the cancellation of unphysical gauge modes in the thermal sums. At higher orders, medium effects generate distinct self–energies in the transverse and longitudinal sectors, $\Pi_T(\omega_n,\vec{p})$ and $\Pi_L(\omega_n,\vec{p})$. In particular, $\Pi_L(0,\vec{p}\,\to\,0)\equiv
m_D^2$ yields Debye screening, while $\Pi_T(0,\vec{p}\,\to\,0)=0$ at leading order \cite{espinosa1993nature,carrington1992effective,arnold1992phase}, consistent with the symmetry breaking $O(1,3)\to O(3)$ by the heat bath.\\

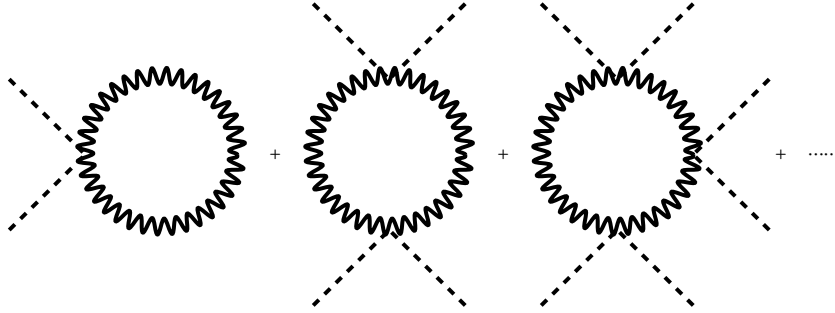
\begin{figure}[htb!]
\centering
\begin{tikzpicture}
\draw[dashed,line width = 1.5pt] (-1,1) -- (0,0);
\draw[dashed,line width = 1.5pt]  (-1,-1) -- (0,0);
\draw[line width = 1.5pt, decorate, decoration={snake, amplitude=3pt, segment length=5pt}]  (1,0) circle (1);
\draw[dashed,line width = 1.5pt] (3,2) -- (4,1);
\draw[dashed,line width = 1.5pt]  (5,2) -- (4,1);
\draw[dashed,line width = 1.5pt] (3,-2) -- (4,-1);
\draw[dashed,line width = 1.5pt]  (5,-2) -- (4,-1);
\draw[line width = 1.5pt, decorate, decoration={snake, amplitude=3pt, segment length=5pt}]  (4,0) circle (1);
\draw[dashed,line width = 1.5pt] (6,2) -- (7,1);
\draw[dashed,line width = 1.5pt]  (8,2) -- (7,1);
\draw[dashed,line width = 1.5pt] (6,-2) -- (7,-1);
\draw[dashed,line width = 1.5pt]  (8,-2) -- (7,-1);
\draw[dashed,line width = 1.5pt] (9,1) -- (8,0);
\draw[dashed,line width = 1.5pt]  (9,-1) -- (8,0);
\draw[line width = 1.5pt, decorate, decoration={snake, amplitude=3pt, segment length=5pt}]  (7,0) circle (1);
\begin{scriptsize}
\draw[] (2.5,0) node {$+$};
\draw[] (5.5,0) node {$+$};
\draw[] (9.5,0) node {$+ \quad \cdots\cdots $};
\end{scriptsize}
\end{tikzpicture}
\caption{The diagrammatic representation of the 1-loop contribution from the weak gauge fields to the effective potential at finite temperature following the imaginary time formulation.}
\label{vector-imag-time}
\end{figure}

\noindent
Now we calculate the effective potential of the gauge fields. In Feynman gauge, the gauge field contribution to the effective potential will have a similar structure to that of the scalars in Eq.\eqref{summedscalarint}, with a different number of degrees of freedom ($n_i=2(N_c^2-1)$, $N_c$: is the color factor) that counts only the physical components,
\begin{align}
V_\eff^G(\phi_c) &=n_i\frac{T}{2\Omega} \tz \Big[ \ln(-\partial_\tau^2 - \nabla^2  )\Big],\label{gaugefieldveff}\\
&=  n_i \frac{T}{2}\sum_{n=-\infty}^\infty\int\frac{d\vec{p}}{(2\pi)^3} \ln[\omega_n^2 + \abs{\vec{p}}^2 ]\label{gaugecontrib}\\
&= n_i V_B^{(1)}(\phi_c,T) \label{gaugeveff}
\end{align}
Which has a diagrammatic representation, Fig.\ref{vector-imag-time},  similar to the one in Eq.\eqref{eq22}, where
The evaluation of the summed integral in Eq.\eqref{gaugecontrib} goes exactly similarly to the corresponding ones in the scalar sector [Check Eq.\eqref{bosonicsum}:Eq.\eqref{bosoniccontribu}].\\

\noindent
Aside from the technical complications of quantizing non-Abelian gauge theories, their thermal
dynamics parallel those of scalars in many respects. Two salient differences are worth emphasizing. The gauge sector introduces the gauge parameter(s) $\xi_{W,B}$ through the covariant $R_\xi$ fixing, which explicitly breaks gauge invariance at the level of the Lagrangian and renders off–shell quantities —such as the generating functional and the (finite–temperature) effective potential— gauge dependent. This is not pathological: these objects are by definition off–shell and need not be gauge invariant. Physical observables derived from them (on–shell \(S\)-matrix elements, thermodynamic quantities at stationary points, etc.) must be gauge independent. In particular,
gauge invariance of scattering amplitudes is ensured by Ward–Takahashi/Slavnov–Taylor identities,
while the gauge independence of the effective potential at its extrema is controlled by the Nielsen identity. At any finite truncation (fixed loop order, partial resummations) a residual $\xi$–dependence can remain; when the calculation is performed consistently, this dependence cancels in physical predictions. We return to this point in Sec.~\ref{nilsen-idet}. Unlike the scalars in Eqs.~\eqref{freeenergytrace},  the gauge boson in \eqref{gaugefieldveff} has no explicit mass term 
because the weak–sector Lagrangian \eqref{weaklag} was written in the unbroken phase for clarity.
After electroweak symmetry breaking, $W^\pm$ and $Z$ acquire masses and mix with Goldstone modes; the ’t~Hooft $R_\xi$ gauge–fixing terms remove gauge–Goldstone mixing and give the corresponding ghost fields masses. Then adding the corresponding mass terms at the quadratic
level simply shifts the kernels by $+M_V^2\,\delta^{\mu\nu}$, and the momentum–space forms in
Eqs.~\eqref{momentumZ} become $p_E^2\,\delta^{\mu\nu}-\Big(1-\tfrac{1}{\xi_V}\Big)p^\mu p^\nu
\;\longrightarrow\; \big(p_E^2+M_V^2\big)\delta^{\mu\nu}-\Big(1-\frac{1}{\xi_V}\Big)p^\mu p^\nu, \qquad V\in\{W,B\},
$
so that the free propagators in Eq.\eqref{completepropag} acquire the usual massive denominators $p_E^2\!\to p_E^2+M_V^2$ (and, in the broken phase, should be written in
the mass eigenstate basis $W^\pm,Z,A$). The Abelian sector $B_\mu$ remains massless before symmetry breaking and mixes with $W^3_\mu$ afterward to form the photon $A_\mu$ and the
$Z$ boson in the standard way. These insertions propagate straightforwardly through the Gaussian path integral and subsequent sum–integral expressions, with ghosts remaining free in the Abelian sector and acquiring $\xi$-dependent masses in the non-Abelian broken sector as dictated by the
gauge fixing.

%%%%%%%%%%%%%%%%%%%%%%%%%%%%%%%%%%%%%%%%%%%%%%%%%%%%%%%%
%%%%%%%%%%%%%%%%%%%%%%%%%%%%%%%%%%%%%%%%%%%%%%%%%%%%%%%%

\section{Theoretical Uncertainties in the  effective potential}\label{uncertinitiessec}

While developing the thermal dynamics of scalars, fermions, and gauge fields, several theoretical
uncertainties arise that can substantially affect quantitative predictions. A universal issue across
all sectors is the ultraviolet (UV) divergence of the temperature–independent one–loop 1PI
contribution—the Coleman–Weinberg term $V_\cw^{(1)}(\chi,\bar\mu)$, which requires the
standard zero–temperature renormalization with $T\!=\!0$ counterterms \cite{coleman1973radiative,quiros1998,laine2017basics}. By contrast, infrared (IR) pathologies
are specific to the bosonic sector: the vanishing (zero) Matsubara mode of scalars and gauge
fields enhances long–wavelength contributions and generates IR divergences
\cite{laine2017basics,dolan1974symmetry,carrington1992effective,fendley1987effective,takahashi1985perturbative}.
A further complication is the gauge dependence of off–shell quantities such as the effective potential, introduced by covariant gauge fixing in the electroweak sector; while physical observables are gauge independent (Ward–Takahashi/Slavnov–Taylor identities), the potential
away from its extrema carries explicit $\xi$–dependence and must be handled with care (Nielsen identity) \cite{dolan1974gauge,andreassen2015,patel2011baryon,garny2012gauge}. At high temperature these issues intertwine: the bosonic zero mode drives a breakdown of naive
perturbation theory and can induce spurious imaginary parts of $V_{\rm eff}$ for certain field values,
both symptoms of IR sensitivity \cite{delaunay2008dynamics}. Systematic resummations (e.g.\ ring/daisy, hard–thermal–loop
and dimensionally reduced EFTs) are therefore required to restore a controlled expansion. We
return to these points and their quantitative impact in the next subsections.

\subsection{Divergences}

Both ultraviolet (UV) and infrared (IR) divergences arise in finite–temperature field theory.
The UV divergences originate from the temperature–independent part of the one–loop effective
action—i.e.\ the Coleman–Weinberg contribution $(V^{(1)}_\cw$ in Eq.~\eqref{cwpotential}—and are
precisely the remnants of the zero–temperature theory; the explicitly thermal pieces are UV–finite.
By contrast, IR singularities are specific to the bosonic sector and stem from the vanishing
(“zero”) Matsubara mode of scalars and gauge fields. The standard cures mirror those at
\(T=0\): UV divergences are removed by renormalization with zero–temperature counterterms,
while IR pathologies require appropriate resummations (e.g.\ ring/daisy, hard–thermal–loop,
or dimensional–reduction frameworks).
These procedures, when implemented at finite order, introduce residual theoretical uncertainties:
parametric scale dependence from truncating the perturbative series and explicit
gauge dependence of off–shell quantities (such as $V_\eff$ away from extrema), both of which
can impact the predicted phase–transition dynamics. In the subsections below, we examine the
UV and IR issues in turn, their respective remedies, the induced uncertainties, and practical
techniques to mitigate them.

\subsubsection{Sources and solution of UV-divergence}

From Eqs.~\eqref{cwpotential}, \eqref{fermionveff}, and \eqref{gaugeveff} it is clear that each
Coleman–Weinberg (CW) one–loop contribution is ultraviolet divergent as $p\to\infty$. These
divergences are unphysical and reflect the continuum (infinite) number of degrees of freedom in
QFT. Since they arise entirely from the temperature–independent ($T=0$) sector, they are treated
exactly as in vacuum QFT by renormalization. Among the standard regularization schemes, we
adopt dimensional regularization (DR), which preserves gauge invariance (unlike a hard
momentum cutoff) and is technically convenient.
In DR one analytically continues the loop integrals to $d=4-2\epsilon$ dimensions and introduces
the renormalization scale $\mu$ so that couplings remain dimensionless. A useful master
identity is
\begin{align}
\int\!\frac{d^{d}p}{(2\pi)^d}\,\frac{1}{(p^2+m^2)^{\alpha}}
= \frac{1}{(4\pi)^{d/2}}\,
\frac{\Gamma\!\big(\alpha-\tfrac{d}{2}\big)}{\Gamma(\alpha)}\,
\big(m^2\big)^{\tfrac{d}{2}-\alpha},
\label{feynmantrick}
\end{align}
According to Eq.\eqref{cwpotential}, the CW-potential of all fields in DR is given by\footnote{Here we separated each field's degree of freedom from the definition of $V_\cw$ as in Eq.\eqref{fermionveff}, and Eq.\eqref{gaugeveff} so that the integrand function is going to be identical to the scalar case.},
\begin{align}
    V^{(1)}_\cw = \frac{\mu^{4-d} }{2}\int \frac{d^d p}{(2\pi)^d} \log [p^2 + m^2(\phi_c)].\label{drcw}
\end{align}
Differentiating  Eq.\eqref{drcw} with respect to the squared mass and using the identity in Eq.\eqref{feynmantrick}, we get
\begin{align}
    \frac{dV^{(1)}_\cw}{dm^2(\phi_c)} = f(\varepsilon) \,\, [m(\phi_c)]^{2-2\varepsilon}.\label{dvcw}
\end{align}
Here, we set $d=4-2\varepsilon$ leading to an overall factor of $\varepsilon$ given by
\begin{align}
    f(\varepsilon) = \frac{1}{32\pi^2}\,\, \bar\mu^{2\varepsilon} \,\, \Gamma(-1 + \varepsilon),
\end{align}
where, $\bar\mu^2= 4\pi\mu^2e^{-\gamma_E}$ is the modified scale after absorbing the Euler-Mascheroni constant $\gamma_E$. Integrating Eq.\eqref{dvcw} and using the identities,
\begin{align*}
    A^{\pm \varepsilon} = 1 \pm \varepsilon\log A + \mathcal{O}(\varepsilon^2),\qquad
    \Gamma(-1+\varepsilon) = -\frac{1}{\varepsilon} + \gamma_E - 1 + \mathcal{O}(\varepsilon),
\end{align*}
the regularized CW-potential then becomes
\begin{align}
    V^{(1)}_\cw = \frac{m^4(\phi_c)}{64\pi^2}\left[ \left(-\frac{1}{\varepsilon} + \gamma_E \right) + \left(\log \frac{m^2(\phi_c)}{\bar\mu^2} - \frac{3}{2} \right)\right].\label{regularvcw}
\end{align}
This is the result we also obtain for the fermionic contribution, except with the corresponding fermion degree of freedom ($n_f=-4\times N_c$). However, for the gauge bosons, the remnant finite term becomes $-\frac{5}{6}$ instead of $-\frac{3}{2}$ as only the transverse component of the gauge tensor contributes in the CW-correction, while the longitudinal part is canceled by the corresponding ghosts contribution as we discussed in section \ref{vectorfields}. Also, gauge fields have a different degree of freedom factor $n_i$. The first term on the RHS of Eq.\eqref{regularvcw} is, of course, singular at $\varepsilon \to 0$. This term is proportional to $\lambda \phi_c^4$ where $m^2(\phi_c) = \frac{1}{2}\lambda \phi_c^2$ for $V_\tl(\phi_c) = \frac{1}{2}m^2 \phi_c^2 + \frac{1}{4!}\lambda \phi_c^4$. This singular term is exactly canceled by adding a counterterm to the tree-level potential $V_{\text{\tiny c.t.}} = \frac{\delta \lambda}{4!} \phi_c^4$ so that the coupling $\lambda$ in the tree-level potential is replaced by the renormalized one, $\lambda_\text{\tiny R} = \lambda + \delta \lambda$, such that
\begin{align}
    \delta \lambda = -\frac{m^4(\phi_c)}{64\pi^2}\left(-\frac{1}{\varepsilon} + \gamma_E \right).
\end{align}
Consequently, the UV-divergence is absorbed by renormalizing the quartic coupling in that example through introducing a divergent counterterm that exactly cancels the identical divergent term resulting from the CW-correction. After subtracting the corresponding counterterm contribution, the CW potential becomes UV-finite,
\begin{align}
    V^{(1)}_\cw = \frac{m^4(\phi_c)}{64\pi^2} \left(\log \frac{m^2(\phi_c)}{\bar\mu^2} - \frac{3}{2} \right).
\end{align}
However, this result is still sensitive to the scale $\bar\mu$, which is arbitrary and leads to different values for different choices, introducing the uncertainty of the effect potential dependent on the scale. This is a result of inconsistent truncation of the higher-order loop corrections. At fixed loop order, the effective potential retains a residual dependence on the (arbitrary) renormalization scale $\bar\mu$, because explicit $\ln(m^2/\bar\mu^2)$ terms are not fully canceled by the implicit running of the parameters. In the exact (all-orders) theory, observables extracted from $V_\eff$ are independent of $\bar\mu$; the residual scale dependence at finite order therefore quantifies missing higher-order
corrections. The renormalization-group equation (RGE) enforces this statement and organizes the resummation of the leading logarithms, thereby suppressing the $\bar\mu$-dependence. In Callan–Symanzik form, the RGE equation takes the form
\begin{align}
\left(\bar\mu\frac{\partial}{\partial\bar\mu}
+\sum_i \beta_{\lambda_i}\frac{\partial}{\partial \lambda_i}
+\beta_{m_1^2}\frac{\partial}{\partial m_1^2}
-\gamma\,\phi_c\,\frac{\partial}{\partial \phi_c}\right)
V_{\rm eff}(\phi_c,\bar\mu;\lambda_i,m_1^2)=0,
\label{RGEcallan}
\end{align}
where the sum runs over all dimensionless couplings $\lambda_i$ (gauge, Yukawa, quartic) and $m_1^2$ is the Higgs mass parameter. After all, the physical observables extraced from the effective potential have to be scale independent which is translated into a constraining condition on the effective potential which can be parameterized as \begin{align} \frac{d}{dt} V_\eff(\phi_c(t),\bar\mu(t);\lambda_i(t),m_1^2(t) ) =0. \end{align}
Hence, a set of differential equations for each parameter is obtained,
\begin{align} \frac{d \bar\mu}{dt} = \bar\mu\qquad \hspace{15pt} &\to \qquad \mu_*=\bar\mu(t) = \bar\mu e^t,\\
\frac{d \phi_c }{dt} = -\gamma\phi_c\qquad &\to \qquad \phi_c(t) = \phi_c\,\rho(t),\quad\text{where,}\,\, \rho(t)= e^{-\int_0^t dt'\gamma(\lambda_i(t'))},\\ \frac{d \lambda_i}{dt} = \beta_i(\lambda(t)),&\qquad \frac{dm_1^2}{dt} = \beta_{m_1^2}. \end{align}
Solving these coupled first-order differential equations is equivalent to summing the leading logarithms of the higher $1$ PI loops. In this way the implicit running of $\{\lambda_i,m^2\}$ (and the field-strength renormalization through $\gamma$) largely cancels the explicit $\ln\mu_*^2$ (As solving RGE shifts $\bar\mu\to \mu_*=\kappa \phi_c$) terms from the Coleman–Weinberg correction, reducing the residual scale dependence as implied by Eq.~\eqref{RGEcallan}. The remaining mild $\kappa$-dependence reflects missing higher-order (two-loop 1PI) contributions;
including two-loop $\beta$ and $\gamma$ functions together with the two-loop effective potential further diminishes this residual sensitivity (see Appendix~\ref{RGE-sol} for details).
\\

\noindent
At higher loop corrections to the effective potential, $V_\beta^{(n\geq 2)}$, the UV-divergence seems to mix up the thermal part. This is an artificial mixing that is completely isolated upon reorganizing the higher corrections terms and correctly defining the renormalization conditions (This had been shown in details in \cite{laine2017basics}).

\subsubsection{Sources and solution of IR-divergence}\label{IRproblems}

As noted in Section \ref{sec2}, IR sensitivity at finite temperature arises exclusively from the
bosonic sector due to the presence of the vanishing zeroth Matsubara frequency. A superficial hint comes from the thermal piece of the one–loop potential,
\begin{align}
V^{(1)}_{T,\;B}(\phi_c,T)
= T\!\int\!\frac{d^3p}{(2\pi)^3}\,\ln\!\big(1-e^{-E_p/T}\big),
\label{eq:V1Tboson}
\end{align}
which is more singular at small $E_p$ than its fermionic counterpart. However, the true origin of the IR problem and its connection to the breakdown of perturbative expansion is made explicit by writing the one–loop contribution in the sum–integral form and isolating the bosonic zero mode. Starting from
\begin{align}
    V^{(1)}(\phi_c,T) &= \frac{1}{2}\sumint_{\omega_n}\ln[\abs{\vec{p}}^2 + \omega_n^2 + m^2],\\
    &=  \frac{1}{2} \int dm^2 \sumint_{\omega_n}\frac{1}{\abs{\vec{p}}^2 + \omega_n^2 + m^2},\label{ir-origin}\\
    &= \frac{1}{2} \int dm^2 \sumint_{\omega_n} f(\omega_n,\abs{\vec{p}}^2 ).\label{computeboson}
\end{align}
At high temperature, the $n=0$ mode contribution behaves as 
\begin{align}
    T f_B(\omega_0,\abs{\vec{p}}^2 ) &= \frac{T}{E_p^2},\\
    Tf_F(\omega_0,\abs{\vec{p}}^2 ) &\approx \frac{1}{\pi^2 T}.
\end{align}
Where $B,F$ characterize bosons and fermions, respectively. Since each loop is accompanied by a coupling factor (i.e $\lambda,\,g^2$) and a temperature $T$ from the sum, then we can estimate the dimensionless loop factor to be ($\lambda T/\pi m$) for bosonic loop, and ($g^2/\pi^2$) for fermionic loop. Where the momentum $p$ would be integrated out via the loop measure, returning only the mass term for the boson case. Correspondingly, the higher-order bosonic loops will be approximately proportional to the same factor raised to their corresponding power,
 \begin{align}
\alpha^n_B =  \left( \frac{\lambda T}{\pi m} \right)^n. \label{irboson1}
\end{align} 

\noindent
Eq.\eqref{irboson1} already unravels a dual problem, so they have to be related! Firstly, at high temperature in the interaction theory, the self-interaction coupling will receive a large correction according to Eq.\eqref{irboson1}, which threatens the validity of the perturbative expansion. Secondly, in the massless case, the whole interaction will diverge, indicating a severe IR-divergence. For this reason, Takahashi\cite{takahashi1985perturbative} interpreted the breakdown of perturbative expansion at high temperature as the emergence of IR-divergence. As we know from vacuum QFT (Kinoshita-Lee-Nauenberg), IR-divergence indicates a class of same order missing contributions that have to be resummed correctly to cure that divergence. While the higher fermionic loops grow as
\begin{align}
\quad \alpha^n_F &= \left( \frac{g^2 }{\pi^2} \right)^n.\label{irfermion1}
\end{align}
Evidently, the fermionic loops are temperature-independent and are suppressed more for the higher order contributions according to Eq.\eqref{irfermion1}. So, as long as $g^2 \ll \pi$, fermion loops will be well-defined and well-behaved for all orders in perturbation theory.\\

\noindent
We can now compute the IR-problematic integral in Eq.\eqref{ir-origin}. The summed integral is basically the free 2-point function correlation function in a thermal bath,
\begin{align*}
G^0_\beta(x) &= \expval{\phi(0)\phi(x)}_\beta = \sumint_{\omega_n}\frac{1}{\abs{\bvec{p}}^2+ \omega_n^2 + m^2} 
\end{align*}
Isolating the $n=0$ term, and using the identity in Eq.\eqref{feynmantrick}, we get
\begin{align*}
I^{(0)}(m) &= T \int \frac{d^3 \abs{\vec{p}}}{(2\pi)^3}\frac{1}{\abs{\vec{p}}^2 + m^2},\\
&= - \frac{mT}{4\pi} + \mathcal{O}(\varepsilon)
\end{align*}
Then
\begin{align}
    V_B^{(1)}(\phi_c,T;n=0) &= \frac{1}{2}\int dm^2 I^{(0)}(m),\\
    &= - \frac{Tm^3}{12\pi}
\end{align}
The rest of the modes can be computed similarly,
\begin{align*}
I^{(n\neq 0)}(m) &= 2T\sum_{n=1}^\infty \int \frac{d^3 \abs{\vec{p}}}{(2\pi)^3}\frac{1}{\abs{\vec{p}}^2 + m^2 + \omega_n^2},\\
I^{(n\neq 0)}(m)  &= - \frac{T}{2\pi} \sum_{n=1}^\infty \sqrt{m^2 + \omega_n^2} + \mathcal{O}(\varepsilon).
\intertext{This sum can be evaluated using the regularized zeta sum. In the high temperature expansion, the leading term is $-\pi T/6$, then }
I^{(n\neq 0)}(m) &= \frac{T^2}{24}+ \mathcal{O}(\varepsilon).
\end{align*}
All in all, we then have (setting $\varepsilon \to 0$),
\begin{align}
    V_B^{(1)}(\phi_c,T) &= \frac{m^2T^2}{24}- \frac{Tm^3}{12\pi}.\label{cubictermorigin}
\end{align}

\noindent
Turning on the interaction [cf \cite{laine2017basics}] we will get terms proportional to $\lambda,\, \lambda^2,\, \cdots$, which are related to $V_B^{(1)}(\phi_c,T) $:

\begin{align*}
\mathcal{O}(\lambda^1) &\equiv \frac{3}{4}\lambda \left(\frac{1}{m} \frac{d}{dm}  \right)V_B^{(1)}(\phi_c,T)\\
&=  \frac{3}{4}\lambda \left[\frac{T^2}{12}  - \frac{m T}{6\pi} + \mathcal{O}(m^2)\right],\numberthis\label{1loop1}\\
\mathcal{O}(\lambda^2)&\equiv -\frac{3\lambda^2}{4\times 32}\left(\frac{1}{m} \frac{d}{dm}  \right)^2 V_B^{(1)}(\phi_c,T),\\
&= \frac{3}{16\times 32 \pi}\frac{\lambda^2T}{m},\numberthis\label{ir3}\\
&\hspace{5pt}\vdots\\
\mathcal{O}(\lambda^n) &\equiv c \lambda^n\left(\frac{1}{m} \frac{d}{dm}  \right)^n V_B^{(1)}(\phi_c,T). \numberthis \label{daisy2}
\end{align*}
The result in Eq.\eqref{ir3} shows, undoubtedly, that the IR-divergence is exclusively a result of the vanishing Matsubara mode, as it originates from its characteristic term $Tm^3/12\pi$.
The results in Eq.\eqref{1loop1} and Eq.\eqref{ir3} correspond to 1-loop and 2-loop self-correction diagrams in a thermal bath,
\begin{align*}
\Sigma^{(1)}(\phi_c,T) &=
 \begin{tikzpicture}[baseline=-.3cm]
\draw[very thick, dashed] (1,0) circle (.8);
\draw[very thick, dashed] (-.5,-.8) -- (2.5,-.8);
\end{tikzpicture}\propto \lambda\left( \frac{T^2}{16} - \frac{mT}{8\pi}\right),\hspace{10pt} \\
\Sigma^{(2)}(\phi;T) &=
 \begin{tikzpicture}[baseline=.3cm]
\draw[very thick, dashed] (1,0) circle (.5);
\draw[very thick, dashed] (1,1) circle (.5);
\draw[very thick, dashed] (-.5,-0.5) -- (2.5,-0.5);
\end{tikzpicture} \propto \lambda  \left(\frac{\lambda T}{m} \right) .
\end{align*}

\noindent
This matches the power counting we anticipated earlier. Interestingly, the above induction offers us an algebraic way to sum all the higher-order loops contributing to the IR-divergence. Summing all the higher contributions to order $n$ corresponds to the Daisy diagrams, which can now be evaluated using Eq.\eqref{daisy2},
\begin{align}
\begin{tikzpicture}[baseline=0cm,scale=.65] 
\draw[very thick, dashed] (4,0) circle (1.5);
\draw[thick, dashed] (2,0) circle (.5);
\draw[thick, dashed] (4,2) circle (.5);
\draw[dotted, very thick] (5.7,.9) .. controls (6,.7) and (6,-.8) .. (5.7,-.9);
\draw[thick, dashed] (2.586,1.414) circle (.5);
\draw[thick, dashed] (5.414,1.414) circle (.5);
\draw[thick, dashed] (5.414,-1.414) circle (.5);
\draw[thick, dashed] (2.586,-1.414) circle (.5); 
\draw[very thick, dashed] (1.5,-2.7) -- (4,-1.5);
\draw[very thick, dashed] (4,-1.5) -- (6.5,-2.7);
\end{tikzpicture}
=\quad \frac{(\lambda/2)^2}{(n-1)!} \left(\frac{T^2}{12} \right)^{n-1} \frac{\partial^n \Sigma^{(1)}(\phi;T)}{\partial m^{2n}} .
\end{align}
Summing all of  these corrections from 1-loop to n-loop is known by thermal resummation and it yields  the full loop correction,
\begin{align}
\Sigma^{\text{\tiny full}}(\phi;T) \simeq \frac{\lambda }{2} \left[ \frac{T^2 }{12}- \frac{T }{4\pi} \left( m^2 +\frac{\lambda T^2 }{24}\right)^{\frac{1}{2}} \right] .
\end{align}

\noindent
This result is equivalent to shifting the mass to $\displaystyle{m^2 \longrightarrow m^2 +\frac{\lambda T^2}{24} }$, then the effective mass could be directly obtained from the single 1-loop correction diagram by adopting that transformation.
\begin{align}
m_\eff^2 &= m^2 + \frac{\lambda}{2} \Sigma^{(1)}\left( m^2 +\frac{\lambda T^2}{24},T\right)
\end{align} 
Hence, IR-divergence is cured through the thermal resummation which corrects the mass in such a way that the loop factor does not blow up as $m\rightarrow 0$ and consequently the perturbation theory expansion is saved as long as $\lambda \ll 1$.\\

\noindent
Of course, other diagrams participate at $n=2,3,\cdots,$ more than the ring diagrams included \cite{senaha2020}.
\begin{figure}[htb!]
\centering
\begin{tikzpicture}
\draw[very thick, dashed] (0,-0.5) circle (.8);
\draw[very thick, dashed] (-1.5,-0.5) -- (1.5,-0.5);
\draw[very thick, dashed] (7,-1.5) circle (0.5);
\draw[very thick, dashed] (7,-0.5) circle (.5);
\draw[very thick, dashed] (7,0.5) circle (.5);
\draw[very thick, dashed] (5.75,-2) -- (8.25,-2);
\begin{scriptsize}
\draw[] (3, -0.5) node {$\mathlarger{\mathlarger{\sim \lambda^2 T^2 \ln\left(\frac{m}{T}\right)}}$};
\draw[] (9, -0.5) node {$\mathlarger{\mathlarger{\sim \lambda T^2 \left(\frac{\lambda T}{m} \right)^2}}$};
\end{scriptsize}
\end{tikzpicture}
\caption{Sun set appears at  n=$2$ and the right loop appears at $n=3$.}
\label{loops}
\end{figure}
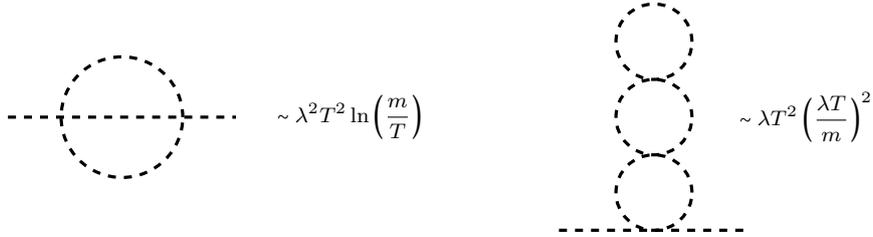
However, they will be dominated by the ring diagrams at high temperature as anticipated by the power counting for the other $n=2,3$ diagrams shown in Fig.\ref{loops} as an example. \\

\noindent
Another approach to control the IR problem of thermal gauge theories is the dimensional reduction (DR), i.e. the construction of a three-dimensional effective field theory (3D EFT) for the long-distance degrees of freedom in a thermal bath. At high temperature, as we have discussed in the previous parts, the Euclidean-time integral is finite, and the fields decompose into Matsubara modes. Nonstatic bosonic modes have frequencies $\omega_n=2\pi nT$ and therefore carry an effective mass of order $|\omega_n|\gtrsim 2\pi T$, while fermions have
$\omega_n=(2n+1)\pi T$ and in particular possess no zero mode. Consequently, for IR observables with spatial momenta $k\ll \pi T$, one can integrate out all fermions and the bosonic modes with $n\neq 0$ (the ``hard'' scale $\sim \pi T$), obtaining an EFT for the static bosonic sector ($n=0$). Consequently, the integrated hard modes renormalize the parameters of the light sector and, in particular, incorporate electric screening: the temporal gauge field zero mode $A_0^{(0)}$ acquires a Debye mass $m_D\sim gT$, and scalar zero modes acquire thermal masses. In this way, the
IR sensitivity associated with the electric sector is treated consistently (equivalently, the relevant daisy/ring contributions are resummed within the EFT framework). The remaining soft fields are $\tau$--independent, so the Euclidean time integral factorizes,
$\int_0^\beta d\tau=\beta$, and the resulting EFT is genuinely three-dimensional with dimensionful couplings, e.g.\ $g_3^2\sim g^2T$. One may further integrate out the \emph{Debye/electric} scale $\sim gT$ (often by integrating out $A_0$), leaving a magnetostatic 3D gauge--Higgs theory governing the ultrasoft scale
$\sim g^2T$. This magnetic sector is genuinely nonperturbative (the Linde problem) and cannot be cured by perturbative screening masses. However, dimensional reduction isolates it into a
super-renormalizable 3D theory, making nonperturbative lattice simulations substantially cheaper and conceptually cleaner than attempting the corresponding calculation directly in the full 4D
finite-temperature theory. As an illustrative example, we follow \cite{kajantie1993lattice} and consider the
finite-temperature SU(2)$+$Higgs theory (with $g'=0$ for simplicity). The Euclidean partition function reads
\begin{align}
\mathcal{Z}
&= \int \mathcal{D}A_\mu^a \, \mathcal{D}\phi \;
\exp\!\left\{-S_E[A_\mu^a,\phi]\right\},
\intertext{where,}
S_E[A_\mu^a,\phi] &= \int_0^\beta d\tau \int d^3 x \,
\left[
\frac{1}{4} F^a_{\mu \nu}F^a_{\mu \nu}
+ (D_\mu \phi)^\dagger(D_\mu \phi)
+ m^2 \phi^\dagger \phi
+ \lambda (\phi^\dagger \phi)^2
\right].
\end{align}
At high temperature, the bosonic nonstatic Matsubara modes ($n\neq 0$) are heavy for
IR physics ($|\mathbf{k}|\ll 2\pi T$). Integrating out all the nonstatic modes to obtain an effective action for the static sector:
\begin{align}
e^{-S_{\rm eff}[A_\mu^{(0)},\phi^{(0)}]}
\;\equiv\;
\int \mathcal{D} A_\mu^{(n\neq 0)}\,\mathcal{D}\phi^{(n\neq 0)}
\exp\!\left\{-S_E\!\left[A_\mu^{(0)}\!+\!A_\mu^{(n\neq 0)},\,
\phi^{(0)}\!+\!\phi^{(n\neq 0)}\right]\right\}.
\label{scaleseparation}
\end{align}
The integration over non-vanishing Matsubara modes is performed perturbatively, and its effect is encoded in the temperature-dependent parameters of the resulting EFT.
However, the remaining vanishing Matsubara modes still contain different infrared scales that may be further separated, following \cite{hirvonen2022intuitive}. While the soft electric sector can still be treated perturbatively, the ultrasoft magnetic sector of non-Abelian gauge theories is intrinsically nonperturbative, and this becomes particularly important near the critical temperature where infrared effects are enhanced. A fully controlled description of this sector therefore requires lattice methods. In this review, we restrict ourselves to the perturbative treatment associated with integrating out the hard non-vanishing Matsubara modes.
In particular, electric screening (Debye and scalar thermal masses) is incorporated in these matched parameters, reproducing the daisy/ring resummation physics of the electric sector. Since the remaining fields are static, the Euclidean time integral factorizes and one may
write
\begin{equation}
S_{\rm eff}[A_\mu^{(0)},\phi^{(0)}]
= \beta \int d^3x \;\mathcal{L}_{\rm eff}(A_i,A_0,\phi),
\end{equation}
where we have renamed the static modes as $A_i\equiv A_i^{(0)}(\mathbf{x})$,
$A_0\equiv A_0^{(0)}(\mathbf{x})$, and $\phi\equiv \phi^{(0)}(\mathbf{x})$.
The corresponding form of the SM 3D electrostatic EFT Lagrangian is
\begin{align}
\mathcal{L}_{\rm eff}(A_i,A_0,\phi)
&= \frac{1}{4}F^a_{ij}F^a_{ij}
+ \frac{1}{2}(D_i A_0)^a(D_iA_0)^a
+ (D_i \phi)^\dagger (D_i \phi)
+ m_3^2 \,\phi^\dagger \phi
+ \lambda_3 (\phi^\dagger \phi)^2
\nonumber\\
&\quad
+ \frac{1}{2}m_D^2\,A_0^aA_0^a
+ h_3 \,A_0^aA_0^a\, \phi^\dagger \phi
+ \frac{1}{4}\lambda_A \,(A_0^aA_0^a)^2
+ \cdots ,
\label{eq:3dEFT}
\end{align}
where the ellipsis denotes higher-dimensional operators suppressed by powers of $(2\pi T)^{-1}$.
We decomposed the static gauge field into temporal and spatial components, $A_\mu^{(0)}=(A_0,A_i)$, because in the dimensionally reduced theory the temporal component $A_0(\mathbf{x})$ behaves as an adjoint scalar field in three dimensions. The parameters of the resulting 3D EFT, $m_3^2,\lambda_3,m_D^2,h_3,\lambda_A,\ldots$, are obtained by matching to the full 4D theory: integrating out the hard modes (all fermions and the bosonic modes with $n\neq 0$)
shifts the coefficients of the operators built from the static fields, which secures their corresponding. A convenient way to formulate matching is in terms of static ($\omega=0$) correlation
functions at soft external momenta $|\mathbf{p}|\ll 2\pi T$. For example, the inverse two-point function of the scalar in the 3D EFT can be written schematically as
\begin{align}
\Gamma^{(2)}_{3}(p) = p^2 + m_3^2 + \Pi_3(p^2),
\label{3d_2pt}
\end{align}
where $\Pi_3$ denotes loop corrections generated from the soft modes within the 3D EFT. In the underlying 4D theory, the corresponding static two-point function admits a
separation into hard and soft contributions,
\begin{align}
\Gamma^{(2)}(p) &= p^2 + m^2 + \Pi_{\rm soft+hard}(p^2),\\
&= p^2 + m^2 +\Pi_{\rm soft}(p^2) + \Pi_{\rm hard}(p^2)
\label{4d_2pt}
\end{align}
where $\Pi_{\rm hard}$ collects contributions involving at least one hard scale ($n\neq 0$ Matsubara frequency and/or momenta $\sim 2\pi T$). 
\begin{align}
\Pi_\mathrm{hard}(p^2) = \qquad 
\begin{tikzpicture}[baseline =0.01cm]
\draw[line width = 1pt, dashed] (0,0) circle(0.75);
\draw[line width = 1pt, dashed] (-1.5,-0.75) -- (1.5,-0.75);
\begin{scriptsize}
\draw[] (-1.35,-1) node {$n=0$};
\draw[] (1.35,-1) node {$n=0$};
\draw[] (0.5,0.9) node {$n\neq0$};
\end{scriptsize}
\end{tikzpicture}
\end{align}

\noindent
Since $\Pi_{\rm hard}(p^2)$ receives contributions only from hard scales
(nonzero Matsubara modes and fermions), it is analytic at $p^2=0$ (No IR-divergence) and can be expanded as
\begin{align}
\Pi_{\rm hard}(p^2)=\Pi_{\rm hard}(0)+p^2\,\Pi'_{\rm hard}(0)+\mathcal O\!\left(\frac{p^4}{(2\pi T)^2}\right).
\end{align}
Accordingly, the static ($\omega_n=0$) inverse two-point function in the underlying 4d theory can be written as
\begin{align}
\Gamma^{(2)}(p)
&= p^2+m^2+\Pi_{\rm soft}(p^2)+\Pi_{\rm hard}(p^2) \nonumber\\
&= \bigl[1+\Pi'_{\rm hard}(0)\bigr]\,\left\{ p^2
+\left[m^2+\Pi_{\rm hard}(0)\right]\left[1-\Pi'_{\rm hard}(0)\right]
+\Pi_{\rm soft}(p^2)+\cdots \right\}.
\label{4d_2pt_expanded}
\end{align}
Then matching $\Gamma^{(2)}_3(p)$ in Eq.\eqref{3d_2pt} to the static $\Gamma^{(2)}(p)$ in Eq.\eqref{4d_2pt_expanded} at $p\ll 2\pi T$ amounts to absorbing the hard contributions $\Pi_{\rm hard}(0)$ and $\Pi'_{\rm hard}(0)$ into $m_3^2$ and $Z_\phi$, while the soft part is reproduced by loops within the 3D EFT, $\Pi_3(p^2)=\Pi_{\rm soft}(p^2)$ up to higher-dimension operators.
Thus\footnote{Knowing that the 3D field is related to the 4D field through $\phi_{3d} = \frac{1}{\sqrt{T}}\,Z_\phi^{1/2}\,\phi^{(0)}_{4d},
\,\,\, Z_\phi = 1+\Pi'_{\rm hard}(0)$ [cf.  \cite{kajantie1996generic}].},

\begin{align}
m_3^2 &= \left[m^2+\Pi_{\rm hard}(0)\right] \left[1-\Pi'_{\rm hard}(0)\right]+\cdots. \label{eq:m3_matching_general}
\end{align}

\noindent
For the SU(2)$+$Higgs theory, the one-loop DR relations for the 3d electrostatic theory
are (in the $\overline{\rm MS}$ scheme) \cite{farakos19943d}
\begin{align}
g_3^2 &= g^2(\mu)T \left(1 + \frac{43}{6}\frac{L_s}{(4\pi)^2} g^2 \right),\\
\lambda_3 &= T\left(\lambda(\mu) - \frac{L_s}{(4\pi)^2}\left[ \frac{9}{16}g^4 - \frac{9}{2}\lambda g^2 + 12\lambda^2\right] +\frac{1}{(4\pi)^2}\frac{3}{8} g^4 \right),\\
h_3 &= \frac{1}{4}g^2_3 \left( 1 + \frac{1}{(4\pi)^2} \left[ 12\lambda + \frac{49}{6} g^2 - \frac{1}{3} g^2\right]\right),\\
\lambda_A &= \frac{17}{48\pi^2} g^4(\mu)T,\\
m_D^2 &= \frac{5}{6}g^2(\mu)T^2.
\end{align}
Where $L_s = \log\frac{\mu^2}{(4\pi)^2T^2} + 2\gamma_E$\footnote{The renormalization scale in \cite{farakos19943d} was kept to be just $\mu$ not $\bar\mu$ as we adapted in this paper.}. The key point is that integrating out hard modes generates thermal (Debye) screening in the electric sector, while the magnetostatic sector remains genuinely non-perturbative at the scale $g^2T$ (the Linde problem). Consequently, dimensional reduction implements (and organizes) the same physics as the
familiar ring/daisy resummation of the electric sector. This becomes manifest once one computes the one-loop effective potential in the 3D EFT. Writing the 3d tree-level potential (for constant background fields) as
\begin{align}
V_{3,\,\rm TL}(\phi,A_0) &=
m_3^2\,\phi^\dagger\phi+\lambda_3(\phi^\dagger\phi)^2
+\frac12 m_D^2\,A_0^aA_0^a
+\frac14\lambda_A\,(A_0^aA_0^a)^2
+h_3\,A_0^aA_0^a\,\phi^\dagger\phi ,
\end{align}
the one-loop correction from the static ($n=0$) modes takes the generic form
\begin{align}
V^{(1)}_3(\phi,A_0)
= \frac{T}{2}\sum_i n_i \int\!\frac{d^3p}{(2\pi)^3}\,
\ln\!\bigl[p^2+M_i^2(\phi,A_0)\bigr] ,
\label{V1_3d_generic}
\end{align}
where $i$ runs over the bosonic degrees of freedom present in the 3D EFT (e.g.\ transverse $A_i$, the adjoint scalar $A_0$, and the Higgs/scalar modes), and $n_i$ denotes their multiplicities. In reference \cite{kajantie1993lattice}, this one-loop correction is written explicitly as a sum of logarithms from the $A_i$ loop and the coupled $(A_0,\phi)$ sector. After renormalization (the 3d theory is super-renormalizable and contains a mass counterterm originating from integrating out only the nonstatic modes), the finite field-dependent part gives the characteristic cubic contribution
\begin{align}
V^{(1)}_3(\phi,A_0)
= -\frac{T}{12\pi}\sum_i n_i\,M_i^3(\phi,A_0)\;+\;\text{(const.)}.
\label{cubic_term}
\end{align}
This result is the EFT avatar of the familiar thermal cubic contribution $-\tfrac{T}{12\pi}\,m^3$, and hence of the ring/daisy resummation. In the dimensional reduction framework, integrating out the hard modes generates the appropriate thermal (screening) masses for the static sector-in particular the Debye mass $m_D\sim gT$ for the adjoint scalar $A_0$-and inserting these screened masses into Eq.~\eqref{cubic_term} reproduces the ring-improved cubic terms in a systematic way. Importantly, this screening only resolves the electric IR sensitivity. The
magnetostatic sector associated with the spatial gauge fields $A_i^{(0)}$ remains genuinely non-perturbative at the ultrasoft scale $g^2T$ (the Linde problem). Dimensional reduction isolates this remaining long-distance physics into a super-renormalizable 3D EFT, so that its non-perturbative contribution can be determined efficiently with three-dimensional lattice simulations, which are substantially simpler than treating the full four-dimensional
finite-temperature theory directly.

\subsection{Imaginary contributions}\label{imaginary-discussion}
 There exist three different sources of the imaginary contributions in the effective potential, all of which appear at the small field values from the scalar contributions. Let us explore them in detail.

\subsubsection{Imaginary contribution from $V_{\text{\tiny CW}}^{(1)}$}

\begin{align}
V_{\text{\tiny CW}}^{(1)}(h;\bar\mu) = \sum_{h,G,W^\pm,Z,t} \frac{n_i m_i^4(h)}{64_\pi^2} \left[\ln \left(\frac{m_i^2(h)}{\bar\mu^2}\right) -C_i\right] 
\end{align}
Since the weak gauge bosons and top quark masses can never be imaginary at any field value however small, the imaginary contributions exclusively result from the scalar sector, where
\begin{align*}
m_h^2 (h)= \lambda(3h^2 - v^2),\qquad\qquad m_G^2(h) = \lambda(h^2 - v^2)
\end{align*}
So, for $h<v/3, v$ for Higgs and Goldstones, respectively, CW-correction will generate imaginary contributions. Hence,
\begin{align}
\text{\bf Im}\left[ V_{\text{\tiny CW}}^{(1)}(h;\bar\mu)\right] = \text{\bf Im}\left[ \sum_{h,G} \frac{n_i m_i^4(h)}{64 \pi^2} \,\, \ln \left(- \frac{m_i^2(h)}{\bar\mu^2}\right) \right].
\end{align}

Using the identity $\ln(-A)=\ln(e^{i\pi}A)= \ln (A) + i\pi$, the corresponding imaginary contribution is then
\begin{align}
\text{\bf Im}\left[ V_{\text{\tiny CW}}^{(1)}(h;\bar\mu)\right] =\sum_{h,G} \frac{n_i m_i^4(h)}{64 \pi}\label{vcw-imag}
\end{align}

\subsubsection{Imaginary contribution from $V_{\beta}^{(1)}$}

\begin{align}
V_\beta^{(1)} = T\int \frac{d^3\bvec{p}}{(2\pi)^3} \ln \left( 1- e^{- \frac{\sqrt{p^2 + m^2}}{T}}\right)
\end{align}
Since only $h,\xi$ produces imaginary masses when $m^2 < 0 \to m^2 = -\mu^2$ where $\mu^2 >0$, the imaginary contributions only result from the region where $p^2< \mu^2$,

\begin{align}
\text{\bf Im}\left[V_\beta^{(1)} \right]= \frac{T}{2\pi^2} \text{\bf Im}\left[\int dp\,\, p^2\,\, \ln \left( 1- e^{- i\phi }\right) \Theta (\mu^2 - p^2) \right],
\end{align}
Where, $\phi = \frac{1}{T} \sqrt{\mu^2 - p^2}$. Using the identity $(1-e^{-i\phi}) = 2ie^{-i\phi/2} \sin(\phi/2)$, then\footnote{Where, $\ln(1-e^{-i\phi}) = \ln(2) + \ln \left(\sin\frac{\phi}{2}\right) + \frac{i}{2}(\pi-\phi)$, since $\ln(ie^{-i\phi}) = \ln(e^{i\pi/2}\cdot e^{-i\phi/2}) = \frac{i}{2}(\pi-\phi)$.}

\begin{align}
\text{\bf Im}\left[V_\beta^{(1)} \right] &= \frac{T}{4\pi^2} \int dp \,\,p^2\,\, (\pi - \phi)\,\,\Theta (\mu^2 - p^2),\\
&= \frac{T \mu^3}{12\pi} -\frac{T}{4\pi^2}\int_0^\mu dp \,\,p^2\, \sqrt{\mu^2 - p^2},\\
&= \frac{T \abs{m}^3}{12\pi} - \frac{m^4}{64\pi}. \label{vbeta-imag}
\end{align}
Here we shifted the second integral measure to $p = \mu \sin \rho$ and used $\int_0^{\pi/2} \sin^2 \rho \cos^2 \rho = \pi/16$. It is obvious that when we restore the correct degrees of freedom for Higgs and Goldstones, the second imaginary part in Eq.\eqref{vbeta-imag} will exactly cancel the corresponding term in Eq.\eqref{vcw-imag}. The first term, on the other hand, is the one we obtained in Eq.\eqref{cubictermorigin} that signals the IR-divergence  resulting from the vanishing bosonic Matsubara mode,
\begin{align}
V^{(1)}(h;T,\bar\mu) &= \frac{T}{2} \sum_{n=-\infty}^\infty \int \frac{d^3 \bvec{p}}{(2\pi)^3} \ln(\omega_n^2 +\bvec{p}^2 + m^2),\\
&= \frac{T}{2} \sum_{n'=-\infty}^\infty \int \frac{d^3 \bvec{p}}{(2\pi)^3} \ln(\omega_n^2 +\bvec{p}^2 + m^2)
 + \frac{T}{2} \int \frac{d^3 \bvec{p}}{(2\pi)^3} \ln(\bvec{p}^2 + m^2),\\
&= \frac{T}{2} \sum_{n'=-\infty}^\infty \int \frac{d^3 \bvec{p}}{(2\pi)^3} \ln(\omega_n^2 +\bvec{p}^2 + m^2)
+ \frac{T\abs{m}^3}{12\pi}.
\end{align}
Here $n'\neq 0$. From this approach, one can appreciate how Kinoshita interpreted the imaginary contribution in the thermal correction as a sign of IR-divergence. In section \ref{IRproblems}, it was shown how the IR-divergence is fixed by resumming the daisy rings, which gives an exact opposite term to the temperature-dependent term in Eq.\eqref{vbeta-imag}, curing both the IR-divergence and the remnant imaginary contributions from $m^3$ bosonic contributions at once.

\subsubsection{Imaginary contribution from $V^{(1)}_{\text{\tiny rings}} $}\label{nilsen-idet}

\begin{align}
V^{(1)}_{\text{\tiny rings}} = -\frac{T}{12 \pi} \sum_{h,G,W^\pm,Z} n_i\Big[M^3(h;T) - m^3(h)\Big]. \label{vrings-imag}
\end{align}
Here, $M^3(h;T) = \left( m^2(h) + m_D^2(T)\right)^{3/2}$ is the thermal corrected mass. It is obvious now that the second term in Eq.\eqref{vrings-imag} will exactly cancel the corresponding cubic term in Eq.\eqref{vbeta-imag}, leaving the whole effective potential real.
However, $M^3(h,T)$ can still develop imaginary contributions at some field values and specific temperatures where $\abs{m^2} > m^2_D(T)$, which was already observed in \cite{delaunay2008dynamics}. However, the regions where $M(h;T)$ becomes tachyonic require very small field values when the temperature is very low, so that the Debye mass is smaller than $\mu^2$. This is somehow contradictory, as the only way to get a negative $M^2$ is when  $\abs{m^2} > m_D^2$ and $m^2<0$, which is only true at low temperature and small $\expval{h}$, which is contradictory as the only way to decrease $\expval{h}$ is by increasing temperature. So, the region where $M(h;T)$ becomes tachyonic is not physical, and any resulting imaginary contribution could be safely ignored, and it will be small compared to $Re[V_{eff}]$ anyway.\\
\begin{figure}[htb!]
\centering
\includegraphics[scale=0.5]{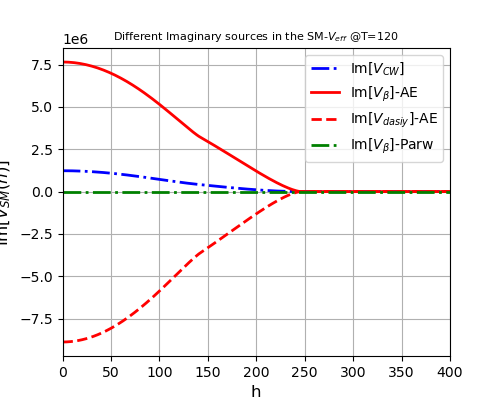}
\includegraphics[scale=0.5]{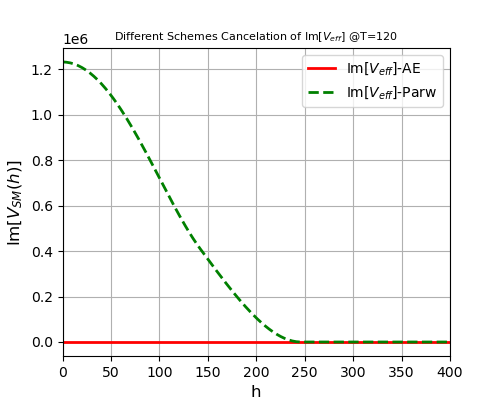}
\includegraphics[scale=0.5]{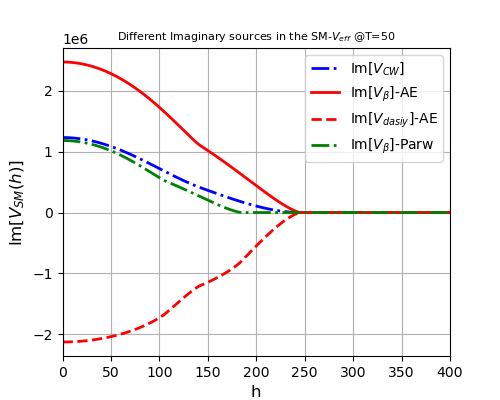}
\includegraphics[scale=0.5]{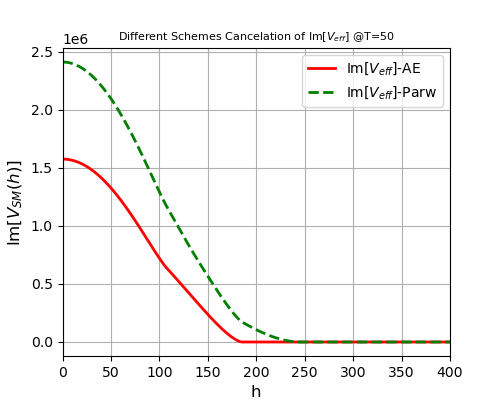}
\includegraphics[scale=0.5]{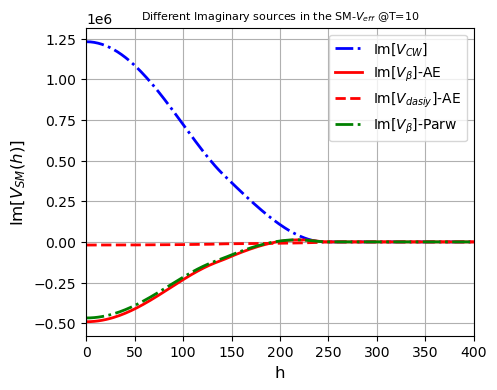}
\includegraphics[scale=0.5]{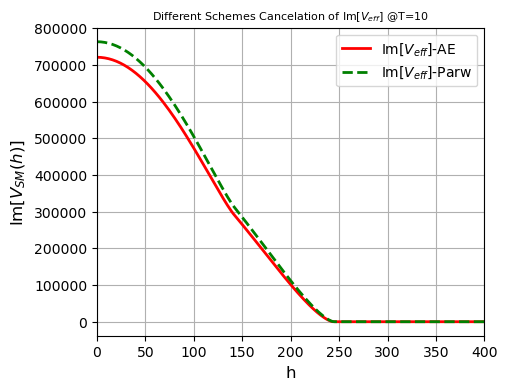}
\caption{The left column shows the weights of Imaginary contributions from the different contributions in $V_\eff$ in the SM at different temperatures. The right column compares the efficiencies AE and Parwani schemes in canceling these terms. The plots also illustrate the fact that these terms only show up at low field values and disappear once $\expval{h}\geq v_{EW}$.}
\label{imagcontrib}
\end{figure}

\noindent
Another perspective on the origin of this remaining imaginary contribution focuses on the thermal correction calculation of $m^2_D(T)$, which is usually evaluated in the high temperature approximation limit, $m/T \ll 1$. Recent studies push these calculations further to account for the sensitivity of the effective mass to higher loop corrections. In \cite{navarrete2025cosmological}, they separate the low momentum modes leading to IR-divergence from the large modes and use EFT for a more accurate resummation of the low mode contributions. In \cite{bittar2025self}, they get the effective mass by solving the gap equation numerically. So, any residual imaginary contribution to the effective potential could be interpreted as an incomplete calculation of the effective mass. The regime followed above, where the daisy contributions are added separately to the effective potential, is known as the Arnold-Espinoza (AE) scheme.
\begin{align}
V_{\text{\tiny eff}} = V_{\text{\tiny TL}} + V_{\text{\tiny CW}}^{(1)} + V_{\beta}^{(1)}  + V_{\text{\tiny rings}}^{(1)}
\end{align}
Another approach is the Parwani scheme, in which the effective potential is defined as
\begin{align}
V_{\text{\tiny eff}} = V_{\text{\tiny TL}} + V_{\text{\tiny CW}}^{(1)} + V_{\beta}^{(1)} 
\end{align}
We do not need to add the ring contribution separately, where we already shift the mass in $V_{\beta}^{(1)} $ $m^2 \to M^2 = m^2 + m_D^2$. At high temperature, the AE scheme is more sufficient in suppressing the imaginary contributions as it is more effective with respect to perturbative counting, unlike the Parwani scheme which successfully removes the imaginary parts from the thermal contribution, but leaves the CW imaginary term uncancelled. At low temperatures, both schemes work almost equally because the whole thermal correction is reduced, and at this limit, the major source of the imaginary part is the CW-contribution that gets partial cancellation from the thermal part(s) until $T\to 0$, where the complete term survives. This comparison has been illustrated for the SM case in Fig.\ref{imagcontrib}.

\subsection{Gauge dependence}\label{gauge-section}

As in vacuum QFT, gauge dependence in finite–temperature field theory does not originate
from thermal averaging itself; it arises from the introduction of gauge–fixing functionals required
to quantize the gauge sector. As discussed in Section~\ref {vectorfields}, covariant $R_\xi$ gauges lead to
gauge–parameter dependence of the generating functional (cf.\ Eqs.~\eqref{gaugeZfinal},
\eqref{momentumZ}, \eqref{modifiedZ0}), and hence of the (off–shell) effective potential obtained
via Eq.~\eqref{freeenergy}. This is expected: $V_{\rm eff}$ is defined off shell (cf.\
Eq.~\eqref{effectiveactionapprox}: scalar fields take constant values, and all the other fields are vanishing, which generates Green's functions with vanishing external momentum), and therefore need not be gauge invariant. Physical observables, however, must be gauge independent. In particular, \(S\)-matrix elements obey Ward–Takahashi/
Slavnov–Taylor identities, and the value of the effective potential at its extrema (vacuum energy, thermodynamic potentials at stationary points) is gauge independent. The latter statement is made precise by the Nielsen identity, which follows from BRST invariance of the gauge–fixed generating functional and holds equally well at finite temperature. Where, BRST symmetry prevents the gauge fixing term and its accompanying ghosts term from contributing to the physical observables when evaluated at the fields' minima according to Eq.\eqref{brst}. This observation is generalized to any physical observable evaluated from the effective potential at any stationary point by the Nielsen identities. Starting from the effective action that is BRST invariant, and using its connection to the effective potential in Eq.\eqref{effectiveactionapprox}, we get the well-known Nielsen identity [Check chapter 7 in \cite{das2023finite}]
\begin{align}
\frac{\partial V_\eff(\phi_c,\xi)}{\partial \xi}
\;=\;
-\,C(\phi_c,\xi)\,\frac{\partial V_\eff(\phi_c,\xi)}{\partial \phi_c}.
\label{Nielsen}
\end{align}
Here $C(\phi_c,\xi)$is a (scheme–dependent) correlation functional determined by the gauge–fixing
and ghost insertions; schematically,
\begin{align}
C(\phi_c,\xi)
\;=\;
-\,\int d^4x\;
\frac{\delta^2 \Gamma_{\rm eff}}{\delta J_{\rm gf}(x)\,\delta J_{\rm gh}(0)}\,,
\label{gf-gh}
\end{align}
where $\Gamma_\eff$ is the effective action and $J_{\rm gf},J_{\rm gh}$ are sources for the
gauge–fixing and ghost operators. Equation~\eqref{Nielsen} immediately implies that the value
of the potential at any stationary point is gauge independent: if $\partial V_\eff/\partial\phi_c=0$
at $\phi_c=\phi_\star$, then $\partial V_\eff(\phi_\star,\xi)/\partial\xi=0$.
The location of the stationary point, by contrast, is gauge dependent. Differentiating the
stationarity condition with respect to \(\xi\) and using Eq.~\eqref{Nielsen} gives
\begin{align}
0=\frac{d}{d\xi}\left.\frac{\partial V_{\rm eff}}{\partial \phi_c}\right|_{\phi_c=\phi_\star}
&= \left.\frac{\partial^2 V_{\rm eff}}{\partial \xi\,\partial \phi_c}\right|_{\phi_\star}
+ \left.\frac{\partial^2 V_{\rm eff}}{\partial \phi_c^2}\right|_{\phi_\star}\,\frac{\partial \phi_\star}{\partial \xi}\\
&= -\,C(\phi_\star,\xi)\left.\frac{\partial^2 V_{\rm eff}}{\partial \phi_c^2}\right|_{\phi_\star}
+ \left.\frac{\partial^2 V_{\rm eff}}{\partial \phi_c^2}\right|_{\phi_\star}\,\frac{\partial \phi_\star}{\partial \xi},
\end{align}
hence
\begin{align}
\frac{\partial \phi_\star}{\partial \xi}=C(\phi_\star,\xi).
\end{align}
Therefore, although \(V_{\rm eff}\) evaluated at a stationary point is gauge independent, the
position \(\phi_\star\) of that point generally is not. This induces a practical gauge ambiguity in
quantities derived from the shape of \(V_\eff\) away from its extrema (e.g.\ critical temperature,
order parameter), which adds uncertainty in the study of the EWPT dynamics. However, this dependence is an artificial one that arises from the trivial truncation of the effective potential. Inaccurate calculations of the effective potential at 1PI loop might not be enough to cancel gauge dependence arising from other contributions beyond 1PI loops. In this sense, the Nielsen identity is equivalent to Ward identities that guarantee gauge independence of the transition matrices only when all contributing diagrams of the same order are included. For that reason, the solution of the $vev$ gauge-dependence is to carefully track and organise the gauge-dependent terms at a given order to have an exact cancellation between them. A convenient bookkeeping device is the loop (or $\hbar$) expansion, which has been introduced in \cite{patel2011baryon,andreassen2015}. In $\hbar$ expnsion, both $C(\phi_c,\xi)$ and $V_\eff(\phi_c,\xi)$ expands as
\begin{align}
V_{\rm eff}(\phi_c,\xi) &= V^{(0)}(\phi_c)+\hbar\,V^{(1)}(\phi_c,\xi)+\hbar^2 V^{(2)}(\phi_c,\xi)+\cdots,\\
C(\phi_c,\xi) &=\hbar\,C^{(1)}(\phi_c,\xi)+\hbar^2 C^{(2)}(\phi_c,\xi)+\cdots,
\end{align}
for which Eq.~\eqref{Nielsen} implies, order by order,
\begin{align}
\frac{\partial V^{(1)}(\phi_c,\xi)}{\partial \xi}
&= -\,C^{(1)}(\phi_c,\xi)\,\frac{\partial V^{(0)}(\phi_c)}{\partial \phi_c},\\
\frac{\partial V^{(2)}(\phi_c,\xi)}{\partial \xi}
&= -\,C^{(2)}(\phi_c,\xi)\,\frac{\partial V^{(0)}(\phi_c)}{\partial \phi_c}
-\,C^{(1)}(\phi_c,\xi)\,\frac{\partial V^{(1)}(\phi_c,\xi)}{\partial \phi_c},
\end{align}
since $V^{(0)}(\phi_c)$ does not depend on the gauge parameter by construction, which sets $C^{(0)}(\phi_c,\xi)=0$. These relations allow one to track and cancel
gauge–dependent pieces consistently at each order, ensuring that physical predictions (e.g., vacuum energy, latent heat, bubble nucleation rate when computed from gauge–invariant observables) are gauge independent.\\

\noindent
Because the derivation of Eq.~\eqref{Nielsen} relies only on BRST invariance and not on any zero–temperature specialization, the same identity holds in the thermal theory. Practically, the only modifications are the standard imaginary–time replacements: \(p^0\to i\omega_n\), and \(\int d^4p/(2\pi)^4 \to T\sum_n\!\int d^3p/(2\pi)^3\). The structure and implications of the Nielsen identity are otherwise unchanged at finite temperature.

%%%%%%%%%%%%%%%%%%%%%%%%%%%%%%%%%%%%%%%%%%%%%%%%%%%%%%%%%
%%%%%%%%%%%%%%%%%%%%%%%%%%%%%%%%%%%%%%%%%%%%%%%%%%%%%%%%%

\section{EWPT from real singlet extension}\label{EWPTsec}

A first-order electroweak phase transition is one of the most probable scenarios to explain the baryon asymmetry in the universe (BAU). According to Sakharov \cite{Sakharov:1967dj}, there has to be some matter-biased generating mechanism in the early universe that leads to baryon number and charge violation. The third condition is that such interactions have to take place in thermal inequilibrium to prevent the sphaleron washout of the excess matter generated. Such a mechanism was originally expected to take place in grand unified theories (GUTs) ($\sim \, 10^{15}$ GeV)\cite{riotto1998theories,cline2006baryogenesis,dine2003origin}, yet other models discussed the possibility of explaining BAU at a later time of EWPT if it was of the first-order type. There FOEWPT generates an inequilibrium environment permitting the survival of the generated asymmetry~\cite{morrissey2012electroweak,cohen1993progress,garbrecht2020there,rubakov1996electroweak,trodden1999electroweak} through suppression of the sphaleron washout mechanism characterized by the sphaleron condition,
\begin{align}
    \frac{E_{\text{\tiny sph}} (T_c)}{T_c} \geq 45,\label{generalconstraint}
\end{align}
where $T_c$ is the critical temperature at which the phase transition was initiated. A simpler constraint that avoids the sphaleron energy calculations and is widely used in the literature is,
\begin{align}
    \frac{v(T_c)}{T_c} \geq 1,\label{specialconstraint}
\end{align}
which is an approximation of the general constraint in Eq.\eqref{generalconstraint} valid for the SM and its real singlet extension (RxSM) \cite{braibant1993sphalerons, ahriche2007criterion}. Unfortunately,  the sphaleron condition (in both forms) is plagued with its own uncertainty as the vacuum expectation value is gauge dependent, as we have discussed in section \ref{gauge-section}, and so is the critical temperature. However, the gauge dependence of $v(T)$ and $T_c$ is often amplified by an inconsistent truncation of the loop expansion, e.g. by mixing different perturbative orders or resummations. Following the discussion in section \ref{gauge-section}, this issue is addressed by organizing the calculation in a consistent $\hbar$ expansion: extrema and thermodynamic quantities are determined order-by-order with the appropriate loop contributions included coherently, which removes spurious gauge artifacts associated with loop mixing and yields gauge-independent results for quantities evaluated at the critical point within the chosen truncation [cf. check \cite{patel2011baryon}]. A complementary strategy is to reformulate the order parameter in terms of a manifestly gauge-invariant (GI) composite operator. In Ref.~\cite{buchmuller1994gauge}, one introduces
\begin{align}
    \sigma_{\rm GI} \equiv \Phi^\dagger\Phi,
\end{align}
and constructs an effective potential $V_{\rm eff}^{\rm GI}(\sigma,T)$ by coupling an external source to $\sigma_{\rm GI}$ and performing the corresponding Legendre transform. Since $\sigma_{\rm GI}$ is constant along gauge orbits (BRST invariant), $V_{\rm eff}^{\rm GI}$ is gauge invariant by construction. The sphaleron criterion may then be expressed using a gauge-invariant order parameter, for example
\begin{align}
    \frac{v(T_c)}{T_c}\;\;\longrightarrow\;\;\frac{\sqrt{\langle\sigma_{\rm GI}(T_c)\rangle}}{T_c},
\end{align}
thereby avoiding reliance on the gauge-dependent location of the minimum of $V_{\rm eff}(\phi,T)$.\\

\begin{figure}[htb!]
    \centering
    \includegraphics[scale=0.5]{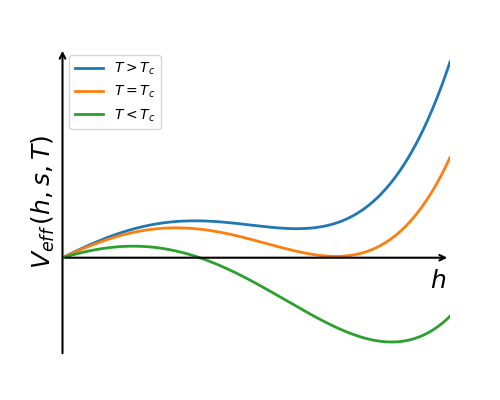}
    \includegraphics[scale=0.5]{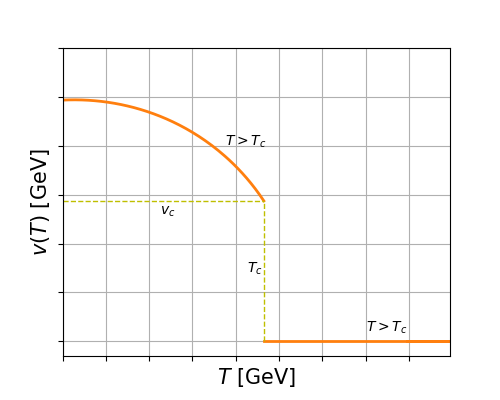}\\
    \includegraphics[scale=0.5]{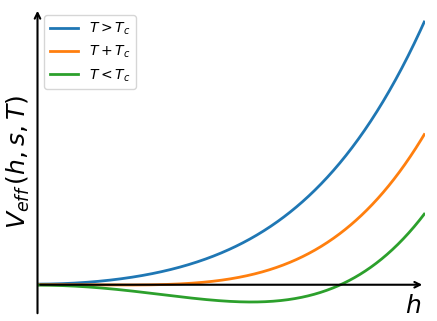}
    \includegraphics[scale=0.5]{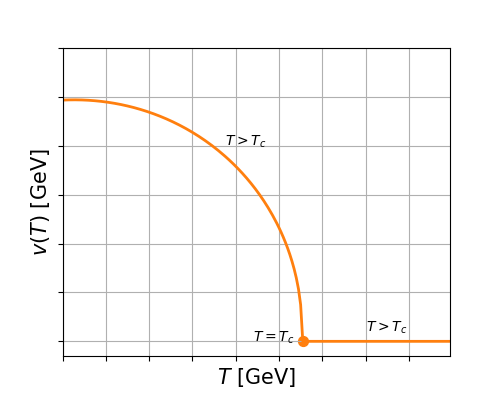}
    \caption{The top row illustrates the potential barrier generated in FOEWP that induces the departure from thermal equilibrium in plasma. On the right, the plot shows the discontinuity of the Higgs $vev$ evolution over temperature in FOEWPT case. Where at $T=T_c$, two degenerate Higgs $vev$s exist, with the broken one satisfying the sphaleron condition $v_c/T_c\geq 1$. The bottom row shows the second-order EWPT case, where the Higgs $vev$ evolves smoothly over temperature, and the whole field decays simultaneously to the broken symmetry once $T<T_c$, due to the absence of the potential barrier.  }
    \label{EWPT_cases}
\end{figure}

\noindent
Now, a gauge invariant calculation of the sphaleron condition in Eq.\eqref{specialconstraint} for the SM case confirms that it cannot generate a first-order EWPT. The only way to generate a barrier between the symmetric and asymmetric Higgs field configurations is through loop corrections, which is only possible if the Higgs boson has a mass of $m_H \leq 50$ GeV. At the critical temperature, the sphaleron condition in Eq.\eqref{specialconstraint} for the SM can be reduced to an upper bound on the Higgs mass \cite{profumo2007singlet,senaha2020},
\begin{align}
    \frac{v(T_c)}{T_c} = \frac{2M^3_W + m^3_Z}{\pi m^2_H}\qquad \longrightarrow \qquad m_h \leq \sqrt{\frac{2m^3_W + m^3_Z}{\pi v_{\text{\tiny EW}}} }.
\end{align}
Lattice calculations had shown that the SM can only lead to a cross-over for $70 \leq m_H \leq 95$ GeV \cite{kajantie1996there}. With the discovery of the Higgs boson at LHC in 2012 with $m_H\sim 125$ GeV, there is no way left to get a FOEWPT from the SM. However, the sphaleron conditions in Eq.\eqref{generalconstraint} and Eq.\eqref{specialconstraint} are sensitive to the beyond SM extensions, which give space for a large phase space volume of extended parameters that allow for the desired FOEWPT scenario. Fig.\ref{EWPT_cases} highlights the differences between a FOEWPT and second order EWPT from the potential barrier separating the symmetric and broken-symmetry Higgs $vev$s, and the behavior of the Higgs $vev$ evolution with temperature.

\noindent
Many beyond SM extensions have been explored in the literature, including real/complex singlet(s) extension, doublet Higgs extension, MSSM extension, and others \cite{barger2009complex, ELLWANGER20101}. One common feature among these different extensions is that they are often of bosonic nature, as the characterizing FOEWPT barrier feature is dominantly controlled by the Higgs field cubic term that can only be produced from the bosonic sector [cf Eq.\eqref{cubictermorigin}]. In this review, we will focus on the real singlet extension (RxSM) since it is one of the most prominent extensions that can generate FOEWPT already at tree level, if it is \zsym-asymmetric, despite being the simplest. The RxSM in its general renormalizable form is represented by,
\begin{align*}
V(H,S) &= -\frac{1}{2} \mu_1^2  (H^\dagger H)  + \frac{1}{4} \lambda _1 (H^\dagger H)^2 + \frac{1}{2} \mu_2^2 S^2  + \frac{1}{4} \lambda_2  S^4 + \frac{1}{2} \alpha S  (H^\dagger H)  \\
& + \frac{1}{4} \lambda_{12} S^2  (H^\dagger H) +  \frac{1}{3} \beta S^3\;. \numberthis \label{reno}
\end{align*}

\noindent
Where the singlet couples directly to the Higgs through the couplings $\alpha, \lambda_{12}$ and indirectly to the other SM particles via mixing with the Higgs, which is dominated by the singlet vacuum expectation value ($vev$). Sending the couplings of the singlet odd terms ($\alpha,\beta$) to zero restores the the \zsym-symmetery of the model, which can still generate a FOEWPT but only through loops corrections, which by definition generate a weaker FOEWPT compared to the \zsym-asymmetric extension for the majority of the parameter space points, which makes testing such a model more challenging. Such an extension could have shifted the Higgs configuration evolution dramatically, depending on the allowed ranges of the impeded free parameters, where the Higgs field would develop richer stationary points due to the coupling with the singlet, where some of them could be degenerate points separated by a maximum stationary point, leading to a FOEWPT. At a later evolutionary time, the Higgs underwent an EWSB, transforming the Higgs field, in the unitary gauge, into $H\to \frac{ 1}{\sqrt{2}}(v+h)$. Assuming that the singlet as well has developed some $vev$ before the Higgs EWSB so that $S\to s + \omega$, the Higgs stationary points will be obtained via the minimization conditions from the coupled equations,
\begin{align}
-\mu_1^2 +\lambda_1 v^2 +\alpha \omega+\frac{1}{2} \lambda_{12}\omega^2 = 0\;,\label{vev1}\\
(2\mu_2^2 + \lambda_{12} v^2) \omega +  2(\beta +\lambda_2 \omega) \omega^2 +\alpha v^2 = 0\;.\label{vev2}
\end{align}
which allows for extra symmetry-breaking Higgs $vevs,$ ($v_b$),
\begin{align}
v_b = \left(\pm \sqrt{\frac{\mu_1^2 - \left( \alpha +\frac{1}{2} \lambda_{12} \omega \right)\omega}{\lambda_1}},\,\,\pm \sqrt{\frac{-2(\mu_2^2 +\beta \omega +\lambda_2 \omega^2)\omega}{\alpha+\lambda_{12}\omega} } \right)\;,\label{xSMvev}
\end{align}
together with the vanishing symmetric one $v_s=0$. Consequently, for different values of the free parameters, the asymmetric $v_b$ gets a better opportunity to satisfy the sphaleron condition in Eq.\eqref{specialconstraint}.At finite temperature, a negative value of the $\alpha$-coupling would shift $v_b$ towards larger magnitudes if $\abs{\alpha} > \frac{1}{2}\lambda_{12}\omega$. Other possible scenarios to satisfy the sphaleron condition would be the suppression of the $\mu_1^2$ coupling due to the thermal correction to the Higgs mass term from the heavy particles' corrections to the Higgs mass at the thermal bath, which in turn decreases the critical temperature, as we will see. \\

\noindent
In the thermal bath, the effective potential becomes,
\begin{align}
    V_\eff(h,s,\bar\mu,T) = V_\tl(h,s) + V_\cw^{(1)}(h,s,\bar\mu) + V_\beta^{(1)}(h,s,T) + V_\rings(h,s,T).\label{vefffinal}
\end{align}
Where $V_\tl(h,s)$ is the tree-level potential in Eq.\eqref{reno} after EWSB, $V_\cw^{(1)}(h,s,\bar\mu),$ $V_\beta^{(1)}(h,s,T),$ $V_\rings(h,s,T)$ are the Coleman-Weinberg 1loop correction, thermal correction and rings (Dasiy) resummed contributions respectively given by,
\begin{align}
    V_\cw^{(1)}(h,s,\bar\mu) &= \sum_{i= h,G,s,W^\pm,Z,t} \frac{n_i}{64\pi^2} m_i^4(h,s)\left( \ln \frac{m_i^2(h,s)}{\bar\mu^2} - r_i\right),\label{combinedCW}\\
    V_\beta^{(1)}(h,s,T) &= \frac{T}{2}\sum_{i= h,G,s,W^\pm,Z,t} n_i \int \frac{d^3p}{(2\pi)^3} \ln \left(1 \mp e^{-\frac{\sqrt{p^2 + m_i^2(h,s)}}{T}} \right),\label{combinedthermal}\\
    V_\rings(h,s,T) &=  -\frac{T}{12\pi}\sum_{i= h,G,s,W^\pm,Z} \Big[m^2(h,s) + \Pi_i(T)\Big]^{\frac{3}{2}} - m^3(h,s).\label{combinedrings}
\end{align}
Where $n_i=1,1,3,6,-12$ is the degree of freedom of Higgs and scalar fields, $G$-goldstones, $Z$-boson, $W^\pm$-boson, and top quark. $r_i$ is the residual finite part from regularizing the UV-divergence of CW-loop, which is $\frac{3}{2}$ for scalars and fermions and $\frac{5}{6}$ for gauge bosons. The plus(minus) sign in Eq.\eqref{combinedthermal} characterizes the fermionic (bosonic) contribution. Eq.\eqref{combinedrings} represents the daisy rings contribution to regularize the IR-divergence that is solely rooted in the bosonic sectors. And $\Pi_i(T)$ is the Debye thermal mass correction [cf \cite{ahriche2007criterion}, \cite{delaunay2008dynamics}],
\begin{align}
    \Pi_h (T)&=\Pi_G =  \frac{1}{16} \left(\frac{3g^2 + g'^2}{4} + y^2_t + 2 \lambda_1 + \frac{1}{3}\lambda_{12} \right) T^2,\\
    \Pi_s (T)&= \frac{1}{12}(3\lambda_s + 4\lambda_{12})T^2,\\
    \Pi^L_W (T)&= \frac{11}{6}g^2T^2, \qquad \Pi^L_B (T)= \frac{11}{6}g'^2T^2\\
    \Pi^T_W (T) &=\Pi^T_B(T) = 0.
\end{align}

\noindent
The sums run over the massive particles alone, despite that all the SM particles will contribute to these corrections. Yet, their contributions, compared to the massive particles, are negligible due to the large mass gap between them. 
The non-analytic behavior of the 1-loop corrections in Eq.\eqref{combinedCW}:\eqref{combinedrings} evidently complicates tracking the Higgs configuration. The parameter set that leads to a FOEWPT has to develop degenerate stationary points separated by a barrier satisfying,
\begin{align}
    &V_\eff(0,\omega,T_c) = V_\eff(v_b,\omega,T_c),\label{degeneracey1}\\
    &\frac{\partial V_\eff(h,s,T_c)}{\partial h}\Big\rvert_{v,\omega}  = \frac{\partial V_\eff(h,s,T_c)}{\partial s}\Big\rvert_{v,\omega} =0,\label{degeneracey2}
\end{align}
which returns a coupled non-linear complicated equations of asymmetric $vevs$ ($v_b,\omega$) that have to be solved numerically in order to extract $v_c,T_c$. Different numerical packages\cite{athron2020phasetracer,wainwright2012cosmotransitions} that can accomplish this task efficiently enough, but unavoidably suffer from the theoretical uncertainties discussed in section \ref{uncertinitiessec}. For this review, we will follow a pedagogical approach to the dynamics of the FOEWPT produced by RxSM, making approximations whenever possible to reduce the problem into an analytically solvable one.\\

\noindent
The masses of the Higgs doublet and the scalar singlet  used in these calculations are obtained from the
\begin{align}
M =
\begin{pmatrix}
M^2_{hh} &  M^2_{hs}\\[7pt]
M^2_{hs} & M^2_{ss}
\end{pmatrix}. \label{masses}
\end{align}
Where,
\begin{align}
    M^2_{hh} &= \frac{\partial^2V_\eff(h,s,0)}{\partial h^2}\Big\rvert_{v},\qquad
    M^2_{ss} &= \frac{\partial^2V_\eff(h,s,0)}{\partial s^2}\Big\rvert_{\omega},\qquad
    M^2_{hs} &= \frac{\partial^2V_\eff(h,s,0)}{\partial h \partial s}\Big\rvert_{v,\omega},
\end{align}
are the potential curvatures obtained at zero temperature. 
Since $M^2_{hs}$ does not vanish due to the interaction terms $\lambda_{12}H^\dagger HS^2,\, \alpha H^\dagger HS$ in Eq.\eqref{reno}, $M$ does not represent the physical masses due to the doublet and singlet scalar mixing. The mass eigenstates are defined by some rotation angle, $\theta$
\begin{align}
\cot \theta = \frac{2M^2_{hs}}{M^2_{hh}-M^2_{ss}+\sqrt{(M^2_{hh}-M^2_{ss})^2+4M^4_{hs}}}. \label{mixingangle}
\end{align}
Diagonalizing $M$ in Eq.\eqref{masses}, gives the physical masses,
\begin{align}
    m_{h,s}^2(v,\omega) = \frac{1}{2} \left[ M^2_{hh} +M^2_{ss} \mp \sqrt{\left(M^2_{hh} - M^2_{ss} \right)^2 - 4M^4_{hs}} \right].
\end{align}
Where we take the lighter mass to be the SM Higgs one, which does reduce to the SM expectation, $m_h = 2\lambda v$, when we set $\lambda_s,\lambda_{12},\mu_s,\alpha,\beta \to 0$.  At high temperature, $v_b(T) < v_{\text{\tiny EW}}$, which is always true since the symmetric Higgs configuration has to be restored at high temperature. Consequently, all the contributing particle masses will be smaller compared to their zero-temperature values. So, it is a viable assumption to adopt the high temperature approximation ($\frac{m_i^2}{T}<1$) to get an analytical value for the non-analytic contribution in Eq.\eqref{combinedthermal}, 
$\mathcal{O}(y^6)$,
\begin{align*}
V_{T,B}^{\text{\tiny HTE}}(v,\omega,T) &= -\frac{\pi^2T^4}{90} + \mathlarger{\mathlarger{\sum}}_{i=G,Z,W^\pm,\atop h,s}n_i\Bigg( \frac{m_i^2T^2}{24} -\frac{m_i^3T}{12\pi} -\frac{m_i^4}{2(4\pi)^2} \left[ \ln\left(\frac{m_ie^{-\gamma_E}}{4\pi T} \right)-\frac{3}{4}\right]\\
& + \frac{m_i^6 \xi(3)}{3(4\pi)^4 T^2} +\mathcal{O}(m^8)\Bigg), \numberthis\label{BHTE}\\[8pt]
V_{T,t}^{\text{\tiny HTE}}(v,\omega,T) &= -\frac{21\pi^2T^4}{180} +\frac{m_t^2T^2}{4} +\frac{6m_t^4}{(4\pi)^2} \left[ \ln\left(\frac{m_te^{-\gamma_E}}{4\pi T} \right)-\frac{3}{4}\right] \\
&- \frac{28m_t^6 \xi(3)}{(4\pi)^4 T^2} +\mathcal{O}(m^8) \numberthis\label{FHTE}
\end{align*}
which is a valid approximation up to $m_i  \lesssim 3T$ as shown in Ref.~\cite{laine2017basics}. On the other hand, the region where $m_i\geq T$ can be neglected as the overall contribution would be exponentially suppressed. However, for precise tracking of the minima evolution, even these exponentially suppressed contributions could affect the minima evolution, leading to excluding more parameter space points, but we are allowed to neglect this effect for this semi-analytical investigation. Evidently, the $m_i^4$ term in Eq.~\eqref{BHTE}, and $m_t^4$ in Eq.~\eqref{FHTE}, almost cancel out the similar term in the CW-contribution in Eq.\eqref{combinedCW}, leaving a residual of $\sim \ln\left( \frac{e^{2\gamma_E}T^2}{\widetilde{\mu}^2} \right)$\footnote{We have referred to this property in section \ref{imaginary-discussion}, where this property was used in canceling the imaginary contributions arising from CW-correction when $m_i^2$ becomes tachyonic for scalars in some regions.}. Qualitatively, that residual part could be checked to have a negligible effect on the Higgs $vev$ evolution. The CW-correction to the Higgs $vev$ corrects the $v^2$-coefficent in Eq.\eqref{vev1} and could be estimated by,

\begin{align}
    \frac{1}{h} \frac{\partial V^{(1)}_\cw(h,s,\bar\mu)}{\partial h} &= \sum_i \frac{n_i \rho_i}{64 \pi^2}\left[ \ln\left( \frac{m_i^2}{\bar\mu^2}\right) -r_i +\frac{1}{2}\right]h^2\;,\label{vcorrection}
\end{align}
where $\rho_i =\frac{m_i}{h}$, is the mass coupling of each particle $i$. This correction will be diluted from the quartic mass term in the thermal corrections in Eq.\eqref{BHTE} and Eq.\eqref{FHTE}. Where, $r_i$ part will exactly cancel and the logarithmic part will almost vanish as $\mathcal{O}\left( \frac{e^{2\gamma_E}T^2}{\bar\mu^2}\right)\sim 1$. Hence the overall shift in Higgs $vev$ from these quartic mass terms will be given by\footnote{$V^{(4)}_\eff(h,s,\bar\mu,T)$ to indicate that we only calculate the shift in Higgs $vev$ from the quartic mass terms in CW-contribution and the high temperature expansion of the thermal correction. }
\begin{align}
 \Delta V_\eff^{(4)}(h,s,\bar\mu) &= V_\eff(h,s,\bar\mu,T) - V_\tl(h,s),
    \intertext{then,}
    \frac{1}{h}\frac{\partial}{\partial h}{\big( \Delta V_\eff^{(4)} (h,s,\bar\mu,T) \big)} &= \frac{n_i \rho_i}{64 \pi^2}\,\, \ln\left( \frac{e^{2\gamma_E}T^2}{\bar\mu^2}\right) \ll 1.
\end{align}
Comparing this correction to the leading $ v^2$-coefficients in Eq.~\eqref{vev1} and Eq.~\eqref{vev2}, it will maximally account for a shift in the Higgs $vev$ of less than $1\%$. This validates dropping all the $ m^4_i$-terms terms resulting from $CW$ and thermal corrections. Consequently, the leading corrections to the effective potential in Eq.\eqref{vefffinal} are the thermal rings contribution and the thermal corrections up to $m^2_i,\,m^3_i$ terms only, which yield the analytic form of the effective potential
\begin{align*}
V_\eff(h,s,T) &=  \frac{1}{2} C(T^2-T_0^2) h^2 - E\, T \, h^3+\frac{1}{2}\alpha h^2s+\frac{1}{4}\lambda_{12}h^2s^2 
+\frac{1}{4}\lambda_{1}h^4+ \frac{1}{2} D(T^2-T_1^2 )s^2\\
&+\frac{1}{3} \beta s^3 +  \frac{1}{4}\lambda_{2}s^4, \numberthis  \label{xsmlag}
\end{align*}
where,
\begin{align}
C &= \frac{1}{4} \left(\frac{1}{4}(3g_1^2+g_2^2)+y^2_t +2\lambda_h +\frac{\lambda_{12}}{3} \right)\;,\label{cvalue}\\
E&= \frac{1}{32\pi} \left( 2g_1^3+ (g_1^2+g_2^2)^{\frac{3}{2}}\right)\;,\\
D&= \frac{1}{12}\left(3\lambda_2 + 4\lambda_{12}\right),\quad T_0^2 = \frac{\mu^2_h}{C},\quad  T_1^2 = \frac{-\mu_s^2}{D}.\label{temp}
\end{align}
The temperature-dependent Higgs $vev$ is now given by
\begin{align*}
v_b(T) &= \Bigg\{\frac{6ET\pm \sqrt{36E^2T^2 - 8\lambda_1 [2C(T^2-T_0^2)+(2\alpha +\lambda_{12} \omega)\omega]}}{4\lambda_1}\;,\\
& \pm\sqrt{\frac{-2[D(T^2-T_1^2) +\beta\omega +\lambda_2 \omega^2]\omega}{\alpha+\lambda_{12} \omega}}\Bigg\}, \numberthis\label{vtemp}
\end{align*}
which is the temperature-dependent version of Eq.~\eqref{xSMvev}. The RxSM correction could be absorbed in $\Omega(T)$, such that,
\begin{align}
    \Omega(T) = \frac{(2\alpha+\lambda_{12} \omega)\omega}{2\lambda_1 T^2 },
\end{align}
Using the relations in  Eq.~\eqref{temp}, the sphaleron condition can be written as, 
\begin{align}
\frac{v_b(T)}{T} = \frac{3E}{2\lambda_1}+\sqrt{\frac{\mu_1^2}{T^2}+\frac{9E^2}{4\lambda^2_1}-\frac{C}{\lambda_1}-\Omega(T)}, \label{spha}
\end{align}
This equation is plotted in Fig.\ref{EWPT_cases}. It reproduces the SM result, $\displaystyle{\frac{v_c}{T_c} = \frac{2E}{\lambda_1} \approx 0.15}$, when dropping the RxSM correction ($\Omega(T) \to 0$), where $\displaystyle{T_c^{\text{\tiny SM}} = \frac{\mu_1}{\sqrt{C\left( 1-\frac{2E^2}{\lambda_1 C}\right)}}}$. Using the degeneracy conditions in Eq.\eqref{degeneracey1} and Eq.\eqref{degeneracey2} on the effective potential in Eq.\eqref{xsmlag}, we get RxSM critical temperature,
\begin{align}
    T_c^{\text{\tiny RxSM}} = \sqrt{\frac{2(\mu_1^2 -\alpha \omega_c) - \lambda_{12} \omega_c^2}{C} }.
\end{align}
Evidently, the RxSM extension can generate a FOEWPT mainly through the correction $\Omega(T)$ in Eq.\eqref{spha}, which is sensitive to the \zsym-asymmetrical extension in Eq.\eqref{reno}. One can say that the portal interactions between the Higgs doublet and the scalar singlet in Eq.\eqref{reno} push the EWPT more towards the first-order type through two competing effects. The portal coupling $\lambda_{12}$ improves the possibility of FOEWPT through increasing the $C$-coefficient in Eq.\eqref{cvalue}, which in turn decreases the corresponding critical temperature. However, large positive values of $\lambda_{hs}$ would suppress the $\Omega(T)$ correction to $v_b(T)$. On the other hand, the $\alpha$-coupling can largely shift $v_b(T)$ towards higher values if it is negative and of a large enough magnitude. That is why we can get much stronger FOEWPT from the \zsym-asymmetrical extension compared to the \zsym-symmetric extension. \\

\begin{figure}
\centering
\includegraphics[scale=0.35]{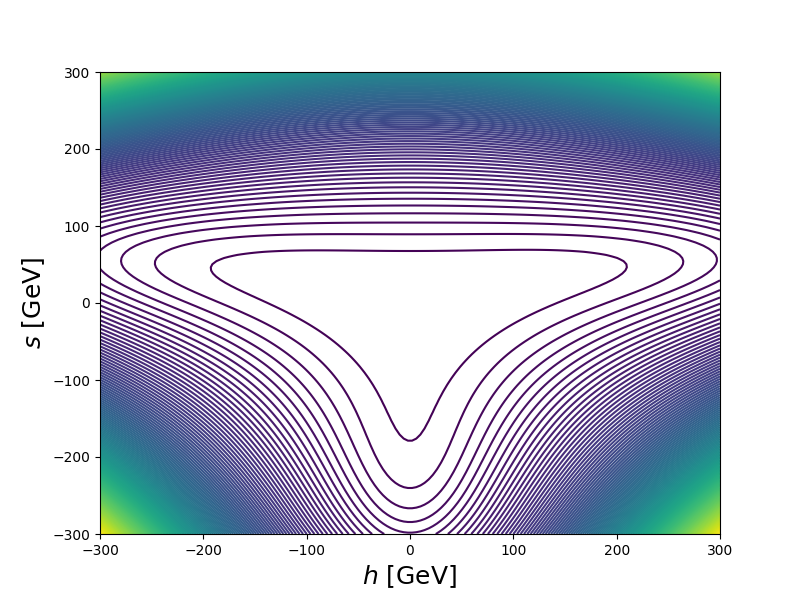}
\includegraphics[scale=0.35]{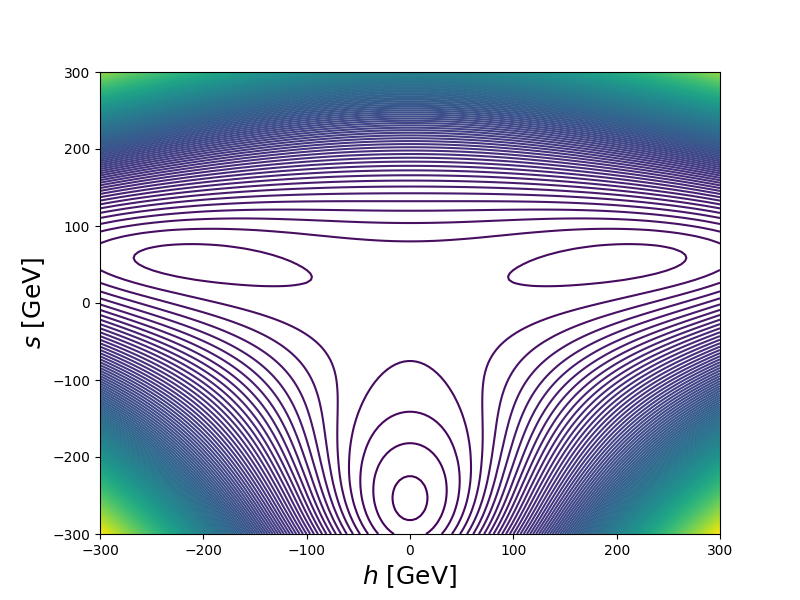}
\caption{Comparison between the effective potential calculations in the HTE-approximation (left) and the complete loop calculations (right). The HTE-approximation does not predict a FOEWPT for that benchmark point in Table \ref{models-comp} due to smoothing out the non-analytic behavior resulting from the CW and thermal corrections.}
\label{HBE_complete_loop}
\end{figure}

\noindent
As discussed earlier, the semi-analytical expressions used above serve primarily as a pedagogical tool to illustrate the qualitative behavior of the RxSM and to identify regions of parameter space that may support a strong FOEWPT. For quantitatively reliable predictions, however, one must evaluate the full finite-temperature effective potential in Eq.~\eqref{vefffinal}, rather than relying on the HTE. The HTE approximation is known to smooth the non-analytic structure of the one-loop potential—most importantly the bosonic thermal contributions responsible for the $-m^3 T$ cubic terms that generate the barrier—and can therefore artificially remove the degeneracy between minima at the critical temperature. This effect becomes increasingly significant for scenarios involving a relatively heavy scalar singlet.
To demonstrate this explicitly, we compute the complete one-loop finite-temperature effective potential of Eq.~\eqref{vefffinal} for the RxSM benchmark point listed in Table~\ref{models-comp}. According to Table~\ref{models-comp}, this parameter set yields a strong FOEWPT with critical values $v_c = 206.58~\mathrm{GeV}$ and $T_c = 100.88~\mathrm{GeV}$, a result clearly confirmed by the contour plot shown in the right panel of Fig.~\ref{HBE_complete_loop}. In contrast, when the same benchmark is analysed using the HTE potential of Eq.~\eqref{xsmlag}, no first-order transition is observed—the thermal barrier is completely smoothed out—demonstrating that the HTE approximation may lead to qualitatively incorrect conclusions. This comparison highlights the necessity of employing the full loop-corrected effective potential when assessing the viability of FOEWPT scenarios in the RxSM.
\vspace{5pt}

\noindent
The full one-loop finite-temperature effective potential in Eq.~\eqref{vefffinal} must be evaluated for any beyond-the-Standard-Model (BSM) scenario under consideration when assessing its capability to realise a strong first-order electroweak phase transition. This requirement is model independent: irrespective of the specific sector extension, reliable predictions for the critical temperature, the order of the transition, and the corresponding critical $vev$s demand the use of the complete loop-corrected potential rather than approximate treatments.\\

\noindent
In Table~\ref{models-comp}, we present a comparison of representative benchmark points that yield a FOEWPT in several well-studied scalar extensions of the Standard Model, including the complex scalar model (CxSM) and the two-Higgs-doublet model (2HDM), alongside the RxSM benchmark discussed above. This comparison highlights how different scalar sectors can accommodate a first-order transition, provided the full thermal effective potential is employed in the analysis.

\begin{table}[tb]
\centering
\begin{tabular}{ccccc} 
\hline
\bf Model & \bf Viable Parameters & $\mathbf{v_b}$  & $\mathbf{T_c}$  & $\displaystyle{\mathbf{\frac{v_c}{T_c}}}$ \\ \hline
 \bf RxSM  &  
    \begin{tabular}{ccccccc}
    $\mu_s$  & $\lambda_h$ & $\lambda_s$ & $\lambda_{hs}$ & $\alpha$ & $\beta$ & $\omega_c$ \\ \hline
    $92.49$ & $0.13$ & $0.85$ & $3.23$ & $-230.1$ & $253.73$ & $54.36$\\
    \end{tabular}  
& $206.58$  &  $100.88$  & $2.05$\\[3pt] \hline 
% \rule{0pt}{1.5\normalbaselineskip}
\bf CxSM\cite{cho2021electroweak} & 
    \begin{tabular}{cccccc}
    $\lambda_h$ & $b_1$  & $b_2$ & $a_1$ & $d_2$ & $\delta_2$ \\ \hline
    $0.511$ & $(107.7)^2$ & $-(178)^2$ & $-6576.17$ & $1.77$ & $1.69$ 
    \end{tabular}  
& $201.5$  &  $106.8$  & $1.89$\\[3pt] \hline 
% \rule{0pt}{1.5\normalbaselineskip}
\bf 2HDM\cite{chaudhuri2021effects} &  
    \begin{tabular}{cccccccc}
    $\lambda_1$ & $\lambda_2$ & $\lambda_3$ & $\lambda_4$ & $\lambda_5$ & $m_{12}$ & $m_H$ & $m_A$ \\ \hline
    $0.258$ & $0.258$ & $-0.23$ & $0.244$ & $0.244$ & $157.3$ & $125$ & $485$
    \end{tabular}  
& $208.25$  &  $160.19$  & $1.3$ \\ \hline
\end{tabular}
\caption{Some valid benchmark points showing the possibility of obtaining a FOEWPT from different BSM extensions. Couplings $\mu_s,\alpha,\beta,\omega_c$ and masses $m_{12},m_H,m_A$ have dimensions of GeV. Similarly are $v_b$ and $T_c$. Couplings $b_1,b_2$ have dimensions of GeV$^2$, while $a_1$ has a dimension of GeV$^3$.}
\label{models-comp}
\end{table}

\subsection{ Different tests for EWPT }
\subsubsection{Colliders approaches}
If a real singlet scalar were responsible for electroweak symmetry breaking, its mass would
naturally be expected in the vicinity of the electroweak scale, \(\mathcal O(10^{2}\text{--}10^{3})~\mathrm{GeV}\)~\cite{ramsey2020electroweak,Papaefstathiou_2021,papaefstathiou2022electro}. 
This energy range is, in principle, accessible at current and next–generation colliders. The LHC
has operated at \(\sqrt{s}=13~\mathrm{TeV}\) without observing evidence for additional scalar states of this
type~\cite{cms2013search}. In a hadron collider, however, only a fraction of the proton–proton center–of–mass
energy is available to the hard scattering, so the production of heavy, weakly coupled singlets
suffers from limited parton luminosities. Several studies indicate that pushing to
\(\sqrt{s}\sim 100~\mathrm{TeV}\) substantially extends the direct–production reach for singlet scalars, both
through increased partonic center–of–mass energy and enhanced vector–boson–fusion rates~\cite{ramsey2020electroweak,papaefstathiou2022electro,Papaefstathiou_2021,buttazzo2018fusing,costantini2020vector}.
Moreover, the complex QCD environment at hadron machines generates large backgrounds that can obscure electroweak signals. These considerations motivate complementary lepton–collider options, in particular a muon collider. Thanks to the muon’s larger mass, synchrotron–radiation losses are strongly suppressed relative to \(e^+e^-\) machines, enabling multi–TeV operation (\(\sqrt{s}\sim 3,\,10,\,30~\mathrm{TeV}\) have been proposed) \cite{Al_Ali_2022,deblas2022physicscase3tev}. In such a setting, a large fraction of the beam energy is delivered to the hard process, and the experimental environment is cleaner than in \(pp\) collisions, facilitating precision studies of scalar resonances and their couplings. Recent sensitivity estimates suggest that a \(\sqrt{s}\sim 3~\mathrm{TeV}\) muon collider could probe sizeable regions of singlet–scalar parameter space via resonant production and vector–boson fusion, with higher energies further extending the reach~\cite{liu2021probing,Aboudonia:2024frg}. 
Therefore, both a \(\sim100~\mathrm{TeV}\) hadron collider and a multi–TeV muon collider (at the \(\sim\!3\)–\(10\)~TeV
scale) offer realistic and technologically plausible avenues for directly testing singlet–scalar
scenarios tied to the electroweak phase transition.\\

 \noindent
In addition to direct production at colliders, a real singlet can be probed indirectly via its
modification of the Higgs self–interaction, an effect that is typically enhanced in \zsym-asymmetric realizations. From the Higgs–singlet interactions in
Eq.~\eqref{reno}, electroweak symmetry breaking (EWSB) induces mixing between the neutral CP–even
states. Writing the gauge–eigenstate fields as \((h,s)\) and the mass eigenstates as \((h_1,h_2)\),
\begin{align}
\begin{pmatrix} h_1\\[4pt] h_2 \end{pmatrix}
=
\begin{pmatrix}
\cos\theta & \sin\theta\\[2pt]
-\sin\theta & \cos\theta
\end{pmatrix}
\begin{pmatrix} h\\[2pt] s \end{pmatrix},
\qquad
c_\theta\equiv\cos\theta,\ \ s_\theta\equiv\sin\theta,
\label{eq:hs-mixing}
\end{align}
with \(\theta\) given by Eq.~\eqref{mixingangle}. The trilinear self–coupling of the SM–like state
(\(h_1\simeq h\) for small \(|\theta|\)) is shifted accordingly,
\begin{align}
\lambda_{hhh}
=\frac{1}{4}\Big[
c_\theta\big(\lambda_1 c_\theta^{2}+\lambda_{12}s_\theta^{2}\big)\,v_{\rm EW}
+(\alpha+\lambda_{12}\,\omega)\,c_\theta^{2}s_\theta
+\frac{4}{3}\big(\beta+3\lambda_2\,\omega\big)\,s_\theta^{3}
\Big],
\label{eq:hhh-RxSM}
\end{align}
where \(v_{\rm EW}\simeq 246~\mathrm{GeV}\), \(\omega\) is the singlet vacuum expectation value, and
\(\{\lambda_1,\lambda_2,\lambda_{12};\,\alpha,\beta\}\) are its a accompanied parameter set. It is convenient to express the deviation from the SM
expectation via
\begin{align}
\kappa_\lambda
\;\equiv\;
\frac{\lambda_{hhh}^{\rm RxSM}-\lambda_{hhh}^{\rm SM}}{\lambda_{hhh}^{\rm SM}}\,.
\end{align}
Current global fits allow sizeable departures; for illustration,
\(-1.4<\kappa_\lambda<7.8\) at \(95\%\)~CL (CMS) and
\(-1.6<\kappa_\lambda<6.6\) at \(95\%\)~CL (ATLAS)~\cite{2025139210,atlas2025study}.
Experimentally, \(\lambda_{hhh}\) is accessed predominantly through double–Higgs production, whose
cross section is at the percent level relative to single–Higgs production, making a precise
measurement challenging at present. High luminosity and/or future energy upgrades are therefore
required to meaningfully constrain (or discover) the singlet–induced modification of the Higgs
self–coupling~\cite{forslund2022high,liu2021probing,carena2020electroweak,no2014probing,profumo2007singlet,Aboudonia:2024frg,profumo2015singlet}.\\

\subsubsection{Gravitational Waves approach}
However, even a discovery of a scalar with couplings characteristic of a first–order EWPT at colliders would not, by itself, prove its role in the thermal history. The most direct cosmological link is a stochastic
gravitational–wave (GW) background whose spectrum is correlated with the same parameter region tested at colliders. Thus, collider indications act as a guide compass for targeted GW searches, helping to prioritize regions of parameter space where the predicted spectra are largest and fall within experimental sensitivity.\\
As the Universe cools below the critical temperature \(T_c\), the barrier between the symmetric and
broken phases diminishes. Thermal fluctuations can then catalyze the transition locally by
nucleating bubbles of the broken phase within the metastable symmetric plasma. The phase
transition effectively begins at the nucleation temperature \(T_n\), which can be appreciably lower than \(T_c\) in scenarios with supercooling. In that case, bubbles nucleated near \(T_c\) may
collapse due to insufficient free energy to overcome surface tension, and the transition proceeds
only once the temperature drops to \(T_n\) so that expanding bubbles are energetically favored.
The nucleation rate per unit volume can be estimated semiclassically as \cite{coleman1977fate,gould2021effective}
\begin{align}
\frac{\Gamma}{V} \sim A(T)\,\,\exp\{-S_E(T)\},\label{nucl_rate}
\end{align}
where, \(S_E=S_3/T\) is the \(O(3)\)-symmetric Euclidean action
\begin{align}
S_3(T)=4\pi\!\int_0^\infty \! dr\, r^2\!\left[\frac{1}{2}\!\left(\frac{d\phi}{dr}\right)^{\!2}\!+V(\phi,T)\right],
\end{align}
evaluated on the bounce solution \(\phi(r)\) \cite{coleman1973radiative,callan1977fate,linde1981fate}. A conventional criterion for percolation is that roughly
one bubble is nucleated per Hubble volume, which, for electroweak–scale temperatures,
translates to
\begin{align}
\frac{S_3(T_n)}{T_n}\;\simeq\; \mathcal O(130\text{--}140)\qquad (T_n\sim 100~\mathrm{GeV}).
\end{align}
Spherically symmetric bubbles do not radiate GWs. Gravitational waves are sourced when
spherical symmetry is broken: e.g.\ by bubble–bubble collisions (detonations), by the
development of sound waves in the plasma when friction deposits energy into the fluid
(deflagrations), and by subsequent magneto–hydrodynamic turbulence \cite{delaunay2008dynamics,ellis2019gravitational,espinosa2010energy}. In all cases, the power
is controlled by the quadrupole of the stress–energy tensor; parametrically, GW amplitudes scale
with the kinetic energy in the bulk motion over the emission timescale. For first–order EWPTs,
the resulting spectrum is well characterized by two macroscopic parameters that can be extracted
from the effective action:
\begin{align}
\alpha \;\equiv\; \frac{\epsilon}{\rho_{\rm rad}}
= \frac{\Delta V - T\,\Delta s}{\rho_{\rm rad}}
= \left.\frac{\Delta V + T\,\frac{\partial V}{\partial T}}{\rho_{\rm rad}}\right|_{T_n},
\qquad
\rho_{\rm rad}=\frac{\pi^2}{30}\,g_*(T)\,T^4,
\end{align}
which measures the released latent heat relative to the radiation energy density and primarily
controls the amplitude; and
\begin{align}
\frac{\beta}{H_n}
\;\equiv\; T_n\,\left.\frac{d}{dT}\!\left(\frac{S_3}{T}\right)\right|_{T_n},
\end{align}
which sets the duration of the transition (thus the characteristic peak frequency), with
\(H_n\) the Hubble rate at \(T_n\).
For detectability with near–future space–based interferometers, one typically requires
\(\alpha=\mathcal O(1)\) and \(\beta/H_n=\mathcal O(10^2)\), yielding peak GW energy densities
\(\Omega_{\rm GW} h^2 \gtrsim 10^{-10}\) in favorable scenarios (see, e.g., reviews \cite{huber2008gravitational,Athron:2023xlk}). In this sense, collider information—whether direct production or precision Higgs–sector deviations—can significantly
sharpen GW search strategies by narrowing the viable ranges of \(\alpha\) and \(\beta/H_n\) tied to the
underlying scalar dynamics.\\

\noindent
The simplified characterization of stochastic GW signals from FO-EWPT in terms of a small set of parameters, such as $\alpha$ and $\beta$ obscures a number of important
theoretical subtleties.  Besides the equilibrium uncertainties discussed in Sec.~\ref{uncertinitiessec}, GW predictions inherit additional uncertainties from the microphysical and hydrodynamic dynamics of bubble expansion in a hot plasma. In particular, the GW power spectrum depends sensitively on the bubble-wall velocity $v_w$, the efficiency with which released vacuum energy is converted
into bulk fluid motion, and on dissipative transport properties of the plasma (shear and bulk viscosities, heat conductivity, diffusion constants) which govern the relaxation and thermalization of the fluid disturbances generated by the expanding wall.
These ingredients describe an intrinsically non-equilibrium system: the moving bubble wall and the associated macroscopic gradients drive the plasma distribution functions away from equilibrium, while microscopic scattering processes act to restore local equilibrium on a finite relaxation time scale. Consequently, a faithful first-principles treatment of friction, relaxation,
and transport necessarily involves real-time finite-temperature quantum field theory, most naturally formulated using the Schwinger--Keldysh (closed-time-path) formalism (Sec.~\ref{realformalism}).\\
This should be contrasted with collider-oriented studies of the EWPT, for which it is typically sufficient to employ the imaginary-time (Matsubara) formalism: the relevant observables, such as the thermal effective potential, thermal masses, and critical temperature, are equilibrium quantities. In GW studies, the imaginary-time formalism is likewise adequate provided
one assumes local thermal equilibrium (LTE) throughout the plasma evolution. LTE requires that the microscopic relaxation (scattering) time be much shorter than the macroscopic time scale associated with bubble growth, so that the distribution functions remain close to their equilibrium form locally.
However, for a strongly first-order EWPT this assumption need not hold parametrically.  A hot relativistic plasma contains processes operating on different time and momentum scales, and the resulting departures from LTE can feed back on the wall through friction and on the fluid through dissipative transport, thereby modifying $v_w$, the hydrodynamic profiles around the bubbles, and ultimately the predicted GW spectrum.  For completeness, we therefore summarize below the dominant sources of bubble--plasma friction and the real-time framework required to compute them systematically (Sec.~\ref{realformalism}).\\

\noindent
Two key structures in the description of bubble nucleation and expansion make the role of out-of-equilibrium dynamics particularly explicit: (i) the prefactor $A(T)$ in the thermal nucleation rate (Eq.~\eqref{nucl_rate}), and (ii) the collision operator in the relativistic Boltzmann equation (Eq.~\eqref{rel_boltzmann}).
The prefactor $A(T)$ is often written in the Langer form as a product of a dynamical growth rate and a statistical fluctuation factor \cite{csernai1992nucleation},
\begin{align}
A(T) = \kappa(T)\,\Omega_0(T),
\end{align}
where $\Omega_0$ accounts for fluctuations about the critical bubble and the associated phase-space volume, while $\kappa$ determines the real-time growth rate of the bubble, and therefore encodes dissipation in the surrounding plasma. For thermal nucleation governed by the three-dimensional critical bubble
$\phi_b(\mathbf{x})$\footnote{We will denote vectors with bold letters instead of arrows in this section for better appearances.}, the statistical factor may be expressed (schematically) as a ratio of fluctuation determinants \cite{ callan1977fate,linde1981fate}
\begin{align}
\Omega_0(T)\ \propto\ \left[ \frac{\text{det}'\left(-\nabla^2 + V_T''(\phi_b,T)\right)}{\text{det}'\left(-\nabla^2 + V_T''(\phi_{\rm false},T)\right)} \right]^{-1/2},\label{Omega0det}
\end{align}
where $V_T(\phi)$ is the finite-temperature effective potential, $\phi_{\rm false}$
denotes the metastable symmetric phase, and the prime indicates the appropriate treatment of zero/negative modes. The dynamical factor $\kappa(T)$ depends on how efficiently the latent heat is
dissipated into the plasma as the critical bubble begins to grow. In viscous hydrodynamic treatments of relativistic nucleation, $\kappa$ can be related to dissipative transport coefficients such as the shear and bulk viscosities, $\eta$ and $\zeta$, together with the surface tension $\sigma$, the enthalpy density difference $\Delta\omega$ across the interface, and the critical radius $R_\ast$,
\begin{align}
\kappa(T)\ \sim\
\frac{4\,\sigma\!\left(\zeta + \tfrac{4}{3}\eta\right)}
{\left(\Delta\omega\right)^2\,R_\ast^{\,2}},
\label{eq:kappaViscous}
\end{align}
up to model-dependent numerical factors and assumptions regarding bubble-growth hydrodynamics.\footnote{This expression is obtained under specific assumptions (\emph{e.g.} near-critical growth limited by viscous damping) and should be viewed as an illustrative example of how real-time dissipation enters the prefactor [cf. \cite{csernai1992nucleation}].}
Microscopically, dissipation and friction also appear in the dynamics of the bubble wall itself. In semiclassical treatments, the scalar background $\phi(x)$ obeys an equation of motion of the form \cite{espinosa2010energy}
\begin{align}
\partial_\mu\partial^\mu \phi(x) + \frac{\partial V_T(\phi)}{\partial\phi} + \sum_i \frac{d m_i^2(\phi)}{d\phi}\,
\int \frac{d^3\mathbf{p}}{(2\pi)^3\,2E_i}\,
\delta f_i(x,\mathbf{p}) = 0,
\label{wallEOMdeltaf}
\end{align}
where $\delta f_i \equiv f_i - f_i^{\rm eq}$ encodes the departure of the plasma distributions from equilibrium induced by the moving wall. Therefore, a quantitative determination of the friction force (the $\delta f_i$ term) requires computing these non-equilibrium deviations. A natural framework to determine $\delta f_i$ is provided by the relativistic Boltzmann equation governing the evolution of the distribution function $f_s(\mathbf{x},\mathbf{p},t)$ for a plasma species $s$,
\begin{align}
\left(\partial_t + \hat{\mathbf{p}}\cdot\nabla_{\mathbf{x}}\right)
f_s(\mathbf{x},\mathbf{p},t) = -\,\mathcal{C}[f], \label{rel_boltzmann}
\end{align}
where the collision operator $\mathcal{C}[f]$ contains the gain and loss terms associated with the relevant microscopic scattering processes. The collision operator $\mathcal{C}[f]$ encodes the microscopic processes that drive the plasma toward equilibrium and therefore constitutes one of the central inputs that must ultimately be determined from real-time finite-temperature dynamics.  In weakly-coupled gauge theories, the correct leading-order collision
kernels involve soft $t$-channel gauge-boson exchange regulated by HTL screening and are most systematically derived using the Schwinger--Keldysh (closed-time-path) formalism.
At leading order, the dominant processes include elastic $2\leftrightarrow 2$ scatterings with a characteristic transport mean-free time $t_{\rm mfp}^{(2\leftrightarrow 2)} \sim 1/{g^4 T}$, which efficiently redistribute momentum and may also change particle species, for example
$q(k_1)\,\bar q(k_2)\to g(k_3)\,g(k_4)$
(species-changing) or $q(k_1)\,\bar q(k_2)\to q(k_3)\,\bar q(k_4)$
(momentum-changing). The corresponding $2\leftrightarrow 2$ contribution to the collision operator for a species $a$ may be written as \cite{arnold2003effective}
\begin{align*}
\mathcal{C}^{(2\leftrightarrow2)}_{a}[f]
&= \frac{1}{4|\mathbf{p}|\,\nu_a} \sum_{bcd} \int_{\mathbf{k},\mathbf{p}',\mathbf{k}'} (2\pi)^4 \delta^{(4)}(p+k-p'-k')\, \big|\mathcal{M}^{ab}_{cd}(p,k;p',k')\big|^2
\\[2pt]
&\times \Big[ f_a(\mathbf{p})\,f_b(\mathbf{k}) \big(1\!\pm\! f_c(\mathbf{p}')\big) \big(1\!\pm\! f_d(\mathbf{k}')\big)
- f_c(\mathbf{p}')\,f_d(\mathbf{k}') \big(1\!\pm\! f_a(\mathbf{p})\big) \big(1\!\pm\! f_b(\mathbf{k})\big) \Big],
\numberthis\label{eq:C22}
\end{align*}
where $\nu_a$ is the degeneracy factor and $\mathcal{M}^{ab}_{cd}$ denotes the transition amplitude for $ab\leftrightarrow cd$.
A second class of processes consists of collinear $1\leftrightarrow 2$ splittings, including bremsstrahlung and in-medium pair annihilation.  While such processes are kinematically forbidden in vacuum for strictly massless particles, they become allowed and parametrically important in a thermal medium due to thermal masses and repeated soft gauge interactions.  Although the splitting amplitude is nominally $\mathcal{O}(g^2)$ suppressed relative to
elastic scattering, the collinear region produces an enhancement of order $1/g^2$, yielding an effective rate 
$t_{\rm mfp}^{(1\leftrightarrow 2)} \sim 1/{g^4 T}$,
comparable to the $2\leftrightarrow2$ contribution.  The quantitative treatment of these splittings requires resumming the interference effects associated with multiple soft scatterings occurring during the formation time of the emitted radiation - the Landau-Pomeranchuk-Migdal (LPM) effect. Accordingly, the full collision operator relevant for transport phenomena may be
decomposed schematically as
\begin{align}
\mathcal{C}[f] = \mathcal{C}^{(2\leftrightarrow 2)}[f]
+ \mathcal{C}^{(1\leftrightarrow 2)}[f].
\end{align}
In the collinear approximation, the $1\leftrightarrow 2$ contribution for species $a$ can be written (up to convention-dependent normalisation factors) as \cite{arnold2003effective}
\begin{align*}
\mathcal{C}^{(1\leftrightarrow 2)}_{a}[f]
&= \frac{(2\pi)^3}{2|\mathbf{p}|^2 \,\nu_a}
\sum_{b,c} \int_{0}^{\infty} \! dp'_{\parallel}\, dk_{\parallel}\;
\delta\!\left(p_\parallel - p'_\parallel - k_\parallel\right)\,
\gamma^{a}_{bc}(p_\parallel; p'_\parallel, k_\parallel)
\\[-3pt]
&\qquad\times
\Big[ f_a(\mathbf{p}) \big(1\!\pm\! f_b(\mathbf{p}'_\parallel)\big)
\big(1\!\pm\! f_c(\mathbf{k}_\parallel)\big) -
f_b(\mathbf{p}'_\parallel) f_c(\mathbf{k}_\parallel)
\big(1\!\pm\! f_a(\mathbf{p})\big) \Big]
\\[5pt]
&\quad +\; \frac{(2\pi)^3}{|\mathbf{p}|^2 \,\nu_a}
\sum_{b,c} \int_{0}^{\infty} \! dp'_{\parallel}\, dk_{\parallel}\;
\delta\!\left(p_\parallel + k_\parallel - p'_\parallel\right)\,
\gamma^{c}_{ab}(p'_\parallel; p_\parallel, k_\parallel)
\\[-3pt]
&\qquad\times
\Big[ f_a(\mathbf{p}) f_b(\mathbf{k}_\parallel)
\big(1\!\pm\! f_c(\mathbf{p}'_\parallel)\big) -  f_c(\mathbf{p}'_\parallel) \big(1\!\pm\! f_a(\mathbf{p})\big)
\big(1\!\pm\! f_b(\mathbf{k}_\parallel)\big) \Big],
\numberthis\label{eq:C12}
\end{align*}
where $p_\parallel$ denotes the momentum component along the direction of the parent hard particle, and the sums account for both bremsstrahlung and pair-annihilation channels.  The splitting/joining functions $\gamma^a_{bc}$ are given in the Arnold--Moore--Yaffe (AMY) real-time formalism by \cite{arnold2003effective,arnold2002photon}
\begin{align}
\gamma^{a}_{bc}(p; p'_\parallel, k_\parallel)
= \big|\mathcal{J}^{(a)}_{p_\parallel + k_\parallel \rightarrow p}\big|^2 \int_{\mathbf{h}} \mathrm{Re}\!\left[
2\,\mathbf{h}\cdot \mathbf{F}_a(\mathbf{h}; p_\parallel, k_\parallel) \right],
\label{eq:gammaAMY}
\end{align}
where $\mathcal{J}^{(a)}$ is the tree-level splitting kernel and
$\mathbf{F}_a(\mathbf{h})$ is the LPM-resummed amplitude describing the interference of multiple soft scatterings during the formation time.  The function $\mathbf{F}_a$ obtained from solving the LPM integral equation \cite{arnold2003effective,arnold2002photon}
\begin{align*}
2\,\mathbf{h} = i\,\delta E(p_\parallel,k_\parallel)\,
\mathbf{F}_a(\mathbf{h}; p_\parallel, k_\parallel)
&+ \int_{\mathbf{q}_\perp} C(\mathbf{q}_\perp)
\Bigg\{ C_a\!\left[\mathbf{F}_a(\mathbf{h})
-\mathbf{F}_a(\mathbf{h}-p_\parallel\,\mathbf{q}_\perp)\right]
+ C_b\!\left[\mathbf{F}_a(\mathbf{h}) -\mathbf{F}_a(\mathbf{h}-k_\parallel\,\mathbf{q}_\perp)\right]
\\
&\qquad\qquad + C_c\!\left[\mathbf{F}_a(\mathbf{h})
-\mathbf{F}_a\!\left(\mathbf{h}+(p_\parallel+k_\parallel)\mathbf{q}_\perp\right)\right] \Bigg\},
\numberthis\label{eq:LPM}
\end{align*}
with $\delta E$ the energy denominator associated with the collinear transition, $C(\mathbf{q}_\perp)$ the soft momentum-transfer kernel, and $C_{a,b,c}$ the quadratic Casimirs of the participating representations.\\

\noindent
Evidently, both the elastic $2\!\leftrightarrow\! 2$ and collinear 
$1\!\leftrightarrow\! 2$ processes depend crucially on real-time finite-temperature propagators. In particular, the dominant $t$-channel contributions to transport are mediated by soft gauge-boson exchange, which must be regulated by hard-thermal-loop
(HTL) screening.  In thermal equilibrium, the corresponding real-time propagators are conveniently expressed in terms of the spectral function,
\begin{align}
G^{>}_{\mu\nu}(Q) = \bigl(1+n_{B}(Q^0)\bigr)\,\rho_{\mu\nu}(Q),
\qquad \rho_{\mu\nu}(Q)
= i\!\left[G_{\mu\nu}^{R}(Q)-G_{\mu\nu}^{A}(Q)\right],
\end{align}
with $G^{R/A}_{\mu\nu}$ the retarded/advanced HTL-resummed propagators. In complete analogy with the contour-ordered scalar propagators in Eqs.~\eqref{gpp}--\eqref{gmm}, the gluon two-point function in the Schwinger-Keldysh (closed-time-path) formalism is described by four contour components $G^{ab}_{\mu\nu}(Q)$ with $a,b\in\{+,-\}$.  The retarded and advanced gluon propagators are obtained as the standard linear combinations of these contour components in Eqs.~\eqref{gpp}--\eqref{gmm}, adjusted by the appropriate Lorentz structure and (adjoint) color indices for the gluon propagator,
\begin{align}
G^{R}_{\mu\nu}(Q) &= G^{++}_{\mu\nu}(Q) - G^{+-}_{\mu\nu}(Q)
 = G^{-+}_{\mu\nu}(Q) - G^{--}_{\mu\nu}(Q), \\[2pt]
G^{A}_{\mu\nu}(Q) &= G^{++}_{\mu\nu}(Q) - G^{-+}_{\mu\nu}(Q)
 = G^{+-}_{\mu\nu}(Q) - G^{--}_{\mu\nu}(Q).
\end{align}
These HTL-resummed real-time correlators provide the screened soft exchange that enters the effective scattering amplitudes $\mathcal{M}^{ab}_{cd}$ appearing in the $2\!\leftrightarrow\!2$ collision operator. Similarly, the soft momentum-transfer kernel, $C(\bvec{q}_\perp)$, entering the LPM-resummed $1\!\leftrightarrow\!2$ splitting functions can be expressed in terms of HTL spectral densities [cf. \cite{arnold2002photon,arnold2001photon}]. In this way, both $\mathcal{C}^{(2\leftrightarrow2)}$ and $\mathcal{C}^{(1\leftrightarrow2)}$ inherit their leading-order structure from real-time HTL-resummed gauge correlators, even though transport ultimately probes only near-equilibrium deviations.
While solving the full nonlinear integro-differential Boltzmann equation is cumbersome, it is not required for hydrodynamic transport coefficients. Instead, one expands the distribution function about thermal equilibrium,
\begin{align}
f_s(\mathbf{x},\mathbf{p},t)
= f_s^{\rm eq}(p) + \delta f_s(\mathbf{x},\mathbf{p},t),
\qquad |\delta f_s|\ll f_s^{\rm eq},
\end{align}
and linearises the Boltzmann equation in $\delta f_s$.  The equilibrium part satisfies detailed balance, $\mathcal{C}[f^{\rm eq}]=0$, while macroscopic hydrodynamic gradients generate a driving term,
\begin{align}
\left(\partial_t+\hat{\mathbf{p}}\cdot\nabla_{\mathbf{x}}\right)f_s^{\rm eq}(p) = S_s(\mathbf{p}),
\end{align}
where $S_s(\mathbf{p})$ is proportional to the appropriate first-order gradient (e.g. the shear tensor, bulk expansion, or a chemical-potential/temperature gradient). The deviation from equilibrium is parametrised as 
\begin{align}
\delta f_s(\mathbf{p}) = - f_s^{\rm eq}(p)\big(1\!\pm\! f_s^{\rm eq}(p)\big)\, \chi_s^{(X)}(\mathbf{p})\,\mathcal{X},
\end{align}
with $\mathcal{X}$ the corresponding hydrodynamic perturbation.  Substituting and keeping only linear terms yields the linearised integral equation
\begin{align}
S_s(\mathbf{p}) = \sum_{s'} \int_{\mathbf{k}}\,
\mathcal{C}^{\rm lin}_{ss'}(\mathbf{p},\mathbf{k})\,
\chi_{s'}^{(X)}(\mathbf{k}),\label{eq:LinearBoltzmann}
\end{align}
where the linearised collision operator retains the full microscopic content,
\begin{align}
\mathcal{C}^{\rm lin} = \mathcal{C}^{(2\leftrightarrow2)}_{\rm lin}
+ \mathcal{C}^{(1\leftrightarrow2)}_{\rm lin}.
\end{align}
Solving Eq.~\eqref{eq:LinearBoltzmann} for the response functions
$\chi_s^{(X)}(\mathbf{p})$ determines the transport coefficients as quadratic functionals. For instance the shear viscosity would be given by \cite{defu2005shear},
\begin{align}
\eta &= \frac{1}{15T} \sum_s \int_{\mathbf{p}}
\frac{p^4}{E_s^2}\, f_s^{\rm eq}(p)\big(1\!\pm\! f_s^{\rm eq}(p)\big)\, \chi_s^{(\eta)}(\mathbf{p}).
\end{align}
Thus, although the Boltzmann equation is solved only in its linearised form, all relevant out-of-equilibrium scattering effects enter through the real-time collision integrals (including HTL screening and LPM resummation).

\section{Summary}

\noindent
Finite-temperature quantum field theory (FTQFT) furnishes the appropriate framework to analyze the electroweak phase transition (EWPT) by promoting vacuum correlation functions to thermal ensemble averages. While the formalism is mature, quantitative predictions retain several systematical uncertinities. Ultraviolet (UV) divergences arise from short-distance sensitivity of loop integrals in a local continuum theory and are renormalized by the same counterterms as at zero temperature; in particular, the temperature-dependent parts of amplitudes are UV-finite. By contrast, infrared (IR) sensitivities are multiplied in bosonic sectors at finite temperature due to long-wavelength modes (e.g.\ zero Matsubara modes and Bose enhancement). In parallel, gauge dependence appears when gauge-dependent intermediates---such as a truncated effective potential---are used as proxies for physical observables.\\

\noindent
We reviewed the imaginary-time (Matsubara) and real-time (Schwinger--Keldysh) formalisms for incorporating thermal effects, and surveyed methods that reorganize perturbation theory in the presence of thermal scales: ring/daisy resummations for scalar sectors, and renormalization-group improvement to reduce residual scale dependence. The effective potential gauge dependence has been discussed together with the master Nielsen identity that fixes this problem order by order in $\hbar$ expansion when it comes to the calculations of critical parameters like vacuum expectation value.\\

\noindent
As an illustrative case study, we examined the real singlet scalar extension (RxSM). The model demonstrates how additional bosonic degrees of freedom can strengthen the transition and realize a first-order EWPT, subject to theoretical consistency (perturbativity, vacuum stability, and thermal EFT validity) and existing experimental bounds. Finally, we summarized complementary probes of a first-order EWPT: direct and indirect collider searches sensitive to modified Higgs self-interactions, exotic decays, and additional scalars; and prospective gravitational-wave signatures from bubble dynamics and plasma effects. Together, these theoretical tools and experimental avenues delineate a realistic program to test electroweak-scale explanations for the BAU while potentially connecting to dark-matter phenomenology.

\section*{Acknowledgments}
CB acknowledges support from the Australian Research Council through projects DP220100643 and LE250100010. MA acknowledges support from the Monash Graduate Scholarship (MGS).

\pagebreak

\appendix

\section{Grassmann representation of the fermionic fields }\label{grassmanalgebra}

In contrast to the bosonic case, the trace in the fermionic case sums only over two states: vacuum ($\ket{0}$), and (anti)particle state ($\ket{\psi},\, \ket{\overline{\psi}}$) due to the anti-commutation relations of the fermionic field creation and annihilation operators
\begin{align}
\{\hat{a},\hat{a}^\dagger \} &= 1,\qquad \{\hat{a},\hat{a} \} = 0,\qquad \{\hat{a}^\dagger,\hat{a}^\dagger \} = 0.\label{fer2}
\end{align}
These relations lead to the famous Pauli exclusion principle, where a fermionic state can not be occupied by two identical states
\begin{align*}
\hat{a}^\dagger \ket{1} &= \hat{a}^\dagger\hat{a}^\dagger\ket{0}
= 0 \numberthis \label{nothing}.
\end{align*}
Consequently, for a single fermionic mode (fixed momentum, spin, and internal quantum numbers), the number operator $\hat n\equiv \hat a^\dagger \hat a$ has eigenvalues $\{0,1\}$. Thus, the single-mode Hilbert space is two-dimensional, spanned by $\{|0\rangle,|1\rangle\!\equiv\!\hat a^\dagger|0\rangle\}$. The full fermionic Fock space is the tensor product over all modes,
\begin{align}
\mathcal{H}_{\text{Fock}}=\bigotimes_{p,s,\alpha}\mathcal{H}_{p,s,\alpha}, 
\end{align}
and is therefore infinite dimensional even though each factor $\mathcal{H}_{p,s,\alpha}$ is two dimensional. In thermal calculations, the trace over $\mathcal H_{\text{Fock}}$ factorizes mode-by-mode. 
The anti-commutation relations imply that squaring an operator (later a field) leads to a vanishing result, which is a property of Grassmann variables. So, in the path integral formulation, the fermionic fields have to obey the Grassmann Algebra, which will lead to the fermionic anti-periodicity property that we referred to earlier. These properties are:
\begin{align}
\psi^2 &= (\psi^*)^2 = 0,\\
\int d\psi &=\, \int d\psi^* = 0 , \qquad \int d\psi \, \psi =\int d\psi^*\, \psi^* = 1,
\intertext{ in addition to the equivalence of the differential and integral operators,}
\int d\psi \, \psi &= \frac{d}{d\psi} \psi = 1.\label{grassmannint}
\end{align}

\noindent
Following \cite{laine2017basics}, we will derive the path integral formulation of the fermion generating functional starting with the modified version of the completeness relation in Grassmann space. We can deduce this relation using the general representation of the fermionic state as a superposition of the two states $\ket{0},\,\ket{1}$,
\begin{align}
\ket{\psi} &= e^{-\psi\hat{a}^\dagger}\ket{0},\\
&= (1- \psi\hat{a}^\dagger)\ket{0},\\
&= \ket{0} -\psi \ket{1}.\label{fieldsuperposition}
\intertext{and similarly,}
\bra{\psi} &= \bra{0} e^{-\hat{a}\psi^*},
\end{align}
Hence,
\begin{align}
\expval{\psi|\psi} &= \expval{0|e^{-\hat{a}\psi^* -\psi \hat{a}^\dagger}|0}\\
&= \expval{0|0} + \expval{1|\psi^* \psi|1} + \expval{1|\mathcal{O}\left((\psi^* \psi)^2\right)|1}\\
&= e^{\psi^* \psi}
\end{align}
Here, we have used $\psi\ket{\psi} = \hat{a}\ket{\psi} = \psi\ket{0}$ and their complex conjugate in the second line as could be derived from Eq.\eqref{fieldsuperposition}.
The completeness relation  in Grassmann space would then be
\begin{align}
\int d\psi_1^* d\psi_2\,\,e^{-\psi_1^* \psi_2} \ket{\psi_2} \bra{\psi_1} &= \int d\psi_1^* d\psi_2 (1-\psi_1^* \psi_2)\Big\{  \Big(\ket{0} - \psi_2\ket{1} \Big)\Big(\bra{0} -\bra{1}\psi_1^* \Big) \Big\}
\intertext{ From Grassmann rules, this expression will be simplified to,  }
\int d\psi_1^* d\psi_2\,\,e^{-\psi_1^* \psi_2} \ket{\psi}\bra{\psi} &=  \int d\psi_1^* d\psi_2\,\, \psi_2\psi_1^* \left[\ket{0}\bra{0} +  \ket{1}\bra{1}\right]\\
 &= \ket{0}\bra{0} +  \ket{1}\bra{1}.\label{th10}
\end{align}
Similarly, we can define the path integral over a queue of operators $\hat{\mathcal{A}}$ to be given from
\begin{align}
\int d\psi_1^* d\psi_2\,\,e^{-\psi_1^* \psi_2} \expval{-\psi|\aop|\psi} 
&=  \int d\psi_1^* d\psi_2\, \left\{ -\psi_1^*\psi_2  \expval{1|\aop|1}  - \psi_1^*\psi_2 \expval{0|\aop|0}  \right\},\\
&= \tz[\aop] .\label{thr11}
\end{align}
Using these modified formulas for the trace and completeness relations, we can evaluate the fermions generating functional in the usual way: inserting an infinite set of intermediate states using the completeness relation, and Wick rotating the time component to Euclidean space. Due to the anti-commutation relation of the fermion field $\psi$, and the Grassmann variable $\theta_i$, we end up with a minus sign in swapping the fields/variables orders, which upon Wick rotation of time will lead to the fermionic anti-commutation property,
\begin{align}
\psi(x,\tau) &= -\psi(x,\tau+\beta)
\end{align}

\noindent
In the evaluation of the Gaussian integral for the fermionic fields, the Grassmann property in Eq.\eqref{grassmannint} changes the Gaussian integral output from $\sqrt{\frac{(2\pi)^n}{\text{det} \,(\bvec{A}} )}$ for the bosonic fields to $\text{det}\, (\bvec{A})$ for the fermionic case, where,
\begin{align}
    \int \, d\overline{\psi}\,d\psi\, e^{-\overline{\psi} a\psi} = \int \, d\overline{\psi}\,d\psi\, (1 - \overline{\psi} a\psi) = a.
\end{align}
Hence, for an n-tuple of the fermionic field $\psi$ and the antifermionic field $\overline{\psi}$, the Gaussian integral would produce
\begin{align}
    \int \, \mathlarger{\mathlarger{\Pi}}_{i=1}^n\,\,d\overline{\psi_i}\,d\psi_i \,\, e^{-\overline{\psi_i} A_{ij}\psi_j} &= \int \, \mathlarger{\mathlarger{\Pi}}_{i=1}^n\,\,d\overline{\psi_i}\,d\psi_i \,\, \sum_{m=0}^n \frac{(-1)^m}{m!} (\overline{\psi_i} A_{ij}\psi_j)^m\label{expanF}\\
    &= \frac{1}{n!} \sum_{\{i_n\}}^n A_{i_1}A_{i_2} \cdots A_{i_{n-1}}A_{i_n}.\\
    &= \text{det}\,(\bvec{A}).\label{gaussian_fermion}
\end{align}

\noindent
Where, only the term with $m=n$ —the one containing all $n$ distinct $\psi_i$ and all $n$ distinct $\bar\psi_i$— survives the Grassmann integral; the antisymmetry of the Grassmann integral produces an overall factor $n!$ that cancels the factor $1/n!$, yielding $\det A$.

\pagebreak

\section{RGE solution \& the scale suppression of the $V_\eff$}\label{RGE-sol}
As an example of the renormalization scale suppression, we work out an example in the SM case. The SM 1PI effective potential in the Landau gauge ($\xi = 0$) is given by
\begin{align*}
    V_\eff(h,\bar\mu) &= -\frac{1}{2}m_1^2h^2 + \frac{1}{4} \lambda h^4 + \frac{1}{64\pi^2} \Bigg\{M_h^4(h)\left[\ln\left(\frac{M_h^2(h)}{\bar\mu^2} \right) - \frac{3}{2}\right]
    + 3M_G^4(h)\left[\ln\left(\frac{M_G^2(h)}{\bar\mu^2} \right) - \frac{3}{2}\right]\\
    &+ 6M_W^4(h)\left[\ln\left(\frac{M_W^2(h)}{\bar\mu^2} \right) - \frac{5}{6}\right] +3M_Z^4(h)\left[\ln\left(\frac{M_Z^2(h)}{\bar\mu^2} \right) - \frac{5}{6}\right]\\
    &-12 M_t^4(h)\left[\ln\left(\frac{M_t^2(h)}{\bar\mu^2} \right) - \frac{3}{2}\right]
    \Bigg\} \numberthis \label{SMeffp}
\end{align*}

\noindent
It is plagued with a twofold scale dependence: the explicit $\bar\mu$-dependence in Eq.\eqref{SMeffp}, and an implicit one from the couplings running, where each CW-correction is accompanied by a UV-divergence as discussed in Eq.\eqref{regularvcw} which is absorbed by renormalizing the corresponding coupling constant through adding an appropriate counter term. This scale dependence is not a physical result as the physical observables have to be independent of the reference point. It is merely a sign of inconsistent truncation of the higher loop corrections. So, once these higher loop corrections are included, the scale dependence shall be eliminated. However, there are infinite classes of these higher-order loop contributions, making their consideration a ridiculous task. Alternatively, the RGE is an intelligent shortcut to sum all the leading logarithms of a fixed order based on the previous physical constraint that physics shall be scale independent.
\begin{align}
    \frac{d}{dt}V_\eff(h,\bar\mu) = 0,\label{RGEconstarint}
\end{align}
which produces a differential equation whose solution effectively suppresses the scale dependence and consequently is equivalent to summing all the leading logarithms for a fixed order. For the SM case, the Callan-Symanzik RGE equation in Eq.\eqref{RGEcallan} would take the form
\begin{align}
    \left(\bar\mu \frac{\partial}{\partial\bar\mu} +\beta_{m_1^2}\frac{\partial}{\partial m_1^2} +\beta_{\lambda}\frac{\partial}{\partial \lambda} +\beta_{g_1}\frac{\partial}{\partial g_1} +\beta_{g_2}\frac{\partial}{\partial g_2} +\beta_{g_s}\frac{\partial}{\partial g_s} +\beta_{y_t}\frac{\partial}{\partial y_t}         - \gamma h \frac{\partial}{\partial h}\right) V_\eff(h,\bar\mu) = 0.\label{RGE2}
\end{align}
The beta and gamma functions are first-order differential equations in the couplings that are obtained from the physical constraint in Eq.\eqref{RGEconstarint},
\begin{align*}
  \frac{d}{dt}V_\eff(h,\bar\mu;m_1^2,\lambda,g_i) &=  \frac{\partial V_\eff}{\partial \bar\mu} \frac{d\bar\mu}{dt} + \frac{\partial V_\eff}{\partial m_1^2} \frac{dm_1^2}{dt} +\frac{\partial V_\eff}{\partial \lambda} \frac{d\lambda}{dt} +\frac{\partial V_\eff}{\partial g_1} \frac{dg_1}{dt} \\
  &+ \frac{\partial V_\eff}{\partial g_2} \frac{dg_2}{dt}+\frac{\partial V_\eff}{\partial g_s} \frac{dg_s}{dt} +\frac{\partial V_\eff}{\partial h} \frac{dh}{dt} . \numberthis   
\end{align*}
Consequently,
\begin{align}
    \frac{d \ln\bar\mu}{d\bar\mu} = 1 \qquad &\Rightarrow \qquad \mu_*=\bar\mu(t)= \bar\mu \,e^t,\\
    \beta_{m_1^2} = \frac{d m_1^2}{dt},\quad\beta_{\lambda} = \frac{d \lambda}{dt},\quad\beta_{g_1} = \frac{d g_1}{dt},\quad &\beta_{g_2} = \frac{d g_2}{dt},\quad \beta_{g_s} = \frac{d g_s}{dt},\quad\beta_{y_t} = \frac{d y_t}{dt},\\
    \gamma = -\frac{1}{h}\frac{d h}{dt} \qquad &\Rightarrow \qquad h(t) = h \,e^{-\int_0^t dt' \gamma (\lambda(t'))} = h\,e^{-\Gamma(t)}.
\end{align}
These beta and gamma functions are extracted from the 1PI loop corrections that renormalize the theory couplings [cf. \cite{cheng1974higgs,gross1973asymptotically}]. They are determined as the coefficients of the UV-divergent term of these 1PI loop corrections. In the $\overline{\bf MS}$ dimensional regularization they are read from the $\frac{1}{\varepsilon}$-term and are given by \cite{ford1993effective},
\begin{align}
    \beta_{m_1^2} &= \frac{m_1^2}{16\pi^2}\left(12\lambda + 6y_t^2 -\frac{9}{10} \left(5g_1^2 + g_2^2 \right) \right) ,\\
    \beta_{\lambda} &= \frac{1}{16\pi^2}\left(\lambda\left(-9g_1^2 -3g_2^2 + 12 y_t^2 \right) + 24\lambda^2 - 6y_t^4 + \frac{3}{8} \left(2g_1^4 + (g_1^2 + g_2^2)^2 \right) \right),\\
    \beta_{g_1} &= - \frac{19}{16 \pi^2}\frac{g_1^3}{6},\\
    \beta_{g_2} &=  \frac{41}{16 \pi^2}\frac{g_2^3}{6},\\
    \beta_{g_s} &= - \frac{7}{16 \pi^2} g_s^3,\\
    \beta_{y_t} &= \frac{y_t}{16\pi^2} \left(\frac{9}{2}y_t^2 -\frac{9}{4} g_1^2 - \frac{17}{12}g_2^2 - 8g_s^2\right),\\
    \gamma &= \frac{1}{16\pi^2}  \left( \frac{9}{4}g_1^2 + \frac{3}{4} g_2^2 - 3y_t^2\right).
\end{align}
Solving these coupled first-order differential equations numerically using the Runge-Kutta method over ranges of the scale $t= \ln\frac{\mu_*}{\bar\mu}$\footnote{Where $\mu_*$ is an arbitrary renormalization scale (e.g., $\mu_*=\kappa\,h$ with $(\kappa=\mathcal{O}(1)$) at which the running couplings are evaluated, using inputs fixed at the $\overline{\mathrm{MS}}$ reference scale $\bar\mu=m_t$}, we get the couplings running as a function of the scale $t$ ($m_1^2(t), \, \lambda(t),\,g_i(t),\, \gamma(t)$). Consequently, the masses introduced in the CW-potential will be redefined according to these new couplings that run with the scale,
\begin{align}
    M_h^2(h;t) &= -m_1^2(t) + 3\lambda(t) h^2(t),\\
    M_G^2(h;t)  &= -m_1^2(t) + \lambda(t) h^2(t),\\
    M_W^2(h;t)  &= \frac{1}{4} g_1^2(t)h^2(t),\\
    M_Z^2(h;t)  &= \frac{1}{4} (g_1^2(t)+g_2^2(t))\, h^2(t),\\
    M_t^2(h;t)  &= \frac{1}{2} y_t^2(t) h^2(t).
\end{align}
Then the RGE-improved SM effective potential becomes
\begin{align*}
    V_\eff(h,\mu_*) &= e^{4\Gamma(t)}\left(-\frac{1}{2}m_1^2(t)h^2(t) + \frac{1}{4} \lambda(t) h^4(t)\right) + \frac{1}{64\pi^2} \Bigg\{M_h^4(h;t)\left[\ln\left(\frac{M_h^2(h;t)}{\mu_*^2} \right) - \frac{3}{2}\right]\\
    &+ 3M_G^4(h;t)\left[\ln\left(\frac{M_G^2(h;t)}{\mu_*^2} \right) - \frac{3}{2}\right]
    + 6M_W^4(h;t)\left[\ln\left(\frac{M_W^2(h;t)}{\mu_*^2} \right) - \frac{5}{6}\right]\\
    &+3M_Z^4(h;t)\left[\ln\left(\frac{M_Z^2(h;t)}{\mu_*^2} \right) - \frac{5}{6}\right]
    -12 M_t^4(h;t)\left[\ln\left(\frac{M_t^2(h;t)}{\mu_*^2} \right) - \frac{3}{2}\right]
    \Bigg\} \numberthis \label{SMeffp2} .
\end{align*}

\begin{figure}[htb!]
\centering
    \includegraphics[width=17cm, height=5cm]{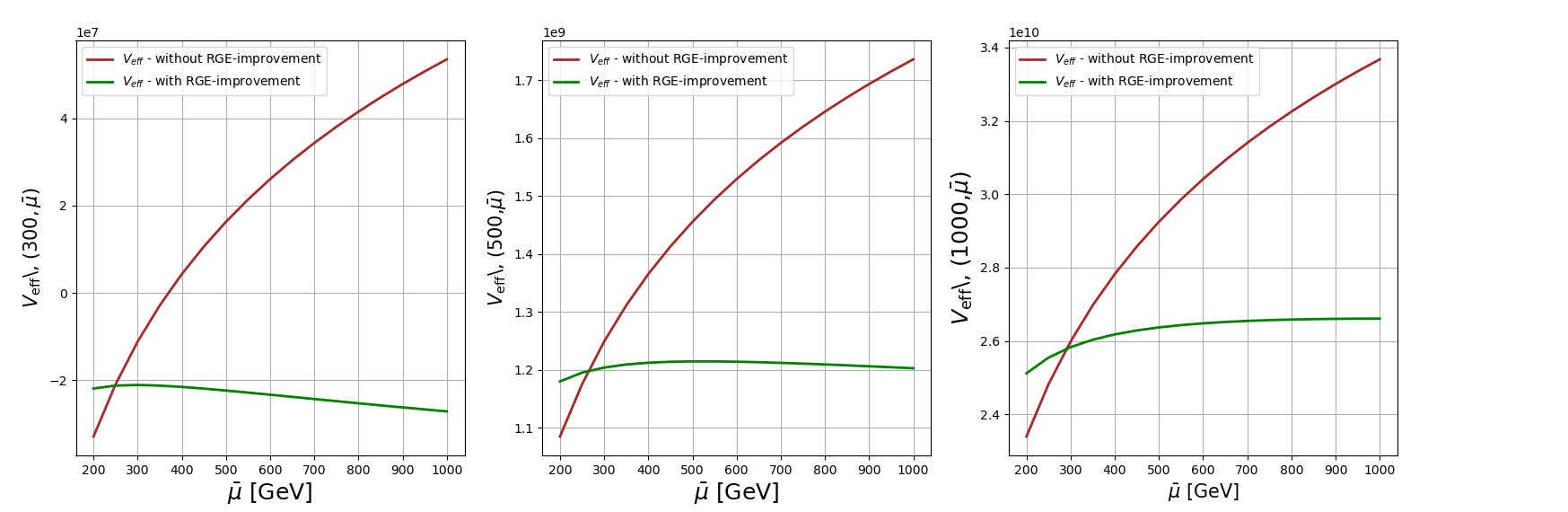}
    \caption{A comparison between the SM effective potential without the RGE enhancement (brown) and the RGE enhanced one (green) at different renormalization scales for three different fixed Higgs field values, $h = 300, 500, 1000$ GeV, respectively.}
    \label{fixed_H}
\end{figure}

\noindent
Here, the implicit scale-running of the  couplings will cancel against the explicit scale dependence in the effective potential as stated in the RGE equation in \eqref{RGE2}. 
\begin{figure}[htb!]
    \centering
    \includegraphics[width=16cm, height=5.5cm]{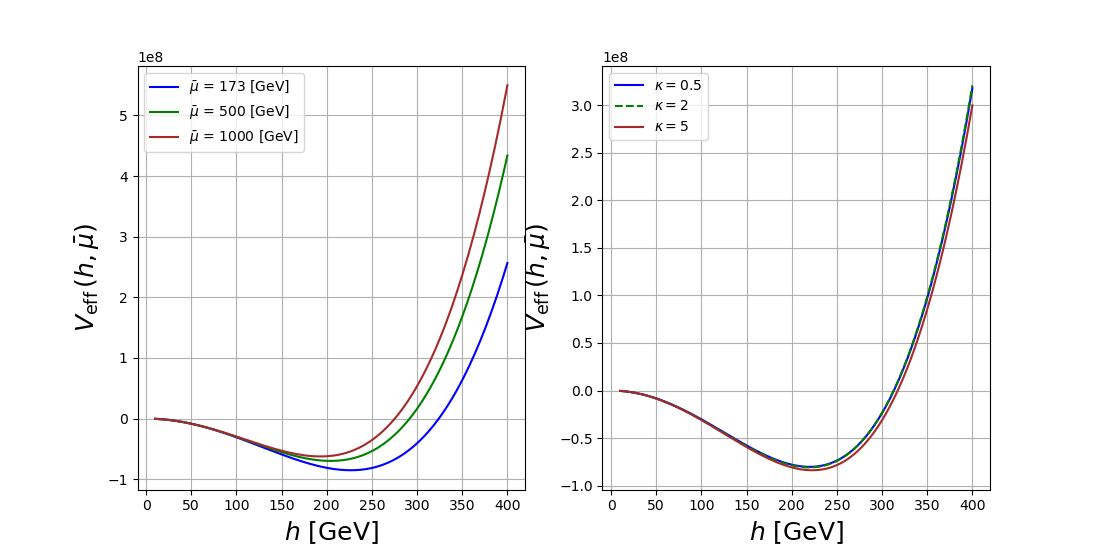}\\
    (A)\hspace{200pt} (B)
    \caption{(A) shows the effective potential in Eq.\eqref{SMeffp} at different renormalization scales of $\bar\mu$ without summing the higher order leading logs contribution of the 1PI loops. As a result, the effective potential exhibits different behaviors for different RG scales. On the other hand, the RGE-solution in Eq.\eqref{SMeffp2} sums these LL-contributions leading to minimal sensitivity for the remaining $\kappa$-dependence as shown in (B).}
    \label{RGE_comp}
\end{figure}
This is evident in Fig.\ref{RGE_comp}, where we show a comparison between the SM effective potential at different fixed renormalization scales with and without RGE enhancement. Evidently, without solving the RGE, the SM is sensitive to the renormalization scale ($\bar\mu$), unlike the RGE improvement, which is now much less sensitive to the value of the arbitrary scale $\mu_*=\kappa\,h$. Still, it shows minimum variation for some $\kappa$ values as shown in Fig.\ref{fixed_H}, which is a sign of missing two-loop contributions. Including these higher-order two-loops and summing their leading logarithms using the RGE equation with the beta and the gamma functions calculated from those two-loop corrections, the minimal remaining $\kappa$-dependence will even be suppressed more and more.

\pagebreak
\bibliographystyle{unsrt}
\bibliography{Reference_list.bib}

\end{document}